\documentclass[%
 aip,
 rsi,
 amsmath,amssymb,
 reprint,
 superscriptaddress,
 floatfix 
]{revtex4-2}

\usepackage[T1]{fontenc}
\usepackage[utf8]{inputenc}
\usepackage[english]{babel}
\usepackage{newtxtext,newtxmath}
\usepackage{upgreek}    
\usepackage{microtype}  

\usepackage[dvipsnames]{xcolor} 
\usepackage{graphicx}

\usepackage{dcolumn}    
\usepackage{bm}         
\usepackage{booktabs}   
\usepackage{multirow}   
\usepackage{tabularx}   
\usepackage{makecell}   

\usepackage{siunitx}
\usepackage{float}       
\usepackage{hyperref}    
\usepackage{cleveref}    

\begin{document}

\sisetup{
  output-decimal-marker = {.},
  per-mode = symbol,
  detect-mode = true,
  separate-uncertainty = true,
  table-align-text-post = false
}

\preprint{Submitted to Review of Scientific Instruments}

\title{Benchmarking Current-to-Voltage Amplifiers for Quantum Transport Measurements}
\author{J. Escorza}
\altaffiliation{Present address: Yachay Tech University, School of Physical Sciences and Nanotechnology, 100119-Urcuquí, Ecuador}
\affiliation{Departamento de Física and Instituto Universitario de Materiales de Alicante (IUMA), Universidad de Alicante, Campus de San Vicente del Raspeig, E-03690 Alicante, Spain.}
\author{G. Pellicer}
\author{T. de Ara}
\altaffiliation{Present address: Institute of Physics, École Polytechnique Fédérale de Lausanne (EPFL), CH-1015 Lausanne, Switzerland.}
\affiliation{Departamento de Física and Instituto Universitario de Materiales de Alicante (IUMA), Universidad de Alicante, Campus de San Vicente del Raspeig, E-03690 Alicante, Spain.}

\author{J. Hurtado-Gallego}
\affiliation{Institute of Condensed Matter and Nanosciences (IMCN/NAPS), Université Catholique de Louvain (UCLouvain), 1348 Louvain-la-Neuve, Belgium.}

\author{E. Scheer}
\affiliation{Department of Physics, University of Konstanz, DE-78457 Konstanz,  Germany.}

\author{C. Untiedt}
\affiliation{Departamento de Física and Instituto Universitario de Materiales de Alicante (IUMA), Universidad de Alicante, Campus de San Vicente del Raspeig, E-03690 Alicante, Spain.}

\author{C. Sabater}
\email[Author to whom correspondence should be addressed: ]{carlos.sabater@ua.es}
\affiliation{Departamento de Física and Instituto Universitario de Materiales de Alicante (IUMA), Universidad de Alicante, Campus de San Vicente del Raspeig, E-03690 Alicante, Spain.}

\date{\today}

\begin{abstract}
Accurate electrical amplification is essential in molecular electronics for measuring conductance through atomic and molecular junctions, where currents often span several orders of magnitude. In this work, we present a systematic design and comparative analysis of four current-to-voltage ($I\text{--}V$) amplifier architectures: single-stage linear, series-linear, logarithmic, and multi-stage cascaded, specifically optimized for break junction (BJ) techniques, including scanning tunneling microscopy (STM-BJ) and mechanically controllable break junctions (MCBJ). Each configuration is evaluated based on sensitivity, noise performance, and dynamic range. Our results characterize the trade-offs between circuit complexity and noise, providing a robust framework and practical guidelines for selecting amplification schemes in quantum transport experiments.
\end{abstract}

\maketitle

\section{\label{sec:intro}Introduction}
A fundamental challenge in molecular electronics\cite{aviram1974molecular,Cuevasbook,Jan2019,natelson2015nanostructures,nitzan2003electron,xin2019, Xiaona24} is the accurate amplification of signals that span several orders of magnitude over very short timescales. This extreme variability occurs independently of the specific transport mechanism, whether it is dominated by coherent tunneling \cite{aviram1974molecular}, where the current ($I$) decays exponentially with length, or by hopping transport \cite{Isma21}.  While some commercial instruments offer high sensitivity, they often require custom-built modifications to achieve the dynamic range and reliability needed for nanoscale data acquisition \cite{Rosenstein,TewariSAB2017,Massee2018,Thermopower,ElkeShot}.

For ohmic conductors, the conductance ($G$) can be determined by the relation $G = I/V$ \cite{ohm1827galvanische}, where $V$ is the applied voltage. For a phase-coherent quantum conductor, the linear conductance is described by the Landauer formalism \cite{Landauer57}: $G = G_\text{0} \sum_i T_i$, where $T_i$ denotes the transmission probability of the $i$-th conduction channel and $G_\text{0} = 2e^2 / h$ is the quantum of conductance \cite{Landauer57,buttiker1986four,vanwees1988quantized,wharam1988one}. In this expression, $e$ represents the elementary charge, $h$ is Planck’s constant, and the factor of 2 accounts for spin degeneracy.
For atomic-sized noble metal contacts (e.g., gold, silver, copper) \cite{Sabater18}, a single conduction channel allows near-perfect transmission ($T \approx 1$), yielding $G \approx G_\text{0} \approx \frac{1}{12\,906~\Omega} \approx 7.75 \cdot 10^{-5}\text{ S}$. When stretched until the metallic bridge breaks, the system enters the tunneling regime. Alternatively, if molecules are present in the surroundings, they can bridge the electrodes, connecting them to form a molecular junction. Here, the transmission probability $T$ often drops exponentially \cite{muller1992,deAra2025measurement}  with the distance ($d$) or molecular length \cite{Reed97,xu2003,Venkataraman2006,Tamara24}: $G \propto e^{-\beta d}$, where $\beta$ is the attenuation coefficient and depends on the molecule. For longer molecules, transport transitions to a hopping regime characterized by a softer inverse dependence on distance ($G \propto 1/d$) and thermal activation ($G \propto e^{-E_a/k_BT}$), where $E_{\text{a}}$ represents the activation energy required for the electron to hop between localized sites along the molecular backbone, $k_{\text{B}}$ is the Boltzmann constant, and $T$ is the absolute temperature \cite{Isma21, Hines2010}. In both regimes, the conductance can drop significantly to $10^{-1} G_\text{0}$ or even $10^{-9} G_\text{0}$ \cite{LathaLow24}. Nevertheless, for the case of conjugated molecules with low connection barriers or resonant tunneling, certain molecular junctions can exhibit conductances approaching the metallic regime  \cite{Smit2002, Kiguchi08, Yelin2016, Edmund24}. 

Bridging the high-transmission metallic regime and the low-transmission molecular junction within a single measurement imposes strict constraints on electronic readouts. The instrumentation must resolve the full conductance range from single-atom contacts to tunneling or molecular junctions, without interruption, handling currents that cover several orders of magnitude. While research groups often rely on either commercial current-to-voltage ($I\text{--}V$) amplifiers\cite{FEMTO,SRSSR570,Keithley428, Otal21, Kumar15} or fully custom-built amplifiers \cite{horowitz2015art, moore2009building,Ornago2023, Pellicer,DeAra22}, the existing literature frequently lacks clarity regarding  which of these amplifiers were used \cite{Sabater12,Kamenetska23}. It is often difficult to discern whether a wide dynamic range stems from the intrinsic design of the amplifier, the number of bits of the data acquisition (DAQ) system, or the use of low-noise bias voltage sources.

Here, we present a side-by-side experimental evaluation of four distinct $I\text{--}V$ architectures: single-stage linear, series-linear, logarithmic, and multi-stage cascaded, all tested under identical break-junction (BJ) conditions. We analyze their circuit designs, derive conductance conversion equations, and quantify their dynamic range, noise performance, and measurement reliability. Ultimately, our results establish practical guidelines for selecting amplification strategies in nanoscale transport experiments, helping to distinguish genuine physical signals from electronic artifacts. This study provides a fundamental framework for the metrology and characterization of nanoscale devices upon which the next generation of quantum technologies is built.

\section{\label{MEtandMAt} Experimental Setups and \texorpdfstring{$I\text{--}V$}{I--V} Amplifiers for Molecular Electronics}

\begin{figure}[!ht]
    \centering
    \includegraphics[width=0.99\linewidth]{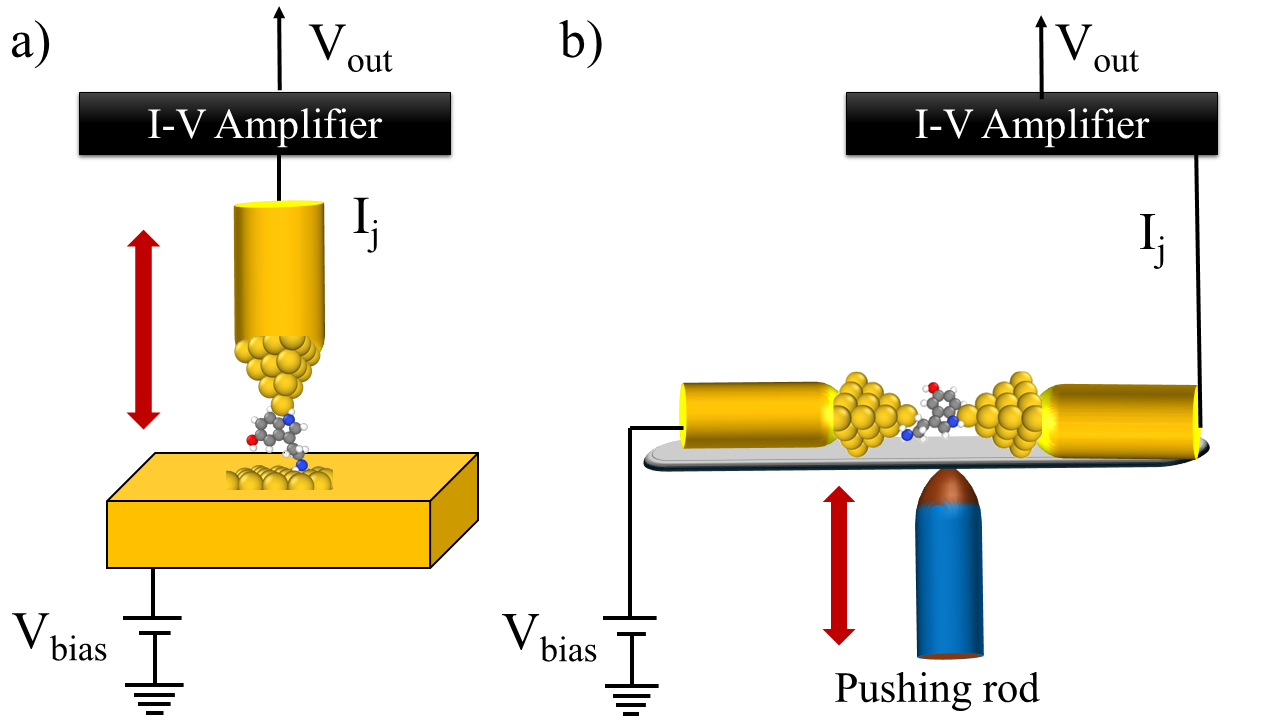}
    \caption{Schematic representations of the experimental setups. (a) STM-BJ and (b) MCBJ configurations showing a single molecule bridged between electrodes; the pushing rod in our setup is a piezoelectric system.}
       \label{fig:setup}
\end{figure}

The two most widely used techniques for measuring quantum transport at the 
single-molecule level are the scanning tunneling microscope in its break junction configuration (STM-BJ) \cite{Pascual1993} and the mechanically controllable break junction (MCBJ) \cite{Krans93, Krans96, muller1992}. Schematic representations of these experimental setups are illustrated in Fig. \ref{fig:setup}. Specifically, panel (a) depicts the STM-BJ configuration, which establishes a junction between a mobile metallic tip and a substrate surface. Panel (b) shows the MCBJ technique, which relies on the precise bending of a flexible substrate to separate two facing electrodes. In these setups, a bias voltage ($V_{\text{bias}}$) is applied to one of the electrodes to drive a junction current ($I_{\text{j}}$) through the sample. This current is subsequently processed by the $I\text{--}V$   amplifier, which scales the input current into a measurable output voltage ($V_{\text{out}}$). The displacement of the electrodes is regulated by a piezoelectric actuator, indicated by red arrows in the schematics.

The experiment is controlled via a DAQ. This system manages both input and output signals: the $V_\text{bias}$ and the piezoelectric voltage ($V_\text{p}$) are generated through the DAQ's analog outputs. $V_\text{p}$ is a triangular (sawtooth) signal that is typically sent to a high-voltage amplifier to be scaled before driving the piezoelectric system, enabling the repeated opening and closing of the junction. Simultaneously, the $V_\text{out}$ from the $I\text{--}V$ converter is fed into one of the DAQ's analog inputs to be recorded and subsequently converted into $G$.

\begin{figure}[!ht]
    \centering
    \includegraphics[width=0.99\linewidth]{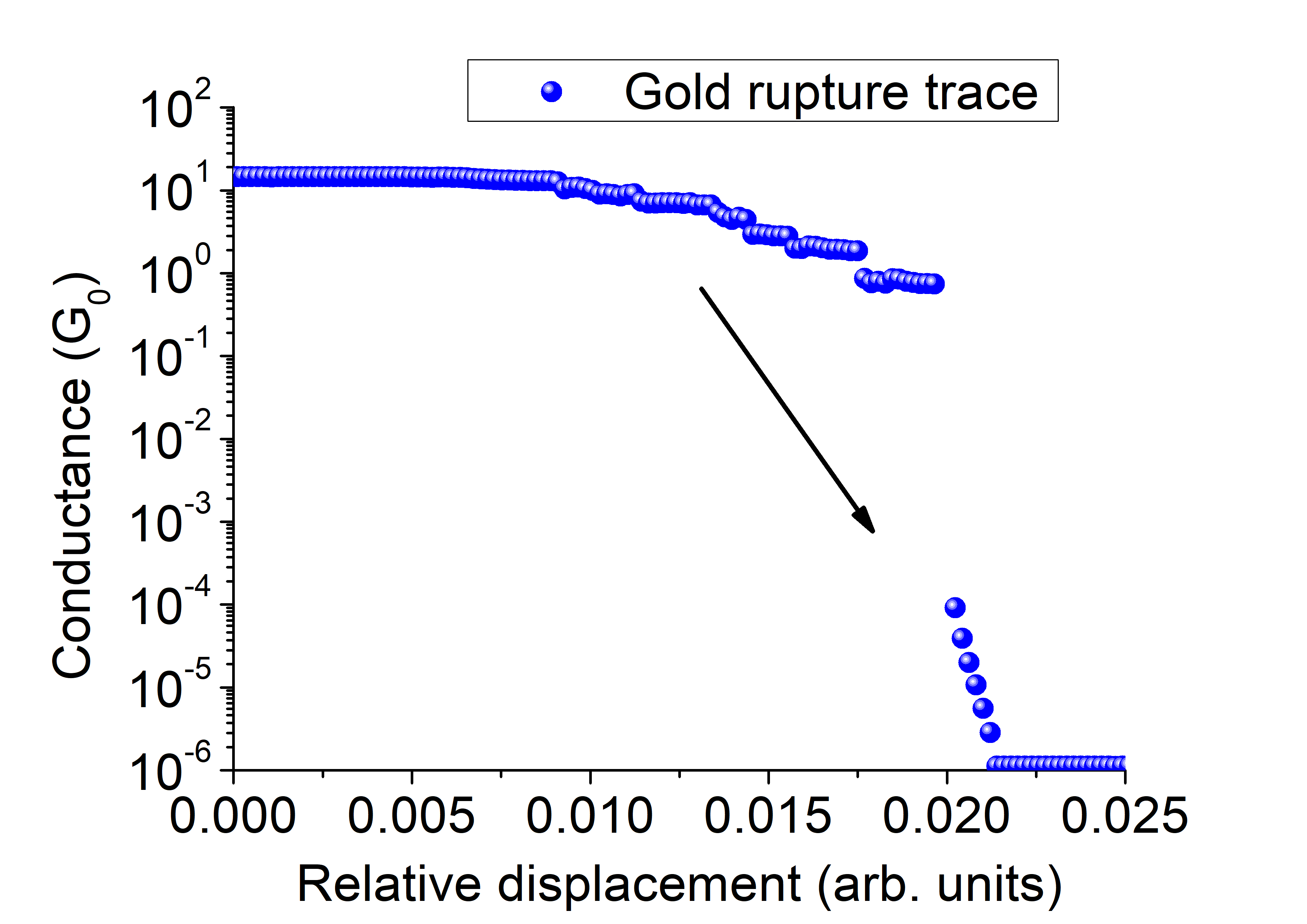}
    \caption{Rupture conductance trace of Au measured at room conditions in MCBJ with a home-made $I\text{--}V$ logarithmic amplifier. The black arrow indicates the direction of the data acquisition flow.}
    \label{fig:Firsttrace}
\end{figure}

Regardless of the setup used (STM or MCBJ), measuring $G$ versus $V_\text{p}$ yields conductance traces. These are termed rupture or formation traces when electrodes are stretched or compressed, respectively. This manuscript focuses exclusively on rupture traces. To ensure that the electron transport remains within the ohmic regime, low voltage biases were applied ($V_{\text{bias}} \leq 100\text{ mV}$). Fig.~\ref{fig:Firsttrace} shows a typical rupture conductance trace of gold (Au) on a log scale, acquired under ambient conditions in an MCBJ setup with a home-made logarithmic $I\text{--}V$ amplifier. There, the metallic junction is stretched and broken in a controlled manner until a single-atom contact forms the metallic bridge (at $1G_0$), after which an exponential decrease in conductance due to tunnel conduction is observed.

\begin{figure}[!ht]
\centering
 \includegraphics[width=1\linewidth]{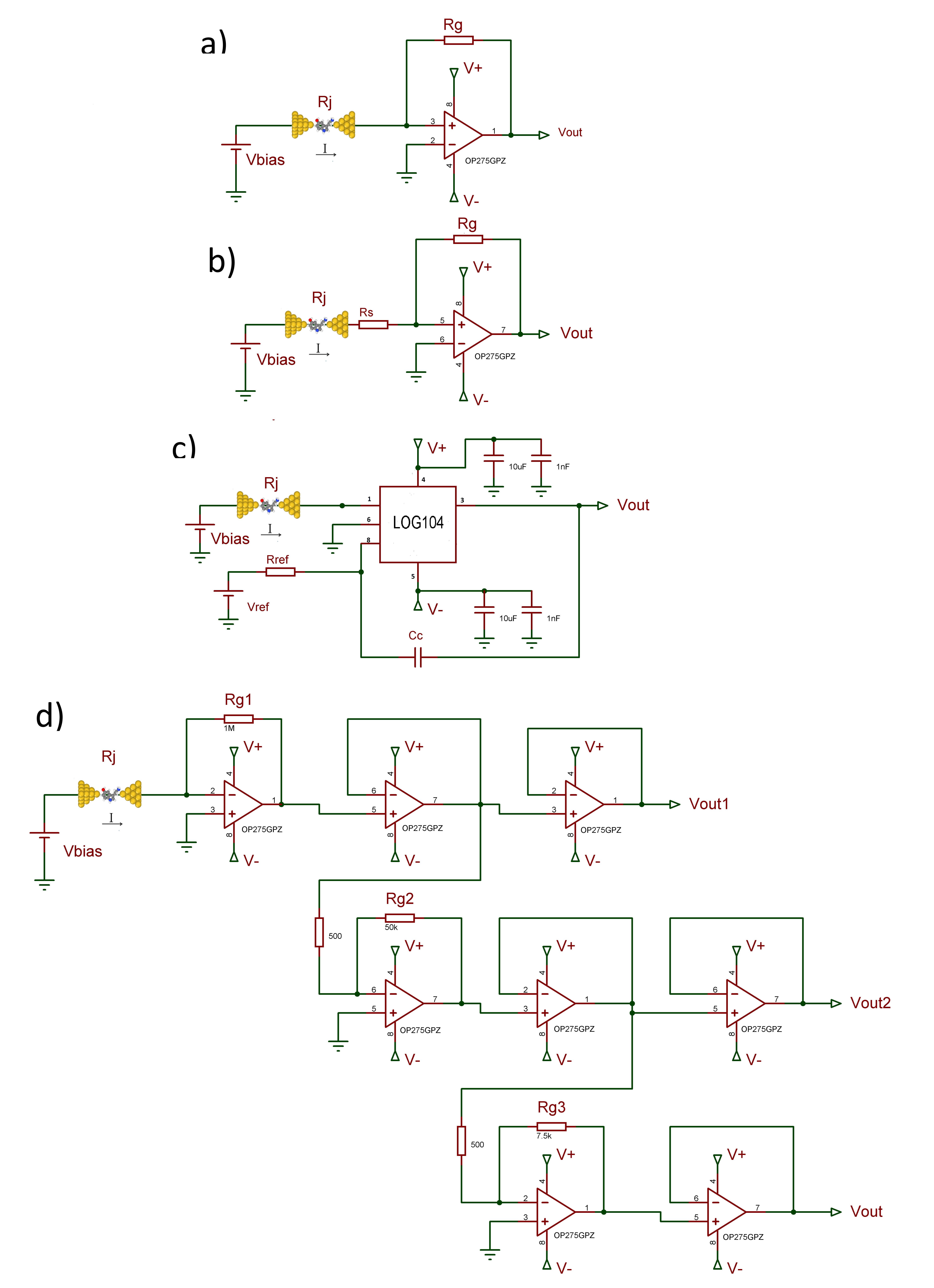}
 \caption{Schematic diagrams of the $I\text{--}V$ amplifiers. (a) ILA, (b) RILA, and (c) ILOGA configurations. The numbers in (c) correspond to the pin configuration of the LOG104 integrated circuit (see Ref.~\cite{DatasheetLog104}). Red parameters highlight the key components and their functional roles. (d) MILAC, providing sequential gains of $10^6$, $10^8$, and $10^9$~V/A. An inverting buffer ($\times (-1$)) is included before the output for optional signal polarity correction.}\label{fig:squemas}
\end{figure}

Accurately measuring conductance in molecular junctions requires solving two key challenges: designing the right $I\text{--}V$ amplifier and converting its output voltage $V_\text{out}$ to conductance units. Here we compare four circuit topologies: three single-stage Op-Amp designs (linear, series-linear, logarithmic) and one multi-stage cascaded architecture for extended dynamic range. These architectures are defined as follows:

The $\textbf{\textit{I--V} Linear Amplifier (ILA)}$ represents the most fundamental topology, utilizing a single operational amplifier (Op-Amp) as a standard transimpedance stage. The configuration shown in Fig.~\ref{fig:squemas}(a) relies on the applied $V_{bias}$ and the junction resistance ($R_\text{j}$) to determine the input current ($I_\text{j}$). The resulting output voltage is determined by the feedback resistor ($R_\text{g}$), which provides the primary amplification gain. The following expression defines the relationship between the output voltage and the circuit parameters:
\begin{equation}
    V_{\text{out}} = -I_{\text{j}} \cdot R_{\text{g}} = -\left( \frac{V_{\text{bias}}}{R_{\text{j}}} \right) \cdot R_{\text{g}}
    \label{eq:ila_gain}
\end{equation}

To extend the measurable conductance range without altering the core hardware, the $\textbf{Resistor + \textit{I--V} Linear Amplifier (RILA)}$ introduces a known resistor ($R_{\text{s}}$) in series between the molecular junction and the amplifier, as illustrated in Fig.~\ref{fig:squemas}(b). In the RILA system, the total current ($I_\text{t}$) entering the amplifier is limited by the sum of the junction resistance ($R_{\text{j}}$) and the series resistance ($R_{\text{s}}$). Consequently,  $V_{\text{out}}$ follows the relationship:

\begin{equation}
    V_{\text{out}} = -I_\text{t} \cdot R_{\text{g}} = -\left(\frac{V_{\text{bias}}}{R_{\text{j}} + R_{\text{s}}}\right) \cdot R_{\text{g}}
    \label{eq:rila_vout}
\end{equation}
This series resistor acts as a voltage divider and a current limiter, effectively protecting the junction and shifting the operating point to prevent saturation during high conductance measurements.

The $ \textbf{\textit{I--V} Logarithmic Amplifier (ILOGA)}$ offers a non-linear approach by converting the input current into a voltage proportional to its logarithm, allowing for signal detection across several orders of magnitude \cite{Ornago2023}. Our implementation utilizes the LOG104 logarithmic amplifier \cite{DatasheetLog104} to compare the $I_\text{j}$ against a fixed reference current ($I_\text{{ref}}$). This generates an output voltage proportional to the base-10 logarithm of their ratio:
\begin{equation}
V_{\text{out}} = k \cdot \log_{10}{\left( \frac{I_\text{j}}{I_\text{{ref}}} \right)}
    \label{eq:iloga_vout}
\end{equation}
where the scale factor $k$ is factory-calibrated to $0.5$ V/decade. The reference current is defined by the reference voltage ($V_{\text{ref}}$) and resistance ($R_{\text{ref}}$). Unlike linear configurations where supply voltages merely dictate hard clipping limits, logarithmic architectures are intrinsically sensitive to their supply lines. Here, the supply voltages  ($V_\text{+}$, $V_\text{-}$) actively establish the operational limits, the low-current stability, and the noise floor of the conversion. According to the LOG104 datasheet \cite{DatasheetLog104}, the permissible input current strictly ranges from \text{10~mA} to \text{100~pA}. This constraint requires a careful configuration of these values to keep the device within its optimal dynamic range. Consequently, the key parameters highlighted in red in Fig. \ref{fig:squemas}(c) specifically $V_{\text{bias}}$, $V_{\text{ref}}$, and $R_{\text{ref}}$ and the supply voltages roles will be discussed in detail in Section \ref{Optimate}.

Finally, the $\textbf{\textit{I--V} Linear Amplifier Cascade (MILAC)}$ is a custom-built, multi-stage architecture designed to bridge the performance gap between linear and logarithmic designs. This system consists of three Op-Amps connected in series, as illustrated in  Fig.~\ref{fig:squemas}(d). Unlike single-stage configurations, this design splits the signal processing into a primary transimpedance stage followed by two subsequent voltage amplification stages. Each one provides an independent output ($V_{\text{out1}}$, $V_{\text{out2}}$, and $V_{\text{out3}}$), allowing the DAQ to record three simultaneous signals with cumulative gains (typically $10^6$, $10^8$, and $10^9$ V/A).  

As is shown in  Fig. \ref{fig:squemas}(d),  the first stage is an inverting transimpedance amplifier (TIA) that converts the $I_\text{j}$ into a voltage. Using a feedback resistor ($R_{G1}$) of 1 M$\Omega$, it achieves an initial gain of $10^{6}$ V/A. The resulting output voltage is described in Equation \ref{ec:VoutMilac1}:
\begin{equation}
V_{out1} = -I_{j} \cdot R_{G1}
\label{ec:VoutMilac1}
\end{equation}

The second stage amplifies the TIA signal using a non-inverting configuration. To preserve signal integrity and to minimize the noise floor, an Op-Amp is utilized due to its ultra-low noise specifications. With a nominal gain ($\text{Gain}_{2}$) of 100, the circuit achieves a cumulative sensitivity of $10^{8}$ V/A. The output voltage is governed by:
\begin{equation}
V_{out2} = V_{out1} \cdot \text{Gain}_{2}
\label{ec:VoutMilac2}
\end{equation}

The final stage provides an additional amplification step to achieve the detection of extremely low conductance levels. By applying a second gain ($\text{Gain}_{3}$) of 10, the system reaches a total cumulative sensitivity of $10^{9}$ V/A. This stage is crucial for resolving signals near the noise floor. The final output voltage is given as:
\begin{equation}
V_{out3} = V_{out2} \cdot \text{Gain}_{3}
\label{ec:VoutMilac3}
\end{equation}

Although not explicitly shown in the simplified schematics of Fig.~\ref{fig:squemas}, the circuit configurations ILA, RILA, and MILAC architectures inherently produce a negative $V_{\text{out}}$ due to their inverting transimpedance nature. Consequently, these stages would strictly require an additional inverting buffer (an $\times(-1)$) to correct the signal polarity for standard data acquisition. In contrast, the ILOGA naturally handles the signal through its internal logic, where the resulting negative sign from the logarithmic conversion is part of its operational framework, rendering an additional hardware inverter unnecessary.

To maintain clarity in the following sections and simplify the presented calculations, we will assume that any negative $V_{\text{out}}$ is corrected to positive. A more detailed discussion regarding the specific implementation of these buffer stages, particularly within the MILAC architecture, is provided in the section \ref{Hard} and supplementary material. 

In terms of data acquisition, the performance of the proposed architectures was evaluated using two different DAQs. The ILA, RILA, and ILOGA configurations were interfaced with an NI PCIe-6363 DAQ \cite{NIPCI6363Specs, ni6363}, in contrast, the MILAC architecture was characterized using an NI PCI-6289 DAQ \cite{NIPCI6289Specs, NIPCI6289Manual}

\section{\label{Hard} Hardware of the \texorpdfstring{$I\text{--}V$}{I--V} Amplifers}

This section provides exhaustive circuit schematics to identify the electronic elements that determine noise, saturation, and the measurable conductance range.

\begin{figure}[!ht]
    \centering
    \includegraphics[width=0.95\linewidth]{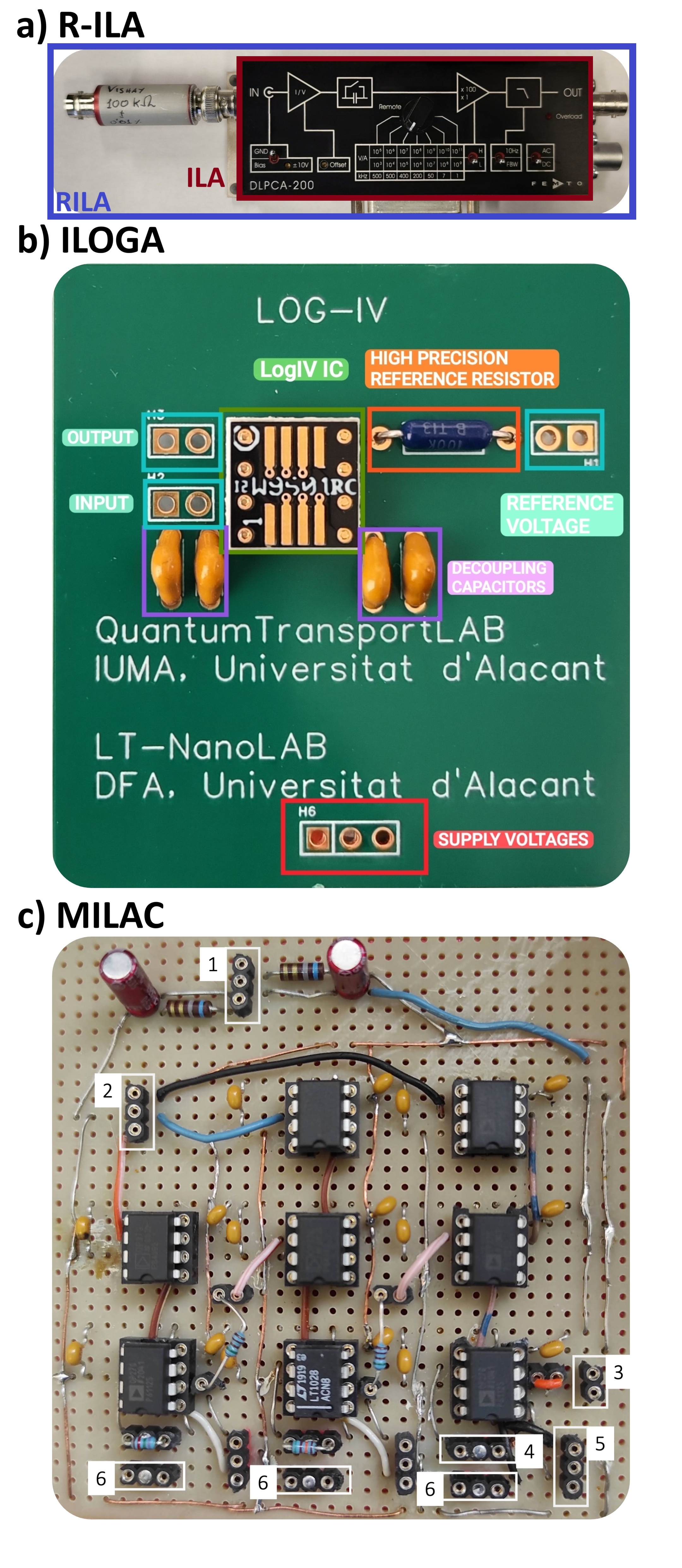}

\caption{Photographic images of the experimental hardware implementations. (a) ILA (red rectangle) and RILA (blue rectangle) setup showing the commercial FEMTO DLPCA-200 amplifier and the gray capsule that can contain the series resistance. (b) Detailed view of the custom-designed ILOGA PCB with color-coded rectangles and labels. (c) Protoboard and connections of the  MILAC architecture and its labels.}
    \label{fig:HardwareOneStage}
\end{figure}

Hardware implementation was divided into two categories: commercial standard solutions and custom-built architectures. For the ILA and RILA, we employed the FEMTO DLPCA-200\cite{FemtoDLPCA200} (see Fig.~\ref{fig:HardwareOneStage} (a)). Conversely, the ILOGA (see Fig.~\ref{fig:HardwareOneStage}~(b)) and MILAC (see Fig.~\ref{fig:HardwareOneStage} (c)) architectures were developed in-house. These custom circuits were assembled on shielded prototyping boards utilizing low-noise sockets and metal film resistors to minimize thermal noise. The entire assemblies are housed in grounded aluminum enclosures acting as a Faraday cage to shield the high-impedance inputs from electromagnetic interference.

The physical implementation of the single-stage architectures is illustrated in Fig.~\ref{fig:HardwareOneStage} (a) and (b). For the ILA configuration, the setup is straightforward: $I_\text{j}$ is fed directly into the BNC input of the FEMTO amplifier, while the resulting $V_\text{out}$ is connected to the DAQ analog input for recording. To implement the RILA architecture, a custom-made resistance box (highlighted by the gray capsule in Fig.~\ref{fig:HardwareOneStage} (a)) is inserted in series between the junction and the amplifier. This gray capsule contains the series resistor, $R_\text{s}$, which limits the total current entering the FEMTO. In the ILA and RILA cases, the amplifier's output remains connected to the DAQ, ensuring a consistent signal path across all linear measurements.

The architecture of the custom-designed ILOGA PCB is presented in Fig.~\ref{fig:HardwareOneStage} (b), where a color-coded scheme identifies the functional regions of the board. The green rectangle, labeled as LogIV IC, designates the mounting site and solder for the chip. These regions contain $10\,\upmu\text{F}$ and $1\,\text{nF}$ capacitors connected in parallel to decouple the chip from power supply noise, as specified in the device's technical documentation and illustrated in the circuit schematics of Fig.~\ref{fig:squemas} (b). The signal interface is organized into distinct functional blocks: the cyan-coded region on the left contains the input pins for the $I_\text{j}$ and the ground connection, while the corresponding output pin $V_\text{out}$ matches the pin 3  of   Fig. \ref{fig:HardwareOneStage} (b). On the right side of the board, a cyan rectangle indicates the $V_\text{ref}$ input, which is essential for the current comparison process. Proper power distribution is managed through the red-highlighted pads. This multi-layer design minimizes parasitic inductance by avoiding surface routing for power distribution. Finally, the reference current signal is established by a high-precision $100\,\mathrm{k\Omega}$ ($0.1\%$) resistor \cite{Vishay}, highlighted in orange.

As illustrated in Fig.~\ref{fig:HardwareOneStage} (c), the MILAC architecture employs a primary high-precision transimpedance stage with a gain of $10^6$~V/A. This is followed by two sequential voltage amplification stages with gains of $100$ and $10$, respectively, achieving cumulative sensitivities of $10^8$ and $10^9$~V/A at the final outputs.  To ensure signal integrity, inverting buffers are integrated between the stages to maintain positive polarity and provide impedance isolation. A detailed circuit schematic, including component values and connectivity, is provided in the Supplementary Material (see Fig.~\ref{fig:3etapas}).
According to the MILAC architecture shown in Fig.~\ref{fig:HardwareOneStage} (c), the physical implementation comprises three columns of aligned Op-Amps, where the OP27GPZ  model \cite{OP27Datasheet} is generally used for amplification and buffering, except for  the second stage, which employs an LT1028  Op-Amp \cite{LT1028Datasheet} due to its superior low-noise characteristics. The board layout includes specific interfaces for operation: The input current is applied via connector labeled as 3, with an initial amplification selectable between 1 and 10~M$\Omega$ (connector~4), however, we have used the first option. A stable $\pm$15~V power supply is provided through the connector labeled as ~1, filtered by an LC circuit to remove low-frequency components such as 50~Hz line noise.  The output voltages ($V_{out,i}$) from each stage are accessible through connector 2, which consists of a three-pin header, with each pin dedicated to a specific output signal. Additionally, connector~5 allows for fine-tuning of the noise offset via a potentiometer, while connector~6 enables the insertion of a capacitor to form an optional RC low-pass filter.

\begin{figure}[!ht]
    \centering
    \includegraphics[width=0.99\linewidth]{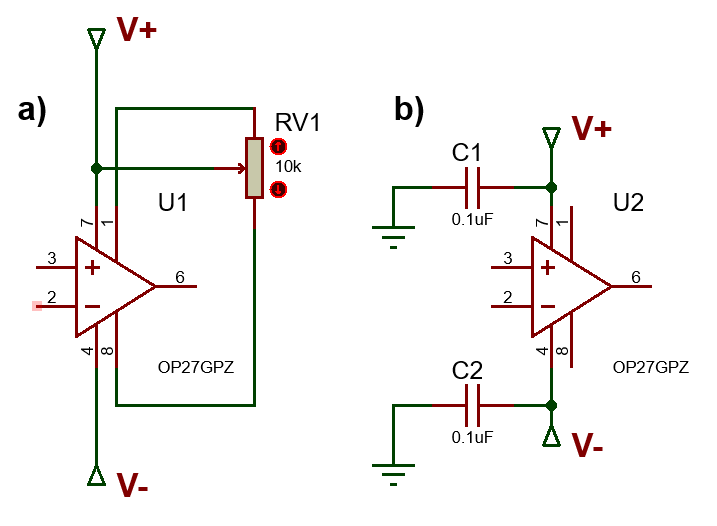}
    \caption{Conditioning circuits for the MILAC architecture: (a) offset nulling stage and (b) power supply high-frequency noise decoupling diagram.}
    \label{fig:polarization}
\end{figure}

Proper signal conditioning is critical for the MILAC configuration to ensure an accurate conductance baseline. While commercial amplifiers, such as the FEMTO unit used for the ILA and RILA setups, feature integrated manual offset correction, the custom multi-stage nature of the MILAC necessitates a dedicated hardware implementation. Although commercial solutions employ similar principles, in this manuscript, we focus on the specific architecture of our circuit to provide full transparency of the signal path. As illustrated in Fig.~\ref{fig:polarization}(a), a dedicated offset nulling circuit is implemented. This adjustment is performed via a $10\,\mathrm{k\Omega}$ potentiometer labeled as RV1, which allows for the calibration of the amplifier's baseline to ensure signal integrity across the high-gain stages. Furthermore, Fig.~\ref{fig:polarization}(b) details the power supply decoupling strategy designed to suppress high-frequency noise. This configuration employs $0.1\,\mu\mathrm{F}$ ceramic bypass capacitors, labeled as C1 and C2, connected between the positive ($V+$) and negative ($V-$) voltage rails and the Op-Amp.

As previously described, the signal path in the MILAC architecture splits after each buffer stage: one path is routed directly to the analog-to-digital converter (ADC) channels ($V_\text{out,i}$ for $i=\{1,2,3\}$), while the other serves as the input for the subsequent amplification stage. Each output is bounded by specific saturation limits and noise floors. To overcome these individual constraints and reconstruct a single broad-range signal, a software-based merging algorithm is required. Specifically, the lower detection limit of stage $i$ must be seamlessly transitioned into the high-sensitivity region of stage $i+1$. This process requires a calibration to convert the raw voltage outputs into conductance units ($G_\text{0}$). The detailed calibration procedure and the mathematical conversion of $V_\text{out}$ to $G$ are discussed in Section \ref{MAtG}. The Supplementary Material includes the LabVIEW front panel and subVI required to stitch the three signal segments together seamlessly without discontinuities (see Figs. \ref{fig:MILCAStich} in the Supplementary Material).

\section{\label{MAtG} CONVERTING OUTPUT VOLTAGE TO CONDUCTANCE}
In all four setups, the DAQ system acquires the amplifier's $V_{\text{out}}$. This section explains how to convert this signal into $G$. For simplicity, the following ILA and RILA calculations assume a positive value. This applies regardless of the actual polarity or the presence of an inverting stage.

\textbf{ILA}

Following Equation \ref{eq:ila_gain}, the DAQ measures $V_{\text{out}}$ using a single analog input channel. The measurement range is $\pm 10\,\text{V}$. The junction conductance $G_j$ (in units of $G_0 = 1/12906 \text{ S}$) is expressed as:

\begin{equation}
G_\text{j} = \frac{1}{R_\text{j}} = \left( \frac{V_{\text{out}} \cdot 12906}{V_{\text{bias}} \cdot R_\text{g}} \right) [\Omega \cdot G_\text{0}]
\label{eq:Gj_norm}
\end{equation}

\begin{figure}[htp]
    \centering
    \includegraphics[width=0.99\linewidth]{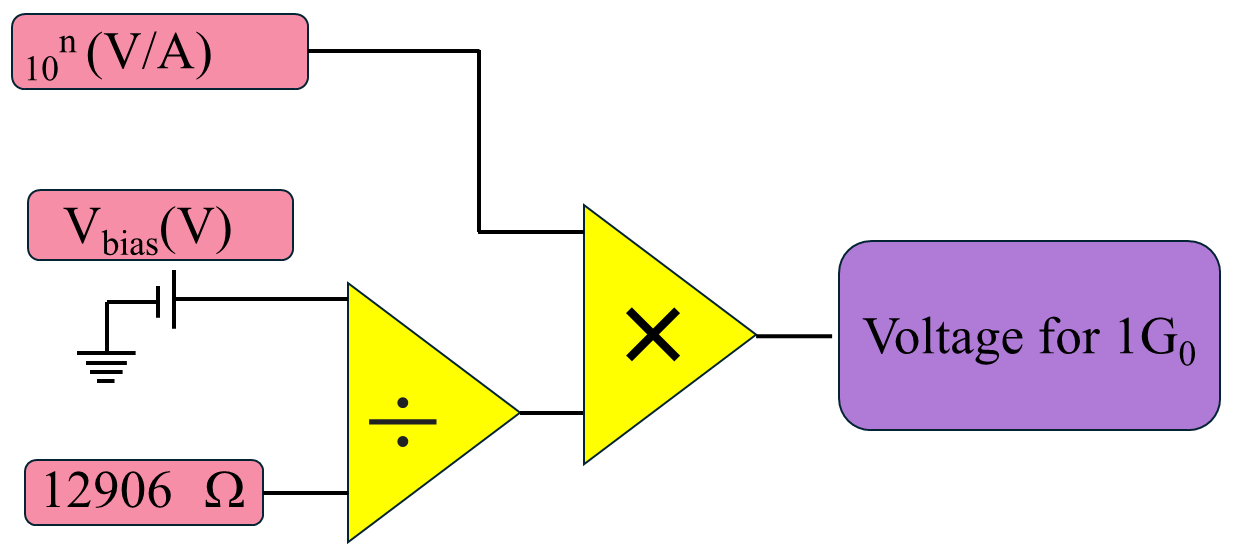}
    \caption{Calculation flow to determine the  $V_\text{out}$) in volts for a conductance of $1G_{0}$. The algorithm incorporates the $V_\text{bias}$, the gain ($10^{n}$ V/A), and the quantum of resistance constant ($12906\ \Omega$) as primary input parameters.}
    \label{fig:WhereisG}
\end{figure}

A practical approach is often more intuitive when using commercial instrumentation. The gain is technically defined by $R_\text{g}$. However, devices like the FEMTO amplifier specify a nominal amplification factor in powers of ten ($10^n$ V/A). Figure \ref{fig:WhereisG} illustrates this calculation workflow. This method helps researchers easily establish a reference `Voltage for $1\,G_\text{0}$'. This benchmark allows for quick identification of expected voltage levels on an oscilloscope or DAQ interface. It also ensures that measurements stay within the linear range and below the 10 V saturation threshold of the acquisition system. Finally, Table \ref{tab:vbias_amplification} provides a systematic roadmap. It details specific bias and gain combinations to validate the amplifier’s response across various conductance regimes.

\begin{table}[htp]
    \centering
    \caption{ $V_\text{out}$ (V) for a \(1\,G_\text{0}\) conductance at various \(V_{\text{bias}}\) and amplification factors. Values in red denote voltages at which the amplifier would saturate.}
    \label{tab:vbias_amplification}
    \setlength{\tabcolsep}{10pt} 
    \begin{tabular}{c c c c}
        \toprule
        \multirow{2}{*}{\textbf{\(V_{\text{bias}}\) (V)}} & \multicolumn{3}{c}{\textbf{\(V_{\text{out}}\) (V) at Amplification Factor}} \\
        \cmidrule(l){2-4}
         & \(\mathbf{10^{5}}\) & \(\mathbf{10^{6}}\) & \(\mathbf{10^{7}}\) \\
        \midrule
        0.01 & 0.774 & 7.74 & \textcolor{red}{77.4} \\
        0.05 & 3.87  & \textcolor{red}{38.7} & \textcolor{red}{387}  \\
        0.10 & 7.74  & \textcolor{red}{77.4} & \textcolor{red}{774}  \\
        0.50 & \textcolor{red}{38.7}  & \textcolor{red}{387}  & \textcolor{red}{3870} \\
        \bottomrule
    \end{tabular}
\end{table}
\bigskip

\newpage
\textbf{RILA}

To extend the measurable conductance range and prevent system saturation when measuring low-resistance junctions, the RILA architecture introduces a known resistor $R_s$ in series, as described in Equation \ref{eq:rila_vout}. The DAQ records $V_{\text{out}}$ using a single analog input channel with a $\pm 10$ V range. To accurately determine $G_{\text{j}}$, the total voltage drop across both the molecular junction and the series resistor must be considered. The resulting expression is: 

\begin{equation}
G_\text{j} = \frac{1}{R_\text{j}} = \frac{1}{\left( \frac{V_{\text{bias}}}{V_{\text{out}}} \cdot R_\text{g} - R_s \right)}
\label{eq:GjRext}
\end{equation}
This expression reveals a potential divergence in the denominator when both terms are equal. To prevent this singularity, we define a saturation voltage ($V_{\text{sat}}$):

\begin{equation}
V_{\text{sat}} = \frac{V_{\text{bias}} \cdot R_\text{g}}{R_\text{s}}
\label{eq:Vsat}
\end{equation}

By isolating $V_{\text{bias}}$ from the saturation voltage definition in Equation \ref{eq:Vsat} and substituting it into the general conductance relationship \ref{eq:GjRext} , the expression is reformulated to describe the $G_\text{j}$ as a function of the limit $V_{\text{sat}}$. By incorporating the necessary conversion units, the final normalized expression becomes:

\begin{equation}
G_\text{j} = \frac{1}{R_\text{j}} = \left( \frac{1}{(\frac{V_\text{sat}}{V_\text{out}} - 1)} \right) \cdot \frac{12906 }{R_\text{s}}[\Omega \cdot G_\text{0}]
\label{eq:Gj_RILA_norm}
\end{equation}

Identifying $V_{\text{sat}}$ defines the point where the conductance would theoretically diverge as $R_{\text{j}} \rightarrow 0$. To prevent these numerical singularities, the software assigns a fixed value of $\sim13\,G_{0}$ once the output reaches saturation. This practical approach avoids calculation errors for metallic contacts. We also incorporated a fine-adjustment parameter, $R_{\text{correct}}$, to increase measurement resolution within this setup. The specific implementation is shown in the \texttt{LabVIEW} block diagram in Fig.~\ref{fig:labviewRILA}, and further details are available in the Supplementary Material.

Combining $R_{\text{s}}$, $V_{\text{sat}}$, and $R_{\text{correct}}$ allows for analyzing the behavior as $V_{\text{sat}}$ approaches $V_{\text{out}}$ (where $R_{\text{j}} \rightarrow 0$).This approach expands the applicable range of gain resistor values, effectively mapping the measurable $G_\text{j}$ across a wider dynamic window than previously possible with the standard ILA framework. 

\bigskip
\textbf{ILOGA}

Unlike the linear ILA and RILA architectures, the ILOGA is a non-linear design that operates by comparing two current signals. Using the LOG104 integrated circuit, it maps currents spanning several orders of magnitude into a single measurable voltage range. According to Equation \ref{eq:iloga_vout}, the output depends on the ratio between the junction current ($I_{\text{j}} = V_{\text{bias}} / R_{\text{j}}$) and the reference current ($I_{\text{ref}} = V_{\text{ref}} / R_{\text{ref}}$). By applying a scale factor of $0.5$ V/decade, the model is expressed as:

\begin{equation}
V_{\text{out}} = 0.5 \cdot \log_{10} \left( \frac{V_{\text{bias}} \cdot R_{\text{ref}}}{V_{\text{ref}} \cdot R_{\text{j}}} \right)
  \label{Voutparametes}
\end{equation}

By substituting $G_{\text{j}} = 1/R_{\text{j}}$ into Equation \ref{Voutparametes} and following the normalization procedure used for the previous architectures, the conductance in units of $G_{0}$ is expressed as:
\begin{equation}
G_{\text{j}} = \left( 10^{2 \cdot V_{\text{out}}} \cdot \frac{V_{\text{ref}}}{V_{\text{bias}} \cdot R_{\text{ref}}} \right) \cdot 12906 \quad [\Omega \cdot G_{0}]
\label{eq:Gj_ILOGA_final}
\end{equation}

\bigskip
\textbf{MILAC}

This system comprises three Op-Amps connected in series, partitioning signal processing into a primary transimpedance stage followed by two voltage amplification stages. Since the MILAC architecture relies on inverting transimpedance configurations, it inherently generates a negative output ($-V_{\text{out}}$). As established in the previous sections, all outputs are polarity-corrected via inverting buffers. Consequently, each stage delivers an independent output ($V_{\text{out1}}$, $V_{\text{out2}}$, and $V_{\text{out3}}$), enabling the DAQ to capture three simultaneous signals with cumulative gains. The conductance for each stage is then defined as:

\begin{equation}
G_\text{j,i} = \frac{V_\text{out,i}}{V_\text{bias} \cdot G_\text{total,i}} \cdot 12906 [\Omega \cdot G_\text{0}]
\end{equation}

\noindent Here, the total gains for each stage are defined as:
\begin{itemize}
    \item Stage 1 ($V_\text{out1}$): Uses a $1\text{ M}\Omega$ feedback resistor ($R_{G1}$) providing an initial gain of $10^6\text{ V/A}$.
    \item Stage 2 ($V_\text{out2}$): Applies a secondary gain of 100, reaching $10^8\text{ V/A}$. This stage employs an LT1028 Op-Amp for its superior low-noise characteristics.
    \item Stage 3 ($V_\text{out3}$): Adds a final gain of 10, achieving a total cumulative sensitivity of $10^9\text{ V/A}$.
\end{itemize}

\section{\label{Optimate} Optimization and Characterization of All \texorpdfstring{$I\text{--}V$}{I--V} AMPLIFIERS.} 

Throughout this work, we distinguish between the nominal conductance range suggested by circuit specifications and the credible measurable range. This operational window is defined by the limits where measurements become susceptible to the electronic noise floor, DAQ digitization, and algorithmic artifacts. Furthermore, its reliability is contingent upon temporal constraints, such as RC filter delays from parasitic impedances, DAQ sampling rates, and the intrinsic settling times of the operational amplifiers

\subsection{ILA and RILA Configurations}
Due to their inherent simplicity, the ILA and RILA configurations do not require exhaustive characterization. We describe below the calibration procedure for the FEMTO current amplifier, which serves as the experimental benchmark. This step validates the amplifier's response using the specific measurement connections.

\begin{itemize}
    \item Input Isolation: Disconnect all external measurement cables. Attach a blank BNC cap, an insulating cap, or a high-impedance open-circuit connector to the FEMTO amplifier's input BNC. The primary objective is to maximize input impedance and ensure the absolute minimum input current ($I_{\mathrm{in}} \approx 0$).
    \item Offset Nulling: Select the specific gain range planned for the experiment (e.g., $10^8~\mathrm{V/A}$). Using a small non-conductive screwdriver, carefully turn the front-panel \texttt{OFFSET} trimpot until $V_{\mathrm{out}}$ is as close to zero as possible ($V_{\mathrm{out}} \approx 0$ V). This step compensates for the amplifier's internal bias and thermal drift.
    \item Noise and Drift Check: Monitor the  $V_{\mathrm{out}}$ signal on a digital oscilloscope or the (DAQ) system. The signal must be symmetrically distributed around the $\approx 0~\mathrm{V}$ baseline.
    \item Gain Verification: Replace the insulating cap with a known $R_{\mathrm{cal}}$. Apply a known $V_{\mathrm{bias}}$ to the circuit. Measure the resulting  $V_{\mathrm{out}}$ and verify that the measured conductance is the same as the nominal.
  
\end{itemize}

The ILOGA and MILAC configurations require a more rigorous analysis. Consequently, both architectures are comprehensively described and characterized in Sections \ref{SectionVB} and \ref{SectionVC}, respectively.

\subsection{ILOGA Configuration\label{SectionVB}}
\begin{figure}[!ht]
    \centering
    \includegraphics[width=0.99\linewidth]{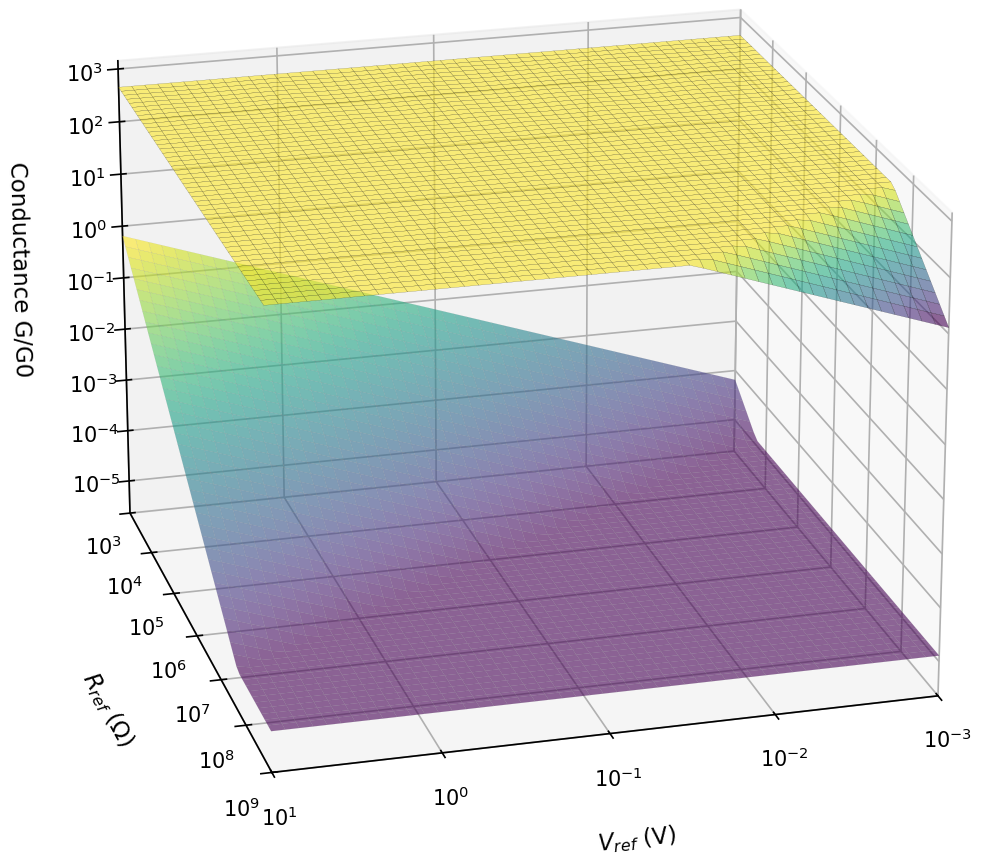}
\caption{Upper and lower nominal conductance boundaries for the ILOGA architecture, determined by the selection of $V_{\text{ref}}$ and $R_{\text{ref}}$.}
    \label{fig:GILOGANOMINAK}
\end{figure}

Optimizing the ILOGA setup requires a systematic evaluation of its core operational parameters to ensure a broad dynamic range and high measurement fidelity. As illustrated in Fig. \ref{fig:squemas}(c), this architecture operates by comparing the $I_{\text{j}}$ with  $I_{\text{ref}}$, where the latter is precisely defined by the ratio between the $V_{\text{ref}}$ and resistance $R_{\text{ref}}$. However, the LOG104 amplifier is bound by strict physical constraints: the input currents must remain strictly within the $100\text{ pA}$ to $10\text{ mA}$ range to maintain linearity.
To safely navigate these boundaries across several orders of magnitude without device saturation or falling into the noise floor, this section analyzes the fundamental pillars of the configuration. We first address the selection of $R_{\text{ref}}$ to establish the measurable conductance limits, followed by the tuning of $V_{\text{ref}}$ to minimize noise effects.  Then, we examine the influence of the power supply voltages ($V_{+}, V_{-}$), demonstrating how an asymmetric topology significantly expands the effective conduction window. To illustrate these constraints, Fig. \ref{fig:GILOGANOMINAK} presents the nominal conductance limits assuming a representative $V_{\text{bias}} = 100\text{ mV}$. This operational window is bounded by the $10\text{ mA}$ saturation threshold and the $100\text{ pA}$ noise, highlighting how the interplay between $V_{\text{ref}}$ and $R_{\text{ref}}$ establishes the definitive boundaries of the nominal measurable conductance region.

However, this preliminary analysis overlooks the influence of the device's power supply rails. Consequently, it is imperative to conduct a systematic study that, by fixing certain parameters, evaluates the impact of variations in the reference $R_{\text{ref}}$, $V_{\text{ref}}$, and the supply voltages ($V_\text{+}, V_\text{-}$). Rather than being limited to a nominal description, this analysis aims to determine the actual measurable range of the system. Therefore, the following subsections explore these fundamental pillars in detail to establish the definitive operational window of the ILOGA configuration.

\subsubsection{The role of the reference resistance.} 

\begin{figure}[!ht]
    \centering
    \includegraphics[width=0.99\linewidth]{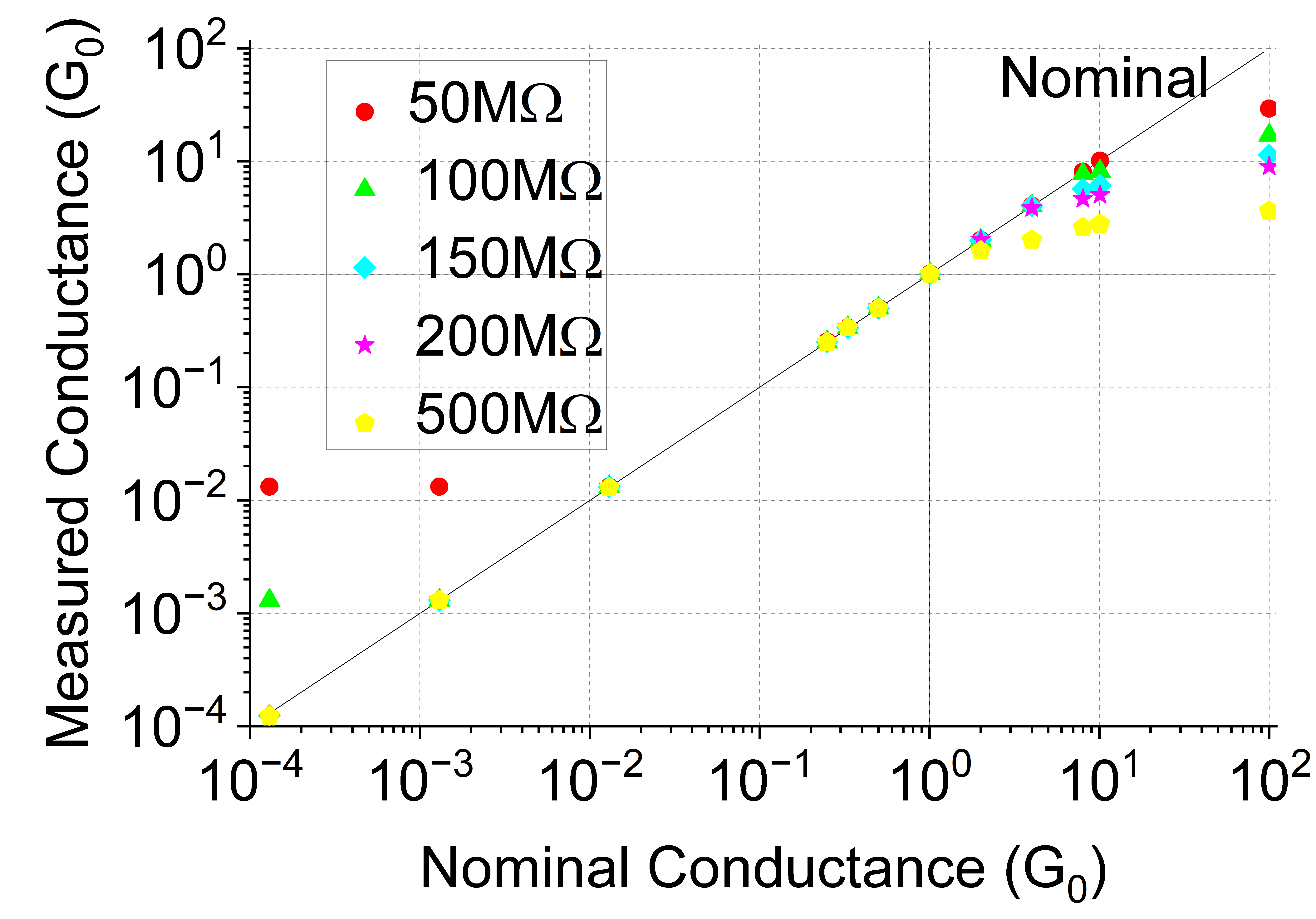}
    \caption{Measured versus nominal conductance, changing $R_\text{ref}$. Data collected from Table \ref{Table:RRef}. }
    \label{fig:R_Ref}
\end{figure}

To evaluate the optimal $R_{\text{ref}}$ and the measurable conductance limits of the logarithmic $I\text{--}V$ amplifier, we conducted a systematic study of its upper and lower saturation bounds. The supply voltages were fixed at $V_\text{+} = 4$ V and $V_\text{-} = -4$ V, which define these saturation limits. Using a DAQ card, we applied a $V_{\text{bias}} =  100\text{ mV}$. Consequently, the $I_{\text{ref}}$ was controlled by $V_{\text{ref}}$ and $R_{\text{ref}}$ implemented in the circuit, with evaluated $R_{\text{ref}}$ values spanning from 50 M$\Omega$ to 500 M$\Omega$. To simulate the junction resistance ($R_\text{j}$), we utilized high-precision nominal resistors (0.01\% tolerance) ranging from 100 $\Omega$ to 100 M$\Omega$. This range allowed us to test conductance values from approximately $10^2\, G_\text{0}$ down to $10^{-5}\, G_\text{0}$. Repeated measurements were performed using the MCBJ setup, and the collected current data were processed using a custom program developed in \texttt{LabVIEW} \cite{labview}. Finally, a statistical analysis was performed on the measured conductance distributions for each $R_{\text{ref}}$ and nominal resistance combination. To quantify the precision and accuracy of the system, the mean of each distribution was compared to its corresponding nominal value. These deviations are summarized in the following table and visually compared in Fig.~\ref{fig:R_Ref} (expressed in units of $G_\text{0}$).

As demonstrated in Fig.~\ref{fig:R_Ref} and Table~\ref{Table:RRef} (Supplementary Material), $R_{\text{ref}}$ values above 500 M$\Omega$ restrict the accessible conductance limits, effectively excluding the $G_0$ at a constant $V_{\text{ref}}$ of 100 mV. This exclusion is highly detrimental for molecular electronics measurements. Conversely, reference resistances below 150 M$\Omega$  fail to significantly extend the dynamic range compared to the RILA architecture, rendering them unsuitable for our objectives.

Consequently, we selected an  $R_{\text{ref}} = 150~\text{M}\Omega $. With this configuration, the system successfully covers a dynamic range spanning from approximately 10 $G_\text{0}$ down to $10^{-4}$ $G_\text{0}$, making it highly suitable for measuring across multiple orders of magnitude, in the conductance range of interest here.

\subsubsection{The role of the voltage reference.} 

With the optimal $R_{\text{ref}}$ established, we investigated the effect of $V_{\text{ref}}$ on the amplifier's dynamic range by testing voltages from 10 µV to 10 V. Following the established protocol, we utilized the MCBJ setup to measure the nominal resistors and statistically analyzed the deviation of the measured averages from their nominal values. Notably, the data reveals that for $V_{\text{ref}}$ exceeding 2 V, the measurable range is severely limited for conductance values below $G_\text{0}$.

This limitation is attributed to increased input currents that lead to measurement saturation. In contrast, for voltages below 1 mV, the results not only exhibit a reduced operational range but also show increased randomness. This behavior is likely due to the dominance of electrical noise at very low current levels.

These observations demonstrate that the optimal window for precise and reliable measurements lies between 10 mV and 1 V, effectively avoiding both high-voltage saturation and low-voltage noise dominance. Consequently, a $V_{\text{ref}}$ of 100 mV was selected for all subsequent experiments, providing an ideal balance between measurement accuracy and dynamic range. The dataset supporting this selection is illustrated in Fig.~\ref{fig:Vrefsweep} (also in Table~\ref{tab:conductance_tableVref} of the Supplementary Material).

\begin{figure}[!ht]
    \centering
    \includegraphics[width=0.99\linewidth]{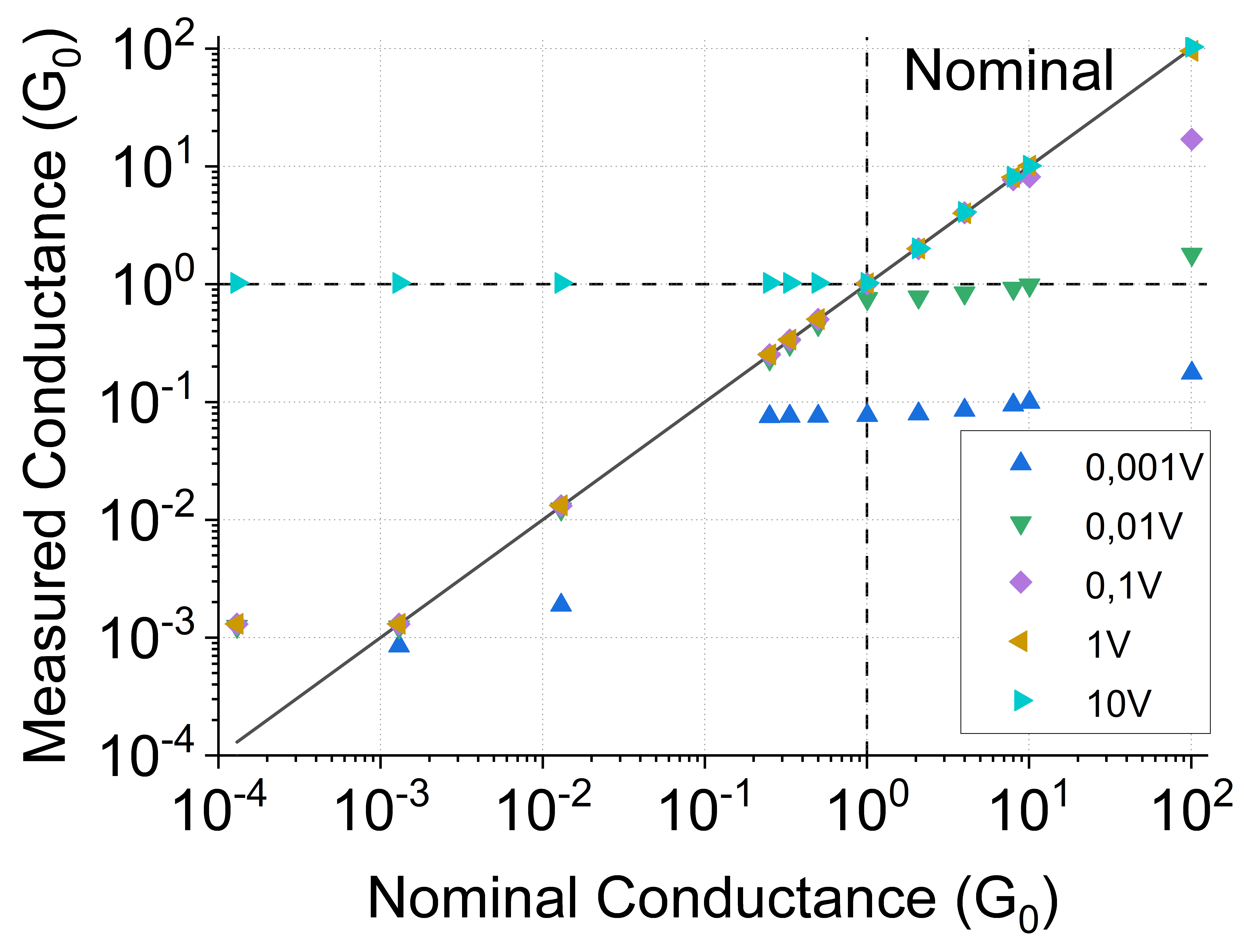}
    \caption{Measured versus nominal conductance changing the parameter $V_{\text{ref}}$. Data collected from Table \ref{tab:conductance_tableVref}.}
    \label{fig:Vrefsweep}
\end{figure}

\subsubsection{The role of the power supply and the advantage of the asymmetry.}  
As briefly noted in Section \ref{MEtandMAt}, while the linear architectures (ILA, RILA, MILAC) operate reliably with standard, often symmetrical power supplies (e.g., $\pm 15$ V) without altering their low-current response, logarithmic amplifiers are highly sensitive to supply voltages. They require precise, and sometimes asymmetric, biasing to prevent offsets, instability, and premature saturation at the low-conductance limits. To ensure measurement robustness, we characterized the system under controlled supply conditions. Additionally, leakage currents were suppressed by implementing low-noise operational amplifiers and precision reference voltages.

\begin{figure}[!ht]
    \centering
\includegraphics[width=0.99\linewidth]{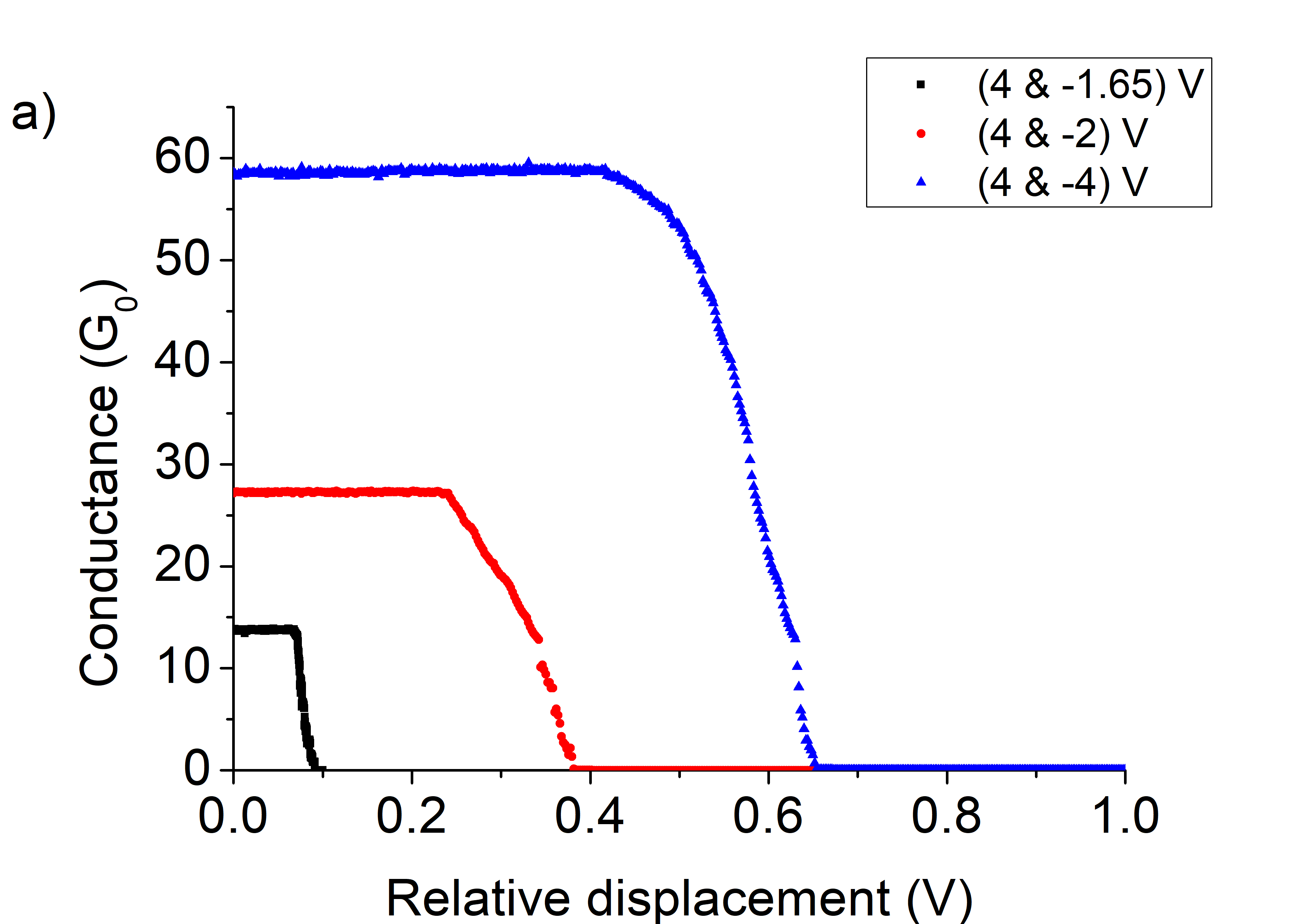}
\includegraphics[width=0.99\linewidth]{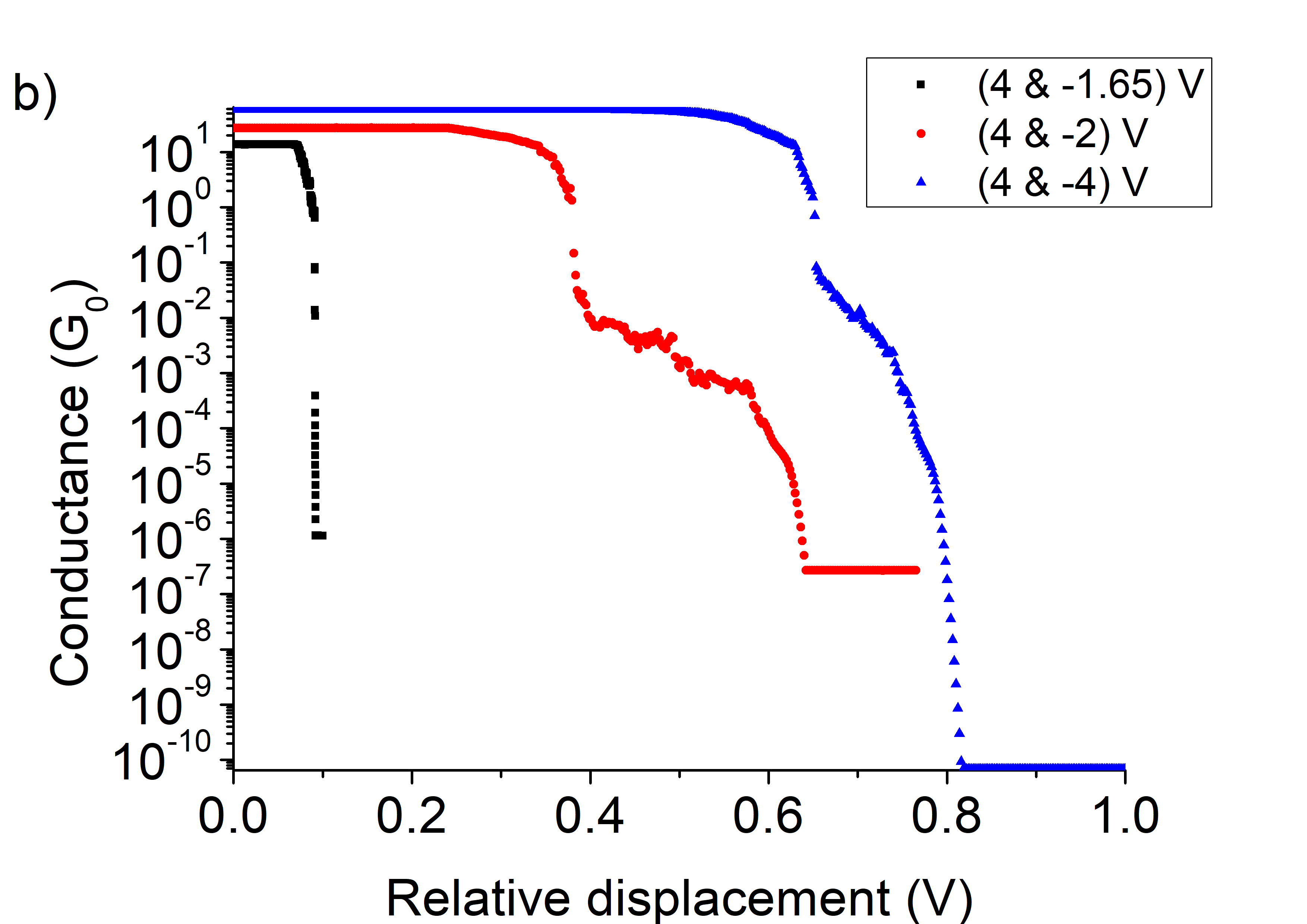}
    \caption{ Traces at distinct supply voltages plotted on (a) linear and (b) logarithmic scales.}
    \label{fig:Asymetric}
\end{figure}

The operational conductance range of the chip is highly dependent on its power supply. To characterize this, we recorded $I\text{--}V$ traces while maintaining a fixed positive supply voltage ($V_\text{+} = 4$ V) and varying the negative supply voltage ($V_\text{-}$) across -4 V, -2 V, and -1.65 V. As shown in Fig.~\ref{fig:Asymetric}(a), reducing the magnitude of $V_\text{-}$ significantly narrows the conduction window. For instance, at $V_\text{-} = -2$ V, the upper conductance limit drops to 30 $G_\text{0}$, and it further contracts to 14 $G_\text{0}$ when applying -1.65~V.

Fig.~\ref{fig:Asymetric}(b) presents the same data on a logarithmic scale to detail the lower conductance regime. At $V_\text{-} = -4$ V, conductance appears to reach as low as $10^{-10} G_\text{0}$; however, the region between $10^{-6} G_\text{0}$ and $10^{-3} G_\text{0}$ is an experimental artifact rather than actual tunneling behavior. A similar artifact is observed at $V_\text{-} = -2$ V, where the lower limit settles at $10^{-3} G_\text{0}$. For $V_\text{-} = -1.65$ V, while the theoretical lower limit is $10^{-6} G_\text{0}$, the effective limit is $10^{-4} G_\text{0}$, as will be discussed later. Based on these findings, the optimal supply configuration was established at $V_\text{+} = 4$ V and $V_\text{-} = -1.65$ V.

\begin{figure}[!ht]
    \centering
\includegraphics[width=1\linewidth]{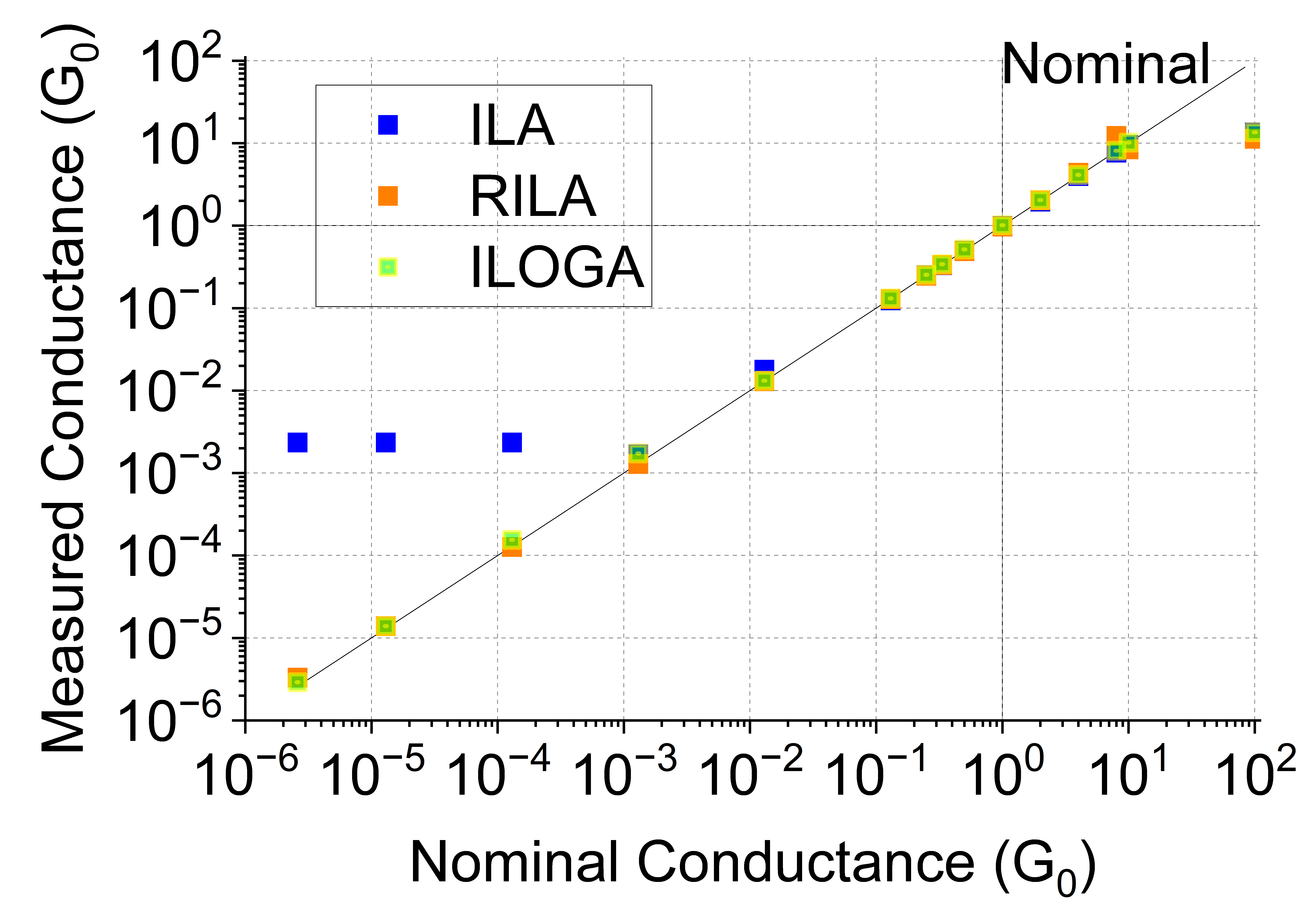}
    \caption{Comparison of the conductance measured depending on the methodology (ILA, RILA, ILOGA) versus nominal conductance. Data collected from Table \ref{tab:conductance}.}
    \label{fig:GIVComparison}
\end{figure}

Using these optimal parameters for the ILOGA, we compared its results against the nominal values and those obtained via the ILA and RILA setups. As shown in Fig.~\ref{fig:GIVComparison}, this comparison clearly illustrates the valid operational range for each method.

\subsection{MILAC Configuration \label{SectionVC}}

\begin{figure}[!ht]
\centering
\includegraphics[width=0.98\linewidth]{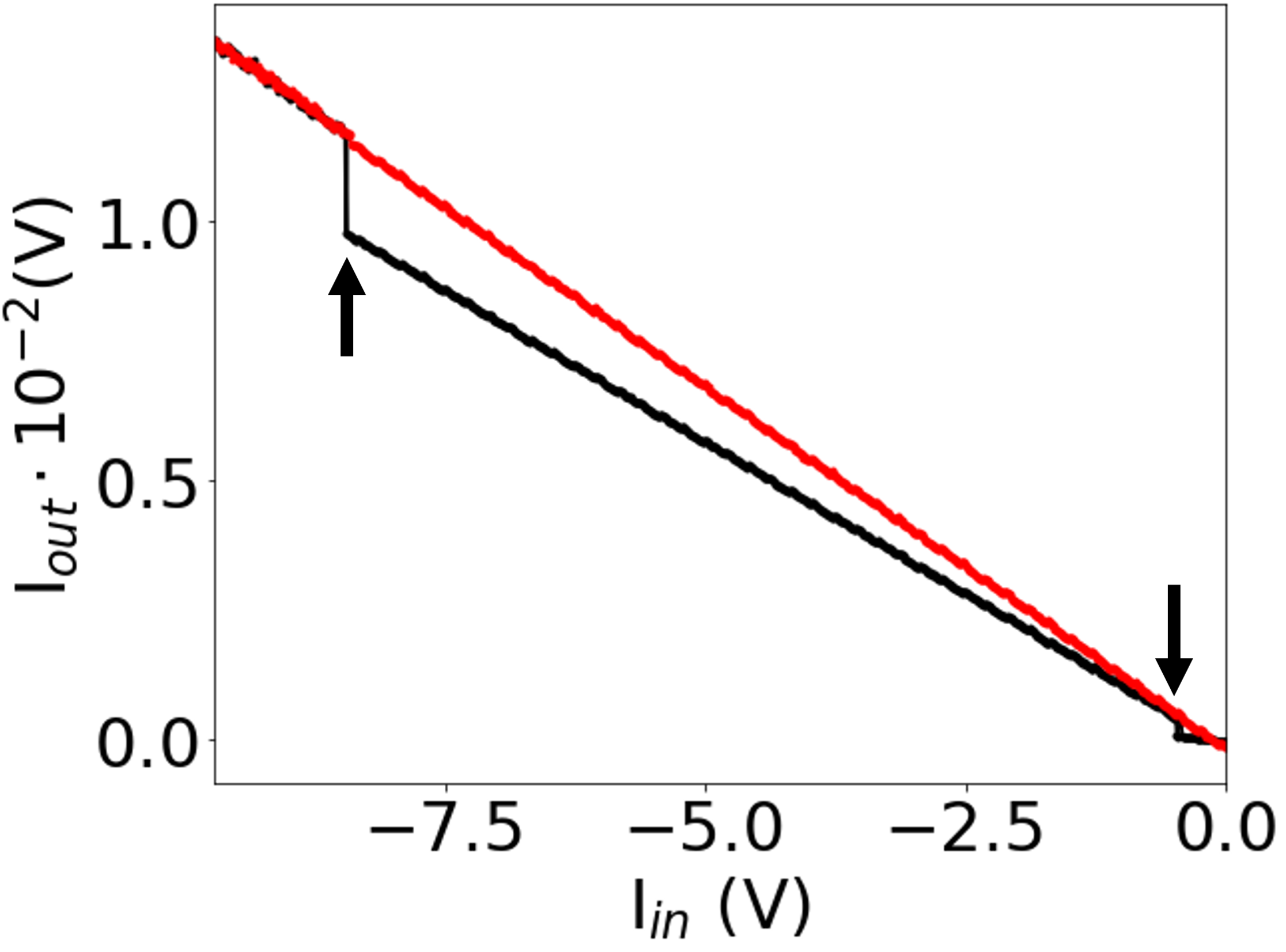}
\caption{Evaluation of the combined reading considering the output signal from each stage. In black, incorrect parameters for merging the outputs reveal the need for calibration to concatenate the signals seamlessly and prevent gaps. The red curve depicts an ideal concatenated signal achieved through proper calibration.}
\label{fig:calibration}
\end{figure}

To validate the amplifier design and calibrate the signal-merging algorithm, we measured the current through a known test resistor. By recording the output voltages ($V_{\text{out},i}$) from each stage simultaneously, we can plot their response in the regions where their measurement ranges overlap, as shown in Fig. \ref{fig:calibration}. As detailed in Section \ref{Hard}, reconstructing a single, continuous conductance trace requires precise knowledge of the gain ratios and the specific overlap points between stages. By fine-tuning these parameters, we rectified the discontinuities where signals meet, achieving the seamless calibration represented by the red curve. In contrast, the black curve illustrates the uncorrected case, where mismatched slopes and offsets create gaps and noise, leading to significant measurement errors.

\begin{figure}[!ht]
\centering
\includegraphics[width=.95\linewidth]{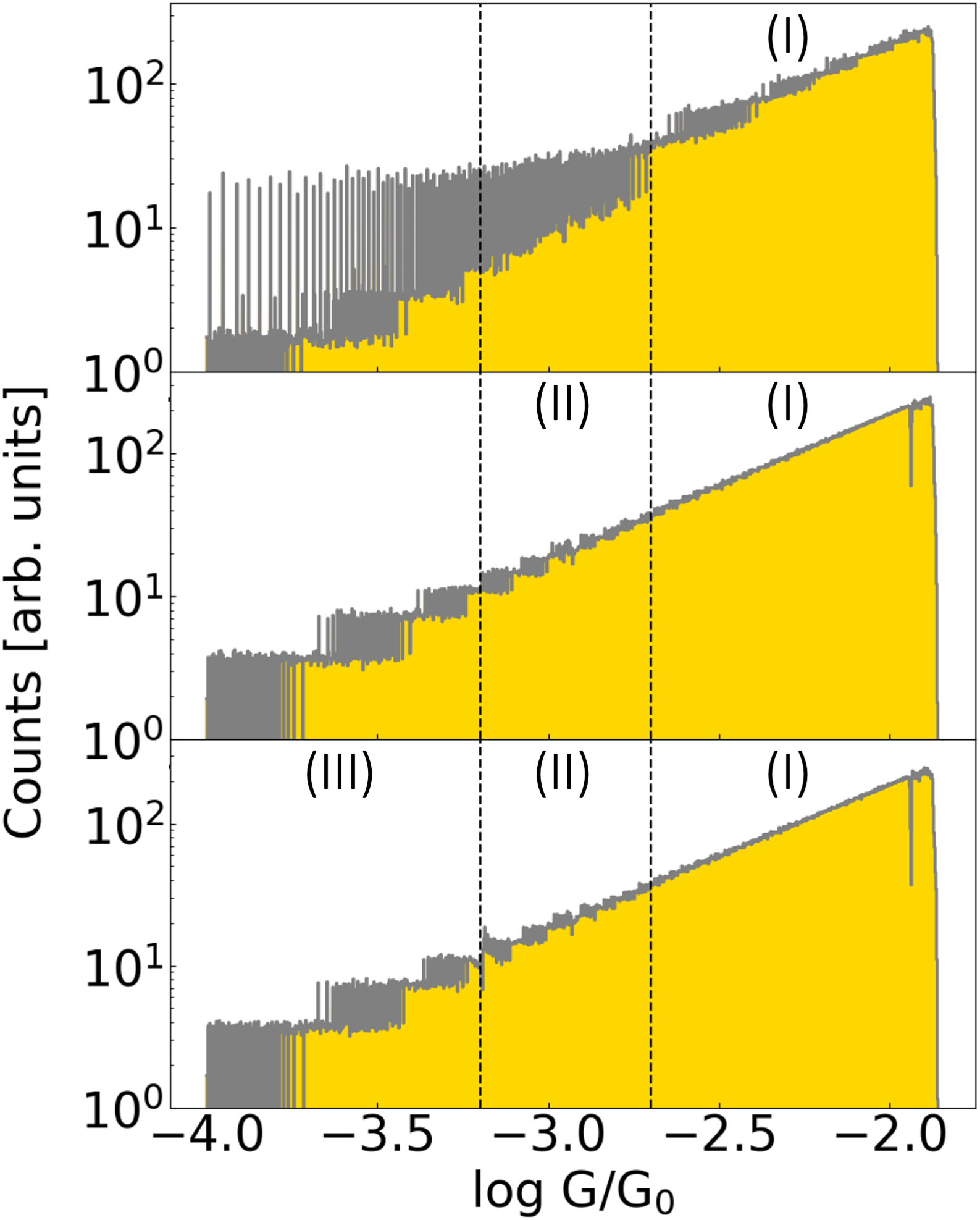}
\caption{Stacked histograms (yellow bins with gray outlines) of the measured current converted to conductance for each active amplifier stage. Successive stages progressively reduce the electrical noise floor down to approximately $10^{-5} G_\text{0}$. Roman numerals denote the first, second, and third amplifier stages.}
\label{fig:stagesHist}
\end{figure}

Once the calibration is completed, Fig. \ref{fig:stagesHist} presents a statistical analysis to assess the amplifier's performance at each stage. Arranged sequentially from top to bottom, these histograms provide an overview of the signal evolution as each stage is activated. Crucially, the data reveal a significant reduction in the noise floor.

Molecular signatures are typically observed at lower conductance values, particularly within the tunneling regime. However, background tunneling contributions and parasitic capacitance, introduced by the amplifier, the surrounding circuitry, and the sample itself, can obscure these intrinsic features, necessitating further analysis.

\section{EXPERIMENTAL PERFORMANCE IN GOLD ATOMIC CONTACTS}

Whichever amplifier is selected, it inherently defines the system's dynamic range. As a result, the acquired conductance traces and statistical histograms will exhibit distinct profiles depending on the setup, especially at low conductance levels.

\begin{figure}[!ht]
    \centering
    \includegraphics[width=0.99\linewidth]{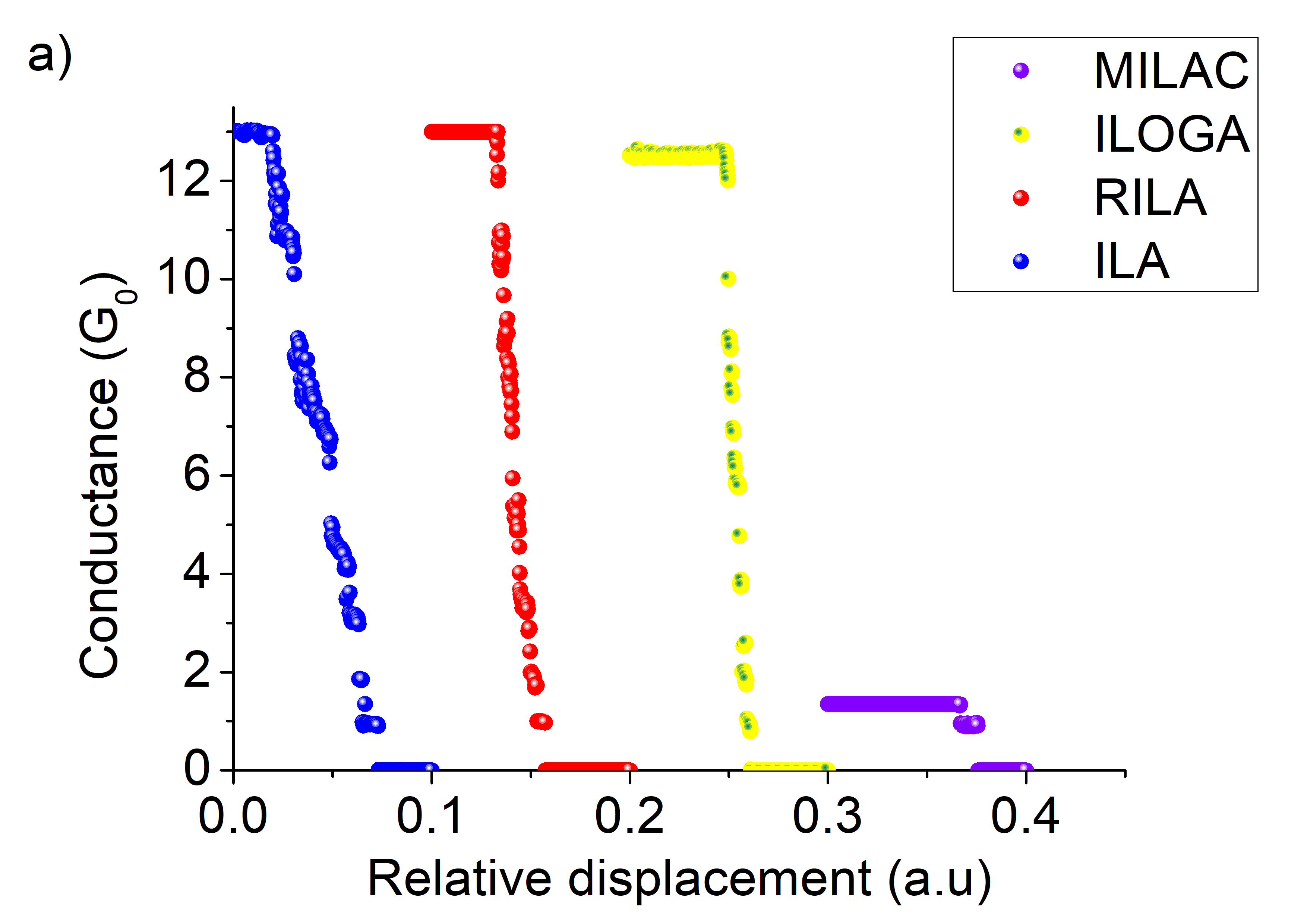}
    \includegraphics[width=0.99\linewidth]{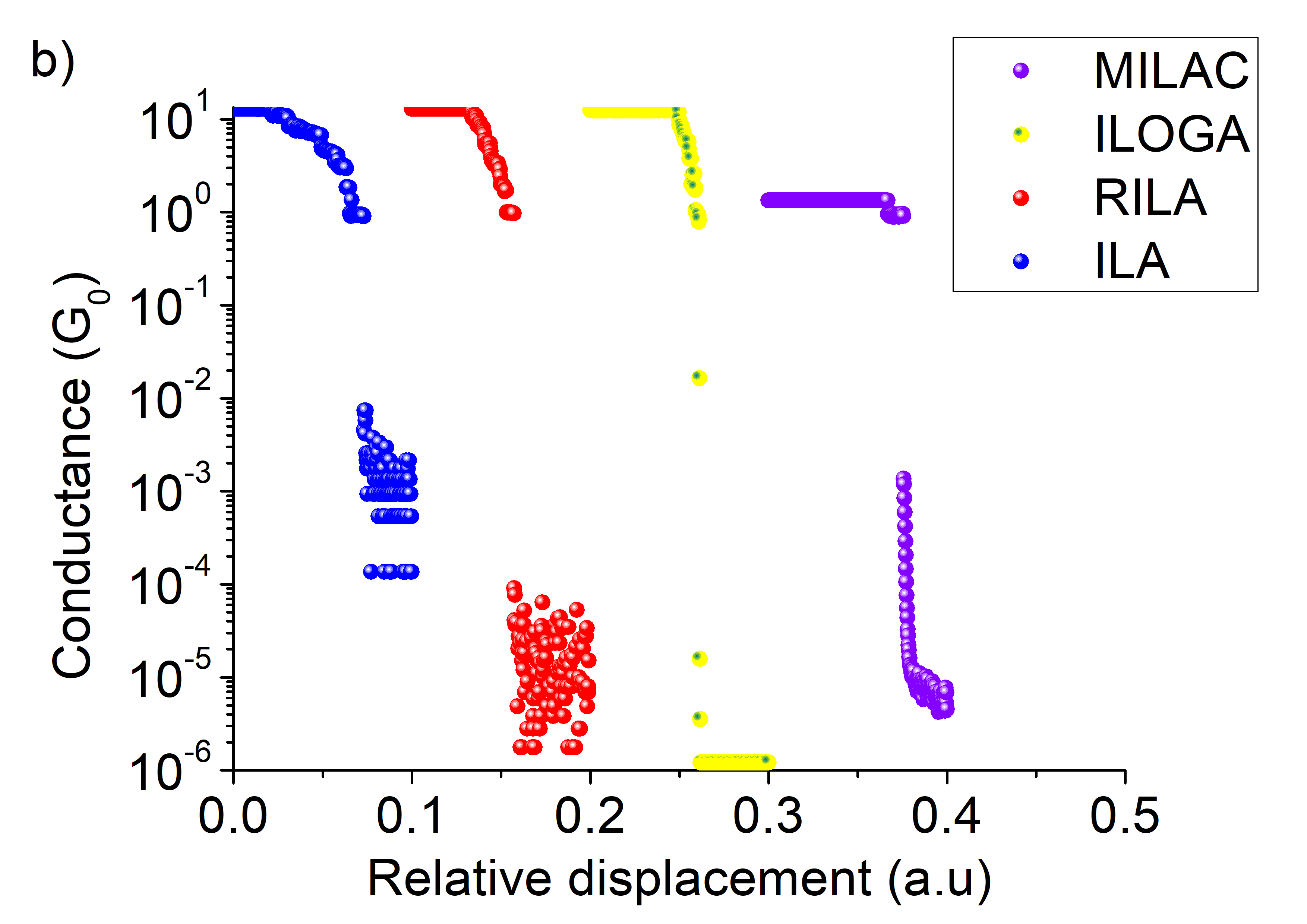}   
    \caption{Rupture conductance traces of Au measured at room conditions using the four amplification architectures, plotted on (a) linear and (b) logarithmic scales.}
    \label{fig:Traceslineallog}
\end{figure}

As shown in Fig.~\ref{fig:Traceslineallog}, representative breaking traces were recorded using the four distinct amplification architectures: ILA (blue), RILA (red), ILOGA (yellow), and MILAC (purple). The conductance is plotted directly in units of $G_\text{0}$ as a function of the applied piezo voltage, displayed on a linear scale in panel (a) and a logarithmic scale in panel (b). These traces span from $I\text{--}V$ amplifier saturation at approximately 13 $G_\text{0}$ down to junction rupture (instrumental noise floor, which in the case of ILOGA is $10^{-6} G_\text{0}$). Although minor variations exist between individual traces, these are expected due to the unique atomic rearrangements inherent to each breaking event and do not represent significant functional differences between the amplifiers.

\begin{figure}[!h]
\centering
\includegraphics[width=0.99\linewidth]{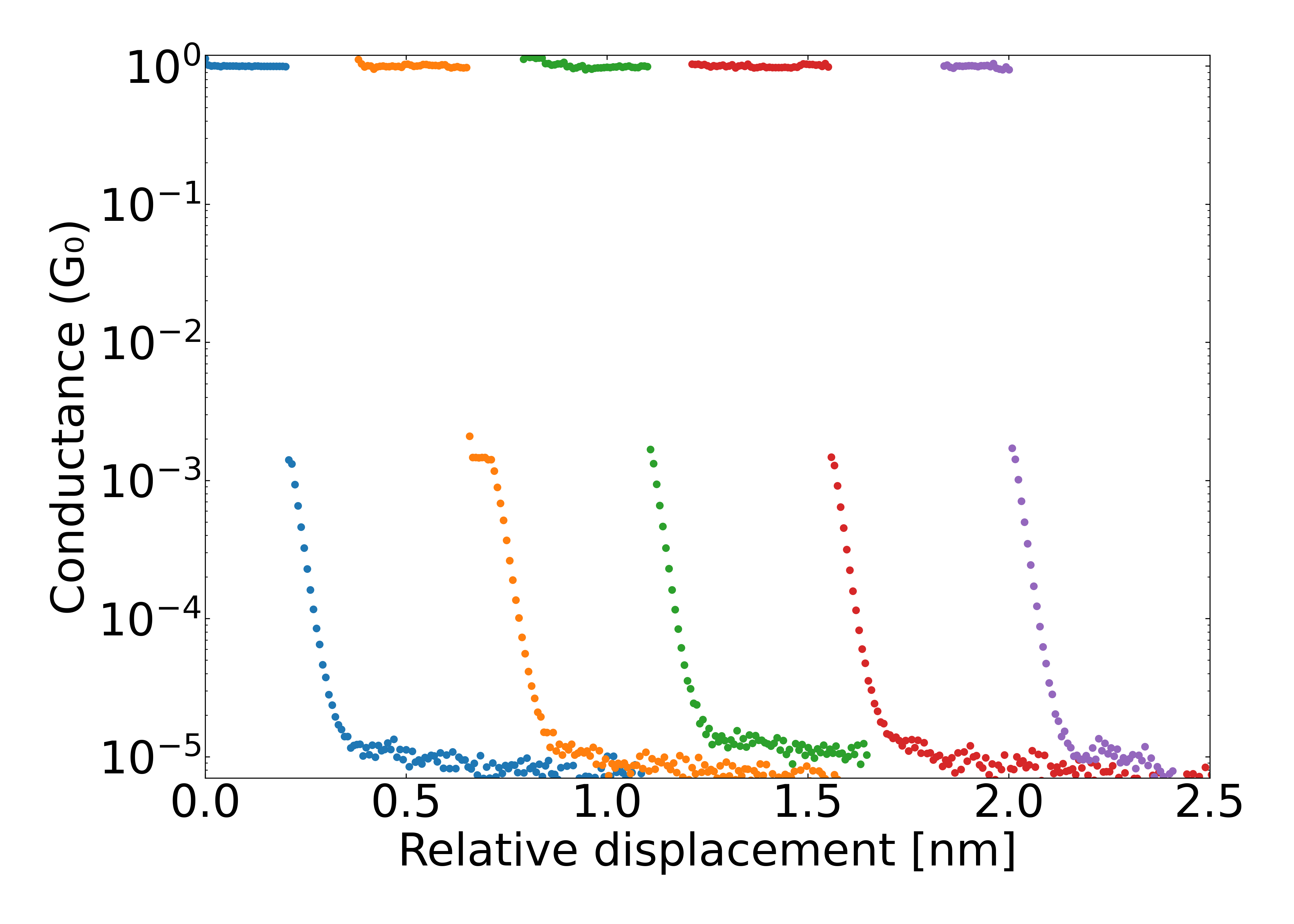}
\caption{Representative breaking traces measured with MILAC without any software filter. Bare Au exhibiting tunneling and non-linear decay attributed to parasitic capacitance and RC time constants.}
\label{fig:goldtraceimpurity}
\end{figure}

To compare the high and low conductance regimes, Fig.~\ref{fig:Traceslineallog}~(b) presents the same traces on a logarithmic scale. As shown, the ILA is limited to approximately $10^{-2}\,G_\text{0}$, whereas the RILA extends the measurement range by two orders of magnitude, reaching a noise floor just below $10^{-4}\,G_\text{0}$. In the case of the ILOGA, the logarithmic compression allows for a theoretically minimal level down to $10^{-6}\,G_\text{0}$. Finally, the MILAC architecture, operating from a maximum of $1.2\,G_\text{0}$ down to a noise floor of $10^{-5}\,G_\text{0}$. The MILAC was designed to measure metallic atomic contacts as well as molecules bridging between the electrodes. In this context, Fig. \ref{fig:goldtraceimpurity} provides practical demonstrations of the amplifier's performance. In these experiments, the current was monitored, and the conductance was computed using a constant bias of 100\text{ mV} and 10$^6$ V/A for the initial amplification under room conditions. The representative breaking traces acquired with the MILAC architecture span six orders of magnitude, with a noise floor around $\sim10^{-5}\,G_0$. As shown in Fig.\ref{fig:goldtraceimpurity}, the traces exhibit a non-linear decay on the logarithmic scale immediately following the junction rupture. This curvature deviates from the ideal linear tunneling slope and is attributed to parasitic capacitance and RC time constants inherent to the multi-stage amplifier design. 

While individual traces provide a unique view of one rupture process, the standard method for obtaining a wide overview is the construction of conductance histograms. To ensure comparability across all architectures, the counts are normalized to unity at the $1\,G_0$ peak. This approach aligns the characteristic metallic features of the gold junctions, allowing for a direct assessment of the detection limits and noise floors of each amplifier. Furthermore, by representing the conductance on a logarithmic scale, we can effectively resolve low-transmission regimes. This approach allows us to experimentally validate the detection limits and operational ranges of each amplification architecture.

\begin{figure}[htb]
    \centering
    \includegraphics[width=0.99\linewidth]{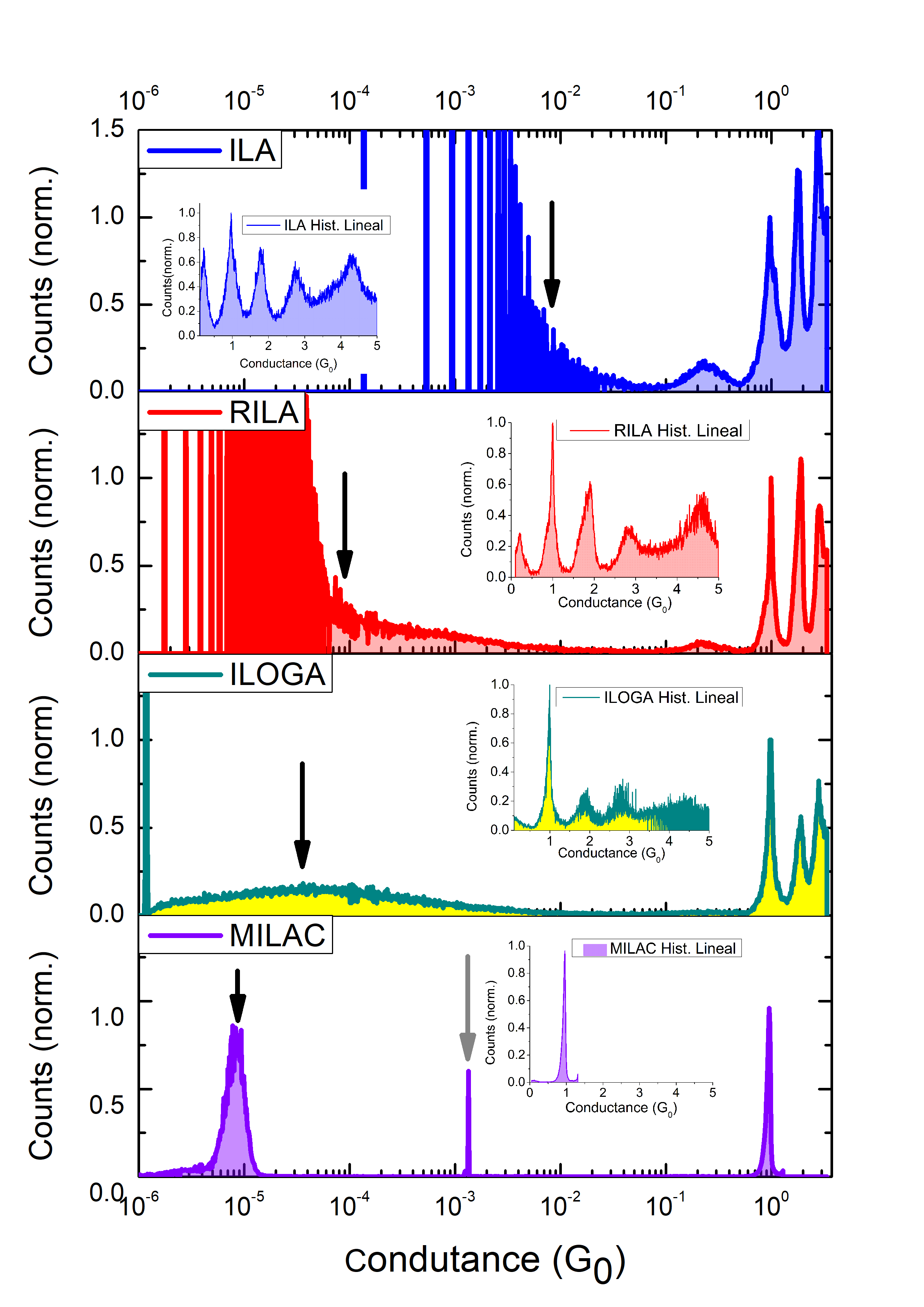}
    \caption{Normalized conductance histograms on a logarithmic scale for the ILA (blue), RILA (red), ILOGA (yellow), and MILAC (purple) configurations. Insets in each panel display the corresponding distributions on a linear scale. The black arrows indicate the minimum credible conductance; in the MILAC panel, the right gray arrow specifically highlights an artifact.}
    \label{fig:Histos}
\end{figure}

Normalized conductance histograms on a logarithmic scale are shown in   Fig.~\ref{fig:Histos}. For the ILA , the resolution limit is $10^{-2}\,G_\text{0}$, with a secondary peak near $10^{-1}\,G_\text{0}$ indicating contamination. The RILA  extends this limit to $10^{-4}\,G_\text{0}$ while resolving similar contamination features. In these linear architectures, the valid measurement range is clearly delimited by the black arrows; beyond these points, the loss of the characteristic tunneling slope indicates that the signal is dominated by the instrumental noise floor. Regarding the ILOGA, although raw data suggests a floor of $10^{-6}\,G_\text{0}$, the credible dynamic range is above at $10^{-4}\,G_\text{0}$. In a logarithmic conductance histogram, pure tunneling transport must manifest as a constant slope; the deviation from this slope is marked here by the black arrow. Finally, the MILAC architecture, characterized by its multi-stage design and high cumulative gain, exhibits an inherent sensitivity to RC time constants arising from the interplay between the junction's tunneling resistance and parasitic capacitances. These electronic effects must be carefully corrected, particularly in the overlap regions between stages, where they may coincide with conductance plateaus originating from environmental contamination\cite{Pellicer}. The methodology for identifying contaminated traces, compensating for RC distortions, and obtaining the final filtered histograms is described in detail in the Supplementary Material (see Fig.~\ref{fig:MILACraw}). According to Fig.~\ref{fig:Histos}, the MILAC architecture reaches a saturation limit of approximately $1.2~G_0$, allowing it to accurately resolve the $1~G_0$ plateau during the formation of gold atomic contacts. In the low-conductance regime, the sharp line near $10^{-3}~G_0$ is a numerical artifact of the stage-merging process, while features at $10^{-5}~G_0$ correspond to the amplifier's internal noise floor.

\section{Discussion}
The selection of an $I\text{--}V$ architecture inherently defines the system's dynamic range and, consequently, the profile of the acquired conductance traces.  As summarized in Table~\ref{tab:sumary}, each configuration offers distinct trade-offs between complexity and performance. The ILA is the most robust and straightforward approach, ideal for high-conductance metallic regimes, but its limited dynamic range ($10^{1}$ to $10^{-2}\,G_{0}$) restricts its utility for molecular junctions. By introducing a series resistor, the RILA effectively extends this range toward lower conductance ($10^{1}$ to $10^{-4}\,G_{0}$) and reduces saturation risks without requiring hardware modifications. However, the RILA introduces non-linear sensitivity near saturation and requires additional software routines for data conversion. For experiments requiring statistical analysis across several orders of magnitude, the ILOGA provides a powerful, cost-effective alternative. By compressing wide current ranges into a logarithmic voltage output, it achieves a credible range down to $10^{-4}\,G_{0}$. However, as specified in the LOG104 documentation, the device's settling time is signal-dependent, increasing significantly as the input current decreases. This inherent dynamic leads to a reduction in bandwidth at the lower end of the conductance range, potentially introducing time-dependent artifacts if the piezoelectric ramping speed exceeds the amplifier's response rate during junction rupture. Nevertheless, this compression sacrifices local linearity and defines the practical limits for reliable interpretation across all architectures.

Finally, the MILAC architecture bridges the gap between these designs. By preserving linear conversion across three cascaded stages, it reaches a dynamic range of six orders of magnitude ($1.2$ to $10^{-5}\,G_{0}$). This makes it uniquely suited for time-resolved measurements where both high bandwidth and linearity are critical, despite the higher implementation cost and the complexity of stage-merging algorithms. At very low conductance, factors such as capacitive coupling, parasitic RC effects, and electronic noise floors define the practical limits for reliable interpretation across all architectures. These constraints motivate the classification procedures and diagnostic framework established in this work to distinguish genuine quantum transport from instrumental artifacts. 

\begin{table}[H]
\centering
\caption{Performance comparison of the $I\text{--}V$ conversion architectures. The table details the methodology (with the number of Op-Amps in parentheses), design type (S: Standard; HM: Homemade), and detectable conductance range in units of $G_\text{0}$. It also evaluates the operational difficulty level alongside a summary of the pros and cons list.}
\begin{tabular}{cccccc}
\hline
\textbf{$I\text{--}V$-(OP-AMP)} & \textbf{Type} & \textbf{Range ($G_0$)} & \textbf{Easy/expert} & \textbf{Pro} & \textbf{Cons} \\
\hline
ILA (1)   & S  & $10^{-2}$ to $10^{1}$ & Easy   & \textcolor{blue}{1,2,4}   & \textcolor{red}{3} \\
RILA (1)  & S  & $10^{-4}$ to $10^{1}$ & Easy   & \textcolor{blue}{1,2,3,4} & \textcolor{red}{1,3} \\
ILOGA (1) & HM & $10^{-4}$ to $10^{1}$ & Easy   & \textcolor{blue}{1,2,3,4} & \textcolor{red}{1,5} \\
MILAC (3) & HM & $10^{-5}$ to 1.2 & Expert & \textcolor{blue}{3}       & \textcolor{red}{1,2,3,4,5,6} \\
\hline
\end{tabular}

\vspace{2mm}
\begin{minipage}{0.49\textwidth}
\footnotesize
\textbf{Pros:} \textcolor{blue}{[1]} Straightforward implementation, \textcolor{blue}{[2]} Simplified data acquisition, \textcolor{blue}{[3]} Extended low-conductance range, \textcolor{blue}{[4]} High portability. \\
\textbf{Cons:} \textcolor{red}{[1]} Expertise required, \textcolor{red}{[2]} High component count, \textcolor{red}{[3]} High implementation cost, \textcolor{red}{[4]} Susceptibility to electronic noise, \textcolor{red}{[5]} RC/Time constant delays,  \textcolor{red}{[6]} Requires software filtering of raw data.
\end{minipage}
\label{tab:sumary}
\end{table}

\section{Conclusions}
This study demonstrates that the selection of an appropriate current-to-voltage ($I\text{--}V$) architecture is fundamental for reliable conductance measurements in atomic and molecular junctions. Our comparative analysis reveals that the single-stage linear architecture (ILA) is the most limited system for low conductance, while the series-linear configuration (RILA)  provides high linearity for moderate conductance ($10^{1}$ to $10^{-4}\,G_\text{0}$). In contrast, the logarithmic architecture (ILOGA) extends the measurable range down to $10^{-4}\,G_\text{0}$ through signal compression, though it sacrifices linearity. Finally, the multi-stage cascaded (MILAC) configuration successfully bridges the gap between metallic contacts and the tunneling regime, achieving a dynamic range of six orders of magnitude with a noise floor near $10^{-5}\,G_\text{0}$.

These results highlight three critical experimental trade-offs: (i) the balance between the number of amplification stages and overall system complexity; a higher component count not only increases intrinsic electronic noise but also demands more rigorous calibration and the development of specialized software for signal processing. (ii) The choice between the extensive dynamic range offered by logarithmic architectures and the simplicity of linear systems, while accounting for the critical limitation of logarithmic amplifiers regarding their inability to operate under negative bias voltages. Finally (iii) the critical distinction between an amplifier's nominal limits and its "credible" operational range, where capacitive coupling, RC delays, and intrinsic noise floors dictate the true boundary for resolving genuine quantum transport phenomena.

Beyond the specific characterization of Au junctions, this work establishes a diagnostic framework that enables the molecular electronics community to clearly distinguish between genuine quantum transport phenomena and instrumental artifacts. By defining the operational limits and noise signatures of diverse amplification architectures, we provide the necessary tools to prevent the misinterpretation of electronic features as novel physical effects. Furthermore, this metrological approach is inherently scalable, offering a robust foundation for the study of emerging 2D materials, van der Waals heterostructures, and spintronic devices, where precise high-dynamic-range measurements are critical for the next generation of quantum technologies~\citep{Dion2004, Sierra2021, Ahn2020}. The decision-making roadmap summarized in Table \ref{tab:amplifierselection} serves as a practical guide for selecting the optimal amplification strategy for diverse nanoscale transport platforms.

\begin{table}[ht!]
\centering
\caption{Roadmap for $I\text{--}V$ amplifier selection. Columns list architecture, ideal application (metallic/molecular), credible $G_0$ range, and primary operational limitations.}
\label{tab:amplifierselection}
\begin{tabular}{llll} 
\toprule
\textbf{$I\text{--}V$} & \textbf{Ideal for} & \textbf{Credible Range ($G_0$)} & \textbf{Main Limitation} \\ \midrule
\textbf{ILA}   & Metal        & $\sim 10^{-2}$--$10^{1}$ & Very limited range \\ 
\textbf{RILA}  & Metal / Mol. & $\sim 10^{-4}$--$10^{1}$ & \begin{tabular}[t]{@{}l@{}}Non-linear near \\ saturation\end{tabular} \\ 
\textbf{ILOGA} & Metal / Mol. & $\sim 10^{-4}$--$10^{1}$ & Positive bias only \\ 
\textbf{MILAC} & Metal / Mol. & $\sim 10^{-5}$--$1.2$    & High complexity \\ \bottomrule
\end{tabular}
\end{table}

\section*{Acknowledgments}
This work received financial support from the Generalitat Valenciana through CIDEXG/2022/45. This research is an integral part of the Advanced Materials program, supported by MCIN with funding from the European Union NextGenerationEU (PRTR-C17.I1) and the Generalitat Valenciana (MFA/2022/045). We also acknowledge funding from MICIU/AEI/10.13039/501100011033 and the European Regional Development Fund (ERDF/EU) under project PID2023-146660OB-I00. J.H.$-$G. is grateful to the UCLouvain for the award of an FSR Incoming Post-doc Fellowship. We extend our gratitude to Prof. Nicolás Agraït, and Dr. Rubén López,  from the Universidad Autónoma de Madrid, for their insightful discussions and valuable advices.
\bibliographystyle{aipnum4-2}
\bibliography{IVpaper}

\begin{thebibliography}{59}%
\makeatletter
\providecommand \@ifxundefined [1]{%
 \@ifx{#1\undefined}
}%
\providecommand \@ifnum [1]{%
 \ifnum #1\expandafter \@firstoftwo
 \else \expandafter \@secondoftwo
 \fi
}%
\providecommand \@ifx [1]{%
 \ifx #1\expandafter \@firstoftwo
 \else \expandafter \@secondoftwo
 \fi
}%
\providecommand \natexlab [1]{#1}%
\providecommand \enquote  [1]{``#1''}%
\providecommand \bibnamefont  [1]{#1}%
\providecommand \bibfnamefont [1]{#1}%
\providecommand \citenamefont [1]{#1}%
\providecommand \href@noop [0]{\@secondoftwo}%
\providecommand \href [0]{\begingroup \@sanitize@url \@href}%
\providecommand \@href[1]{\@@startlink{#1}\@@href}%
\providecommand \@@href[1]{\endgroup#1\@@endlink}%
\providecommand \@sanitize@url [0]{\catcode `\\12\catcode `\$12\catcode `\&12\catcode `\#12\catcode `\^12\catcode `\_12\catcode `\%12\relax}%
\providecommand \@@startlink[1]{}%
\providecommand \@@endlink[0]{}%
\providecommand \url  [0]{\begingroup\@sanitize@url \@url }%
\providecommand \@url [1]{\endgroup\@href {#1}{\urlprefix }}%
\providecommand \urlprefix  [0]{URL }%
\providecommand \Eprint [0]{\href }%
\providecommand \doibase [0]{https://doi.org/}%
\providecommand \selectlanguage [0]{\@gobble}%
\providecommand \bibinfo  [0]{\@secondoftwo}%
\providecommand \bibfield  [0]{\@secondoftwo}%
\providecommand \translation [1]{[#1]}%
\providecommand \BibitemOpen [0]{}%
\providecommand \bibitemStop [0]{}%
\providecommand \bibitemNoStop [0]{.\EOS\space}%
\providecommand \EOS [0]{\spacefactor3000\relax}%
\providecommand \BibitemShut  [1]{\csname bibitem#1\endcsname}%
\let\auto@bib@innerbib\@empty
\bibitem [{\citenamefont {Aviram}\ and\ \citenamefont {Ratner}(1974)}]{aviram1974molecular}%
  \BibitemOpen
  \bibfield  {author} {\bibinfo {author} {\bibfnamefont {A.}~\bibnamefont {Aviram}}\ and\ \bibinfo {author} {\bibfnamefont {M.~A.}\ \bibnamefont {Ratner}},\ }\href {https://doi.org/https://doi.org/10.1016/0009-2614(74)85031-1} {\bibfield  {journal} {\bibinfo  {journal} {Chemical Physics Letters}\ }\textbf {\bibinfo {volume} {29}},\ \bibinfo {pages} {277} (\bibinfo {year} {1974})}\BibitemShut {NoStop}%
\bibitem [{\citenamefont {Cuevas}\ and\ \citenamefont {Scheer}(2017)}]{Cuevasbook}%
  \BibitemOpen
  \bibfield  {author} {\bibinfo {author} {\bibfnamefont {J.~C.}\ \bibnamefont {Cuevas}}\ and\ \bibinfo {author} {\bibfnamefont {E.}~\bibnamefont {Scheer}},\ }\href {https://doi.org/10.1142/10598} {\emph {\bibinfo {title} {Molecular Electronics}}},\ \bibinfo {edition} {2nd}\ ed.\ (\bibinfo  {publisher} {WORLD SCIENTIFIC},\ \bibinfo {year} {2017})\BibitemShut {NoStop}%
\bibitem [{\citenamefont {Evers}\ \emph {et~al.}(2020)\citenamefont {Evers}, \citenamefont {Koryt\'ar}, \citenamefont {Tewari},\ and\ \citenamefont {van Ruitenbeek}}]{Jan2019}%
  \BibitemOpen
  \bibfield  {author} {\bibinfo {author} {\bibfnamefont {F.}~\bibnamefont {Evers}}, \bibinfo {author} {\bibfnamefont {R.}~\bibnamefont {Koryt\'ar}}, \bibinfo {author} {\bibfnamefont {S.}~\bibnamefont {Tewari}},\ and\ \bibinfo {author} {\bibfnamefont {J.~M.}\ \bibnamefont {van Ruitenbeek}},\ }\href {https://doi.org/10.1103/RevModPhys.92.035001} {\bibfield  {journal} {\bibinfo  {journal} {Rev. Mod. Phys.}\ }\textbf {\bibinfo {volume} {92}},\ \bibinfo {pages} {035001} (\bibinfo {year} {2020})}\BibitemShut {NoStop}%
\bibitem [{\citenamefont {Natelson}(2015)}]{natelson2015nanostructures}%
  \BibitemOpen
  \bibfield  {author} {\bibinfo {author} {\bibfnamefont {D.}~\bibnamefont {Natelson}},\ }\href@noop {} {\emph {\bibinfo {title} {Nanostructures and Nanotechnology}}}\ (\bibinfo  {publisher} {Cambridge University Press},\ \bibinfo {year} {2015})\BibitemShut {NoStop}%
\bibitem [{\citenamefont {Nitzan}\ and\ \citenamefont {Ratner}(2003)}]{nitzan2003electron}%
  \BibitemOpen
  \bibfield  {author} {\bibinfo {author} {\bibfnamefont {A.}~\bibnamefont {Nitzan}}\ and\ \bibinfo {author} {\bibfnamefont {M.~A.}\ \bibnamefont {Ratner}},\ }\href {https://doi.org/10.1126/science.1081572} {\bibfield  {journal} {\bibinfo  {journal} {Science}\ }\textbf {\bibinfo {volume} {300}},\ \bibinfo {pages} {1384} (\bibinfo {year} {2003})},\ \Eprint {https://arxiv.org/abs/https://www.science.org/doi/pdf/10.1126/science.1081572} {https://www.science.org/doi/pdf/10.1126/science.1081572} \BibitemShut {NoStop}%
\bibitem [{\citenamefont {Xin}\ \emph {et~al.}(2019)\citenamefont {Xin}, \citenamefont {Guan}, \citenamefont {Zhou}, \citenamefont {Chen}, \citenamefont {Gu}, \citenamefont {Li}, \citenamefont {Ratner}, \citenamefont {Nitzan}, \citenamefont {Stoddart},\ and\ \citenamefont {Guo}}]{xin2019}%
  \BibitemOpen
  \bibfield  {author} {\bibinfo {author} {\bibfnamefont {N.}~\bibnamefont {Xin}}, \bibinfo {author} {\bibfnamefont {J.}~\bibnamefont {Guan}}, \bibinfo {author} {\bibfnamefont {C.}~\bibnamefont {Zhou}}, \bibinfo {author} {\bibfnamefont {X.}~\bibnamefont {Chen}}, \bibinfo {author} {\bibfnamefont {C.}~\bibnamefont {Gu}}, \bibinfo {author} {\bibfnamefont {Y.}~\bibnamefont {Li}}, \bibinfo {author} {\bibfnamefont {M.~A.}\ \bibnamefont {Ratner}}, \bibinfo {author} {\bibfnamefont {A.}~\bibnamefont {Nitzan}}, \bibinfo {author} {\bibfnamefont {J.~F.}\ \bibnamefont {Stoddart}},\ and\ \bibinfo {author} {\bibfnamefont {X.}~\bibnamefont {Guo}},\ }\href {https://doi.org/10.1038/s42254-019-0022-x} {\bibfield  {journal} {\bibinfo  {journal} {Nature Reviews Physics}\ }\textbf {\bibinfo {volume} {1}},\ \bibinfo {pages} {211} (\bibinfo {year} {2019})}\BibitemShut {NoStop}%
\bibitem [{\citenamefont {Xu}\ \emph {et~al.}(2024)\citenamefont {Xu}, \citenamefont {Gao}, \citenamefont {Emusani}, \citenamefont {Jia},\ and\ \citenamefont {Xiang}}]{Xiaona24}%
  \BibitemOpen
  \bibfield  {author} {\bibinfo {author} {\bibfnamefont {X.}~\bibnamefont {Xu}}, \bibinfo {author} {\bibfnamefont {C.}~\bibnamefont {Gao}}, \bibinfo {author} {\bibfnamefont {R.}~\bibnamefont {Emusani}}, \bibinfo {author} {\bibfnamefont {C.}~\bibnamefont {Jia}},\ and\ \bibinfo {author} {\bibfnamefont {D.}~\bibnamefont {Xiang}},\ }\href {https://doi.org/https://doi.org/10.1002/advs.202400877} {\bibfield  {journal} {\bibinfo  {journal} {Advanced Science}\ }\textbf {\bibinfo {volume} {11}},\ \bibinfo {pages} {2400877} (\bibinfo {year} {2024})},\ \Eprint {https://arxiv.org/abs/https://advanced.onlinelibrary.wiley.com/doi/pdf/10.1002/advs.202400877} {https://advanced.onlinelibrary.wiley.com/doi/pdf/10.1002/advs.202400877} \BibitemShut {NoStop}%
\bibitem [{\citenamefont {Ha}\ \emph {et~al.}(2021)\citenamefont {Ha}, \citenamefont {Planje}, \citenamefont {White}, \citenamefont {Aragonès},\ and\ \citenamefont {Díez-Pérez}}]{Isma21}%
  \BibitemOpen
  \bibfield  {author} {\bibinfo {author} {\bibfnamefont {T.~Q.}\ \bibnamefont {Ha}}, \bibinfo {author} {\bibfnamefont {I.~J.}\ \bibnamefont {Planje}}, \bibinfo {author} {\bibfnamefont {J.~R.}\ \bibnamefont {White}}, \bibinfo {author} {\bibfnamefont {A.~C.}\ \bibnamefont {Aragonès}},\ and\ \bibinfo {author} {\bibfnamefont {I.}~\bibnamefont {Díez-Pérez}},\ }\href {https://doi.org/https://doi.org/10.1016/j.coelec.2021.100734} {\bibfield  {journal} {\bibinfo  {journal} {Current Opinion in Electrochemistry}\ }\textbf {\bibinfo {volume} {28}},\ \bibinfo {pages} {100734} (\bibinfo {year} {2021})}\BibitemShut {NoStop}%
\bibitem [{\citenamefont {Rosenstein}, \citenamefont {Lemay},\ and\ \citenamefont {Shepard}(2015)}]{Rosenstein}%
  \BibitemOpen
  \bibfield  {author} {\bibinfo {author} {\bibfnamefont {J.~K.}\ \bibnamefont {Rosenstein}}, \bibinfo {author} {\bibfnamefont {S.~G.}\ \bibnamefont {Lemay}},\ and\ \bibinfo {author} {\bibfnamefont {K.~L.}\ \bibnamefont {Shepard}},\ }\href {https://doi.org/https://doi.org/10.1002/wnan.1323} {\bibfield  {journal} {\bibinfo  {journal} {WIREs Nanomedicine and Nanobiotechnology}\ }\textbf {\bibinfo {volume} {7}},\ \bibinfo {pages} {475} (\bibinfo {year} {2015})},\ \Eprint {https://arxiv.org/abs/https://wires.onlinelibrary.wiley.com/doi/pdf/10.1002/wnan.1323} {https://wires.onlinelibrary.wiley.com/doi/pdf/10.1002/wnan.1323} \BibitemShut {NoStop}%
\bibitem [{\citenamefont {Tewari}\ \emph {et~al.}(2017)\citenamefont {Tewari}, \citenamefont {Sabater}, \citenamefont {Kumar}, \citenamefont {Stahl}, \citenamefont {Crama},\ and\ \citenamefont {van Ruitenbeek}}]{TewariSAB2017}%
  \BibitemOpen
  \bibfield  {author} {\bibinfo {author} {\bibfnamefont {S.}~\bibnamefont {Tewari}}, \bibinfo {author} {\bibfnamefont {C.}~\bibnamefont {Sabater}}, \bibinfo {author} {\bibfnamefont {M.}~\bibnamefont {Kumar}}, \bibinfo {author} {\bibfnamefont {S.}~\bibnamefont {Stahl}}, \bibinfo {author} {\bibfnamefont {B.}~\bibnamefont {Crama}},\ and\ \bibinfo {author} {\bibfnamefont {J.~M.}\ \bibnamefont {van Ruitenbeek}},\ }\href {https://doi.org/10.1063/1.5003391} {\bibfield  {journal} {\bibinfo  {journal} {Review of Scientific Instruments}\ }\textbf {\bibinfo {volume} {88}},\ \bibinfo {pages} {093903} (\bibinfo {year} {2017})}\BibitemShut {NoStop}%
\bibitem [{\citenamefont {Massee}\ \emph {et~al.}(2018)\citenamefont {Massee}, \citenamefont {Dong}, \citenamefont {Cavanna}, \citenamefont {Jin},\ and\ \citenamefont {Aprili}}]{Massee2018}%
  \BibitemOpen
  \bibfield  {author} {\bibinfo {author} {\bibfnamefont {F.}~\bibnamefont {Massee}}, \bibinfo {author} {\bibfnamefont {Q.}~\bibnamefont {Dong}}, \bibinfo {author} {\bibfnamefont {A.}~\bibnamefont {Cavanna}}, \bibinfo {author} {\bibfnamefont {Y.~S.}\ \bibnamefont {Jin}},\ and\ \bibinfo {author} {\bibfnamefont {M.}~\bibnamefont {Aprili}},\ }\href {https://api.semanticscholar.org/CorpusID:52910935} {\bibfield  {journal} {\bibinfo  {journal} {The Review of scientific instruments}\ }\textbf {\bibinfo {volume} {89 9}},\ \bibinfo {pages} {093708} (\bibinfo {year} {2018})}\BibitemShut {NoStop}%
\bibitem [{\citenamefont {Ludoph}\ and\ \citenamefont {Ruitenbeek}(1999)}]{Thermopower}%
  \BibitemOpen
  \bibfield  {author} {\bibinfo {author} {\bibfnamefont {B.}~\bibnamefont {Ludoph}}\ and\ \bibinfo {author} {\bibfnamefont {J.~M.~v.}\ \bibnamefont {Ruitenbeek}},\ }\href {https://doi.org/10.1103/PhysRevB.59.12290} {\bibfield  {journal} {\bibinfo  {journal} {Phys. Rev. B}\ }\textbf {\bibinfo {volume} {59}},\ \bibinfo {pages} {12290} (\bibinfo {year} {1999})}\BibitemShut {NoStop}%
\bibitem [{\citenamefont {Karimi}\ \emph {et~al.}(2016)\citenamefont {Karimi}, \citenamefont {Bahoosh}, \citenamefont {Herz}, \citenamefont {Hayakawa}, \citenamefont {Pauly},\ and\ \citenamefont {Scheer}}]{ElkeShot}%
  \BibitemOpen
  \bibfield  {author} {\bibinfo {author} {\bibfnamefont {M.~A.}\ \bibnamefont {Karimi}}, \bibinfo {author} {\bibfnamefont {S.~G.}\ \bibnamefont {Bahoosh}}, \bibinfo {author} {\bibfnamefont {M.}~\bibnamefont {Herz}}, \bibinfo {author} {\bibfnamefont {R.}~\bibnamefont {Hayakawa}}, \bibinfo {author} {\bibfnamefont {F.}~\bibnamefont {Pauly}},\ and\ \bibinfo {author} {\bibfnamefont {E.}~\bibnamefont {Scheer}},\ }\href {https://doi.org/10.1021/acs.nanolett.5b04848} {\bibfield  {journal} {\bibinfo  {journal} {Nano Letters}\ }\textbf {\bibinfo {volume} {16}},\ \bibinfo {pages} {1803} (\bibinfo {year} {2016})},\ \Eprint {https://arxiv.org/abs/https://doi.org/10.1021/acs.nanolett.5b04848} {https://doi.org/10.1021/acs.nanolett.5b04848} \BibitemShut {NoStop}%
\bibitem [{\citenamefont {Ohm}(1827)}]{ohm1827galvanische}%
  \BibitemOpen
  \bibfield  {author} {\bibinfo {author} {\bibfnamefont {G.~S.}\ \bibnamefont {Ohm}},\ }\href@noop {} {\emph {\bibinfo {title} {Die galvanische Kette, mathematisch bearbeitet}}}\ (\bibinfo  {publisher} {TH Riemann},\ \bibinfo {address} {Berlin},\ \bibinfo {year} {1827})\BibitemShut {NoStop}%
\bibitem [{\citenamefont {Landauer}(1957)}]{Landauer57}%
  \BibitemOpen
  \bibfield  {author} {\bibinfo {author} {\bibfnamefont {R.}~\bibnamefont {Landauer}},\ }\href {https://doi.org/10.1147/rd.13.0223} {\bibfield  {journal} {\bibinfo  {journal} {IBM Journal of Research and Development}\ }\textbf {\bibinfo {volume} {1}},\ \bibinfo {pages} {223} (\bibinfo {year} {1957})}\BibitemShut {NoStop}%
\bibitem [{\citenamefont {B\"uttiker}(1986)}]{buttiker1986four}%
  \BibitemOpen
  \bibfield  {author} {\bibinfo {author} {\bibfnamefont {M.}~\bibnamefont {B\"uttiker}},\ }\href {https://doi.org/10.1103/PhysRevLett.57.1761} {\bibfield  {journal} {\bibinfo  {journal} {Phys. Rev. Lett.}\ }\textbf {\bibinfo {volume} {57}},\ \bibinfo {pages} {1761} (\bibinfo {year} {1986})}\BibitemShut {NoStop}%
\bibitem [{\citenamefont {van Wees}\ \emph {et~al.}(1988)\citenamefont {van Wees}, \citenamefont {van Houten}, \citenamefont {Beenakker}, \citenamefont {Williamson}, \citenamefont {Kouwenhoven}, \citenamefont {van~der Marel},\ and\ \citenamefont {Foxon}}]{vanwees1988quantized}%
  \BibitemOpen
  \bibfield  {author} {\bibinfo {author} {\bibfnamefont {B.~J.}\ \bibnamefont {van Wees}}, \bibinfo {author} {\bibfnamefont {H.}~\bibnamefont {van Houten}}, \bibinfo {author} {\bibfnamefont {C.~W.~J.}\ \bibnamefont {Beenakker}}, \bibinfo {author} {\bibfnamefont {J.~G.}\ \bibnamefont {Williamson}}, \bibinfo {author} {\bibfnamefont {L.~P.}\ \bibnamefont {Kouwenhoven}}, \bibinfo {author} {\bibfnamefont {D.}~\bibnamefont {van~der Marel}},\ and\ \bibinfo {author} {\bibfnamefont {C.~T.}\ \bibnamefont {Foxon}},\ }\href {https://doi.org/10.1103/PhysRevLett.60.848} {\bibfield  {journal} {\bibinfo  {journal} {Phys. Rev. Lett.}\ }\textbf {\bibinfo {volume} {60}},\ \bibinfo {pages} {848} (\bibinfo {year} {1988})}\BibitemShut {NoStop}%
\bibitem [{\citenamefont {{Wharam}}\ \emph {et~al.}(1988)\citenamefont {{Wharam}}, \citenamefont {{Thornton}}, \citenamefont {{Newbury}}, \citenamefont {{Pepper}}, \citenamefont {{Ahmed}}, \citenamefont {{Frost}}, \citenamefont {{Hasko}}, \citenamefont {{Peacock}}, \citenamefont {{Ritchie}},\ and\ \citenamefont {{Jones}}}]{wharam1988one}%
  \BibitemOpen
  \bibfield  {author} {\bibinfo {author} {\bibfnamefont {D.~A.}\ \bibnamefont {{Wharam}}}, \bibinfo {author} {\bibfnamefont {T.~J.}\ \bibnamefont {{Thornton}}}, \bibinfo {author} {\bibfnamefont {R.}~\bibnamefont {{Newbury}}}, \bibinfo {author} {\bibfnamefont {M.}~\bibnamefont {{Pepper}}}, \bibinfo {author} {\bibfnamefont {H.}~\bibnamefont {{Ahmed}}}, \bibinfo {author} {\bibfnamefont {J.~E.~F.}\ \bibnamefont {{Frost}}}, \bibinfo {author} {\bibfnamefont {D.~G.}\ \bibnamefont {{Hasko}}}, \bibinfo {author} {\bibfnamefont {D.~C.}\ \bibnamefont {{Peacock}}}, \bibinfo {author} {\bibfnamefont {D.~A.}\ \bibnamefont {{Ritchie}}},\ and\ \bibinfo {author} {\bibfnamefont {G.~A.~C.}\ \bibnamefont {{Jones}}},\ }\href {https://doi.org/10.1088/0022-3719/21/8/002} {\bibfield  {journal} {\bibinfo  {journal} {Journal of Physics C Solid State Physics}\ }\textbf {\bibinfo {volume} {21}},\ \bibinfo {pages} {L209} (\bibinfo {year} {1988})}\BibitemShut {NoStop}%
\bibitem [{\citenamefont {Sabater}\ \emph {et~al.}(2018)\citenamefont {Sabater}, \citenamefont {Dednam}, \citenamefont {Calvo}, \citenamefont {Fern\'andez}, \citenamefont {Untiedt},\ and\ \citenamefont {Caturla}}]{Sabater18}%
  \BibitemOpen
  \bibfield  {author} {\bibinfo {author} {\bibfnamefont {C.}~\bibnamefont {Sabater}}, \bibinfo {author} {\bibfnamefont {W.}~\bibnamefont {Dednam}}, \bibinfo {author} {\bibfnamefont {M.~R.}\ \bibnamefont {Calvo}}, \bibinfo {author} {\bibfnamefont {M.~A.}\ \bibnamefont {Fern\'andez}}, \bibinfo {author} {\bibfnamefont {C.}~\bibnamefont {Untiedt}},\ and\ \bibinfo {author} {\bibfnamefont {M.~J.}\ \bibnamefont {Caturla}},\ }\href {https://doi.org/10.1103/PhysRevB.97.075418} {\bibfield  {journal} {\bibinfo  {journal} {Phys. Rev. B}\ }\textbf {\bibinfo {volume} {97}},\ \bibinfo {pages} {075418} (\bibinfo {year} {2018})}\BibitemShut {NoStop}%
\bibitem [{\citenamefont {Muller}, \citenamefont {{van Ruitenbeek}},\ and\ \citenamefont {{de Jongh}}(1992)}]{muller1992}%
  \BibitemOpen
  \bibfield  {author} {\bibinfo {author} {\bibfnamefont {C.}~\bibnamefont {Muller}}, \bibinfo {author} {\bibfnamefont {J.}~\bibnamefont {{van Ruitenbeek}}},\ and\ \bibinfo {author} {\bibfnamefont {L.}~\bibnamefont {{de Jongh}}},\ }\href {https://doi.org/https://doi.org/10.1016/0921-4534(92)90947-B} {\bibfield  {journal} {\bibinfo  {journal} {Physica C: Superconductivity}\ }\textbf {\bibinfo {volume} {191}},\ \bibinfo {pages} {485} (\bibinfo {year} {1992})}\BibitemShut {NoStop}%
\bibitem [{\citenamefont {de~Ara}\ \emph {et~al.}(2025)\citenamefont {de~Ara}, \citenamefont {Olivera}, \citenamefont {Sabater},\ and\ \citenamefont {Untiedt}}]{deAra2025measurement}%
  \BibitemOpen
  \bibfield  {author} {\bibinfo {author} {\bibfnamefont {T.}~\bibnamefont {de~Ara}}, \bibinfo {author} {\bibfnamefont {B.}~\bibnamefont {Olivera}}, \bibinfo {author} {\bibfnamefont {C.}~\bibnamefont {Sabater}},\ and\ \bibinfo {author} {\bibfnamefont {C.}~\bibnamefont {Untiedt}},\ }\href {https://arxiv.org/abs/2510.05866} {\bibfield  {journal} {\bibinfo  {journal} {arXiv preprint arXiv:2510.05866}\ } (\bibinfo {year} {2025})},\ \Eprint {https://arxiv.org/abs/2510.05866} {arXiv:2510.05866 [cond-mat.mes-hall]} \BibitemShut {NoStop}%
\bibitem [{\citenamefont {Reed}\ \emph {et~al.}(1997)\citenamefont {Reed}, \citenamefont {Zhou}, \citenamefont {Muller}, \citenamefont {Burgin},\ and\ \citenamefont {Tour}}]{Reed97}%
  \BibitemOpen
  \bibfield  {author} {\bibinfo {author} {\bibfnamefont {M.~A.}\ \bibnamefont {Reed}}, \bibinfo {author} {\bibfnamefont {C.}~\bibnamefont {Zhou}}, \bibinfo {author} {\bibfnamefont {C.~J.}\ \bibnamefont {Muller}}, \bibinfo {author} {\bibfnamefont {T.~P.}\ \bibnamefont {Burgin}},\ and\ \bibinfo {author} {\bibfnamefont {J.~M.}\ \bibnamefont {Tour}},\ }\href {https://doi.org/10.1126/science.278.5336.252} {\bibfield  {journal} {\bibinfo  {journal} {Science}\ }\textbf {\bibinfo {volume} {278}},\ \bibinfo {pages} {252} (\bibinfo {year} {1997})}\BibitemShut {NoStop}%
\bibitem [{\citenamefont {Xu}\ and\ \citenamefont {Tao}(2003)}]{xu2003}%
  \BibitemOpen
  \bibfield  {author} {\bibinfo {author} {\bibfnamefont {B.}~\bibnamefont {Xu}}\ and\ \bibinfo {author} {\bibfnamefont {N.~J.}\ \bibnamefont {Tao}},\ }\href {https://doi.org/10.1126/science.1087481} {\bibfield  {journal} {\bibinfo  {journal} {Science}\ }\textbf {\bibinfo {volume} {301}},\ \bibinfo {pages} {1221} (\bibinfo {year} {2003})},\ \Eprint {https://arxiv.org/abs/https://www.science.org/doi/pdf/10.1126/science.1087481} {https://www.science.org/doi/pdf/10.1126/science.1087481} \BibitemShut {NoStop}%
\bibitem [{\citenamefont {Venkataraman}\ \emph {et~al.}(2006)\citenamefont {Venkataraman}, \citenamefont {Klare}, \citenamefont {Tam}, \citenamefont {Nuckolls}, \citenamefont {Hybertsen},\ and\ \citenamefont {Steigerwald}}]{Venkataraman2006}%
  \BibitemOpen
  \bibfield  {author} {\bibinfo {author} {\bibfnamefont {L.}~\bibnamefont {Venkataraman}}, \bibinfo {author} {\bibfnamefont {J.~E.}\ \bibnamefont {Klare}}, \bibinfo {author} {\bibfnamefont {I.~W.}\ \bibnamefont {Tam}}, \bibinfo {author} {\bibfnamefont {C.}~\bibnamefont {Nuckolls}}, \bibinfo {author} {\bibfnamefont {M.~S.}\ \bibnamefont {Hybertsen}},\ and\ \bibinfo {author} {\bibfnamefont {M.~L.}\ \bibnamefont {Steigerwald}},\ }\href {https://doi.org/10.1021/nl052373+} {\bibfield  {journal} {\bibinfo  {journal} {Nano Letters}\ }\textbf {\bibinfo {volume} {6}},\ \bibinfo {pages} {458} (\bibinfo {year} {2006})}\BibitemShut {NoStop}%
\bibitem [{\citenamefont {de~Ara}\ \emph {et~al.}(2024)\citenamefont {de~Ara}, \citenamefont {Hsu}, \citenamefont {Martinez-Garcia}, \citenamefont {Baciu}, \citenamefont {Bronk}, \citenamefont {Ornago}, \citenamefont {van~der Poel}, \citenamefont {Lombardi}, \citenamefont {Guijarro}, \citenamefont {Sabater}, \citenamefont {Untiedt},\ and\ \citenamefont {van~der Zant}}]{Tamara24}%
  \BibitemOpen
  \bibfield  {author} {\bibinfo {author} {\bibfnamefont {T.}~\bibnamefont {de~Ara}}, \bibinfo {author} {\bibfnamefont {C.}~\bibnamefont {Hsu}}, \bibinfo {author} {\bibfnamefont {A.}~\bibnamefont {Martinez-Garcia}}, \bibinfo {author} {\bibfnamefont {B.~C.}\ \bibnamefont {Baciu}}, \bibinfo {author} {\bibfnamefont {P.~J.}\ \bibnamefont {Bronk}}, \bibinfo {author} {\bibfnamefont {L.}~\bibnamefont {Ornago}}, \bibinfo {author} {\bibfnamefont {S.}~\bibnamefont {van~der Poel}}, \bibinfo {author} {\bibfnamefont {E.~B.}\ \bibnamefont {Lombardi}}, \bibinfo {author} {\bibfnamefont {A.}~\bibnamefont {Guijarro}}, \bibinfo {author} {\bibfnamefont {C.}~\bibnamefont {Sabater}}, \bibinfo {author} {\bibfnamefont {C.}~\bibnamefont {Untiedt}},\ and\ \bibinfo {author} {\bibfnamefont {H.~S.~J.}\ \bibnamefont {van~der Zant}},\ }\href {https://doi.org/10.1021/acs.jpclett.4c01425} {\bibfield  {journal} {\bibinfo  {journal} {The Journal of Physical Chemistry Letters}\ }\textbf {\bibinfo {volume} {15}},\ \bibinfo {pages} {8343}
  (\bibinfo {year} {2024})},\ \bibinfo {note} {pMID: 39110695},\ \Eprint {https://arxiv.org/abs/https://doi.org/10.1021/acs.jpclett.4c01425} {https://doi.org/10.1021/acs.jpclett.4c01425} \BibitemShut {NoStop}%
\bibitem [{\citenamefont {Hines}\ \emph {et~al.}(2010)\citenamefont {Hines}, \citenamefont {Diez-Perez}, \citenamefont {Hihath}, \citenamefont {Liu}, \citenamefont {Bogani}, \citenamefont {Zhao}, \citenamefont {Brunschwig}, \citenamefont {M{\"u}llen},\ and\ \citenamefont {Tao}}]{Hines2010}%
  \BibitemOpen
  \bibfield  {author} {\bibinfo {author} {\bibfnamefont {T.}~\bibnamefont {Hines}}, \bibinfo {author} {\bibfnamefont {I.}~\bibnamefont {Diez-Perez}}, \bibinfo {author} {\bibfnamefont {J.}~\bibnamefont {Hihath}}, \bibinfo {author} {\bibfnamefont {H.}~\bibnamefont {Liu}}, \bibinfo {author} {\bibfnamefont {L.}~\bibnamefont {Bogani}}, \bibinfo {author} {\bibfnamefont {Y.}~\bibnamefont {Zhao}}, \bibinfo {author} {\bibfnamefont {B.~S.}\ \bibnamefont {Brunschwig}}, \bibinfo {author} {\bibfnamefont {K.}~\bibnamefont {M{\"u}llen}},\ and\ \bibinfo {author} {\bibfnamefont {N.~J.}\ \bibnamefont {Tao}},\ }\href@noop {} {\bibfield  {journal} {\bibinfo  {journal} {Journal of the American Chemical Society}\ }\textbf {\bibinfo {volume} {132}},\ \bibinfo {pages} {11658} (\bibinfo {year} {2010})}\BibitemShut {NoStop}%
\bibitem [{\citenamefont {Czyszczon-Burton}\ \emph {et~al.}(2024)\citenamefont {Czyszczon-Burton}, \citenamefont {Montes}, \citenamefont {Prana}, \citenamefont {Lazar}, \citenamefont {Rotthowe}, \citenamefont {Chen}, \citenamefont {Vázquez},\ and\ \citenamefont {Inkpen}}]{LathaLow24}%
  \BibitemOpen
  \bibfield  {author} {\bibinfo {author} {\bibfnamefont {T.~M.}\ \bibnamefont {Czyszczon-Burton}}, \bibinfo {author} {\bibfnamefont {E.}~\bibnamefont {Montes}}, \bibinfo {author} {\bibfnamefont {J.}~\bibnamefont {Prana}}, \bibinfo {author} {\bibfnamefont {S.}~\bibnamefont {Lazar}}, \bibinfo {author} {\bibfnamefont {N.}~\bibnamefont {Rotthowe}}, \bibinfo {author} {\bibfnamefont {S.~F.}\ \bibnamefont {Chen}}, \bibinfo {author} {\bibfnamefont {H.}~\bibnamefont {Vázquez}},\ and\ \bibinfo {author} {\bibfnamefont {M.~S.}\ \bibnamefont {Inkpen}},\ }\href {https://doi.org/10.1021/jacs.4c11241} {\bibfield  {journal} {\bibinfo  {journal} {Journal of the American Chemical Society}\ }\textbf {\bibinfo {volume} {146}},\ \bibinfo {pages} {28516} (\bibinfo {year} {2024})},\ \bibinfo {note} {pMID: 39364997}\BibitemShut {NoStop}%
\bibitem [{\citenamefont {Smit}\ \emph {et~al.}(2002)\citenamefont {Smit}, \citenamefont {Noat}, \citenamefont {Untiedt}, \citenamefont {Lang}, \citenamefont {van Hemert},\ and\ \citenamefont {van Ruitenbeek}}]{Smit2002}%
  \BibitemOpen
  \bibfield  {author} {\bibinfo {author} {\bibfnamefont {R.~H.~M.}\ \bibnamefont {Smit}}, \bibinfo {author} {\bibfnamefont {Y.}~\bibnamefont {Noat}}, \bibinfo {author} {\bibfnamefont {C.}~\bibnamefont {Untiedt}}, \bibinfo {author} {\bibfnamefont {N.~D.}\ \bibnamefont {Lang}}, \bibinfo {author} {\bibfnamefont {M.~C.}\ \bibnamefont {van Hemert}},\ and\ \bibinfo {author} {\bibfnamefont {J.~M.}\ \bibnamefont {van Ruitenbeek}},\ }\href {https://doi.org/10.1038/nature01103} {\bibfield  {journal} {\bibinfo  {journal} {Nature}\ }\textbf {\bibinfo {volume} {419}},\ \bibinfo {pages} {906} (\bibinfo {year} {2002})}\BibitemShut {NoStop}%
\bibitem [{\citenamefont {Kiguchi}\ \emph {et~al.}(2008)\citenamefont {Kiguchi}, \citenamefont {Tal}, \citenamefont {Wohlthat}, \citenamefont {Pauly}, \citenamefont {Krieger}, \citenamefont {Djukic}, \citenamefont {Cuevas},\ and\ \citenamefont {van Ruitenbeek}}]{Kiguchi08}%
  \BibitemOpen
  \bibfield  {author} {\bibinfo {author} {\bibfnamefont {M.}~\bibnamefont {Kiguchi}}, \bibinfo {author} {\bibfnamefont {O.}~\bibnamefont {Tal}}, \bibinfo {author} {\bibfnamefont {S.}~\bibnamefont {Wohlthat}}, \bibinfo {author} {\bibfnamefont {F.}~\bibnamefont {Pauly}}, \bibinfo {author} {\bibfnamefont {M.}~\bibnamefont {Krieger}}, \bibinfo {author} {\bibfnamefont {D.}~\bibnamefont {Djukic}}, \bibinfo {author} {\bibfnamefont {J.~C.}\ \bibnamefont {Cuevas}},\ and\ \bibinfo {author} {\bibfnamefont {J.~M.}\ \bibnamefont {van Ruitenbeek}},\ }\href {https://doi.org/10.1103/PhysRevLett.101.046801} {\bibfield  {journal} {\bibinfo  {journal} {Phys. Rev. Lett.}\ }\textbf {\bibinfo {volume} {101}},\ \bibinfo {pages} {046801} (\bibinfo {year} {2008})}\BibitemShut {NoStop}%
\bibitem [{\citenamefont {Yelin}\ \emph {et~al.}(2016)\citenamefont {Yelin}, \citenamefont {Koryt{\'a}r}, \citenamefont {Sukenik}, \citenamefont {Vardimon}, \citenamefont {Kumar}, \citenamefont {Nuckolls}, \citenamefont {Evers},\ and\ \citenamefont {Tal}}]{Yelin2016}%
  \BibitemOpen
  \bibfield  {author} {\bibinfo {author} {\bibfnamefont {T.}~\bibnamefont {Yelin}}, \bibinfo {author} {\bibfnamefont {R.}~\bibnamefont {Koryt{\'a}r}}, \bibinfo {author} {\bibfnamefont {N.}~\bibnamefont {Sukenik}}, \bibinfo {author} {\bibfnamefont {R.}~\bibnamefont {Vardimon}}, \bibinfo {author} {\bibfnamefont {B.}~\bibnamefont {Kumar}}, \bibinfo {author} {\bibfnamefont {C.}~\bibnamefont {Nuckolls}}, \bibinfo {author} {\bibfnamefont {F.}~\bibnamefont {Evers}},\ and\ \bibinfo {author} {\bibfnamefont {O.}~\bibnamefont {Tal}},\ }\href {https://doi.org/10.1038/nmat4552} {\bibfield  {journal} {\bibinfo  {journal} {Nature Materials}\ }\textbf {\bibinfo {volume} {15}},\ \bibinfo {pages} {444} (\bibinfo {year} {2016})}\BibitemShut {NoStop}%
\bibitem [{\citenamefont {Deng}\ \emph {et~al.}(2024)\citenamefont {Deng}, \citenamefont {González}, \citenamefont {Zhu}, \citenamefont {Anderson},\ and\ \citenamefont {Leary}}]{Edmund24}%
  \BibitemOpen
  \bibfield  {author} {\bibinfo {author} {\bibfnamefont {J.-R.}\ \bibnamefont {Deng}}, \bibinfo {author} {\bibfnamefont {M.~T.}\ \bibnamefont {González}}, \bibinfo {author} {\bibfnamefont {H.}~\bibnamefont {Zhu}}, \bibinfo {author} {\bibfnamefont {H.~L.}\ \bibnamefont {Anderson}},\ and\ \bibinfo {author} {\bibfnamefont {E.}~\bibnamefont {Leary}},\ }\href {https://doi.org/10.1021/jacs.3c07734} {\bibfield  {journal} {\bibinfo  {journal} {Journal of the American Chemical Society}\ }\textbf {\bibinfo {volume} {146}},\ \bibinfo {pages} {3651} (\bibinfo {year} {2024})},\ \bibinfo {note} {pMID: 38301131},\ \Eprint {https://arxiv.org/abs/https://doi.org/10.1021/jacs.3c07734} {https://doi.org/10.1021/jacs.3c07734} \BibitemShut {NoStop}%
\bibitem [{\citenamefont {{FEMTO Messtechnik GmbH}}(2025)}]{FEMTO}%
  \BibitemOpen
  \bibfield  {author} {\bibinfo {author} {\bibnamefont {{FEMTO Messtechnik GmbH}}},\ }\href {https://www.femto.de/en/current-transimpedance-amplifiers} {\enquote {\bibinfo {title} {Current amplifiers},}\ } (\bibinfo {year} {2025}),\ \bibinfo {note} {accessed: 2026-03-22}\BibitemShut {NoStop}%
\bibitem [{\citenamefont {{Stanford Research Systems}}(2025)}]{SRSSR570}%
  \BibitemOpen
  \bibfield  {author} {\bibinfo {author} {\bibnamefont {{Stanford Research Systems}}},\ }\href {https://www.thinksrs.com/products/sr570.html} {\emph {\bibinfo {title} {SR570 Low-Noise Current Preamplifier}}} (\bibinfo {year} {2025}),\ \bibinfo {note} {accessed: 2026-02-21}\BibitemShut {NoStop}%
\bibitem [{\citenamefont {{Keithley Instruments}}(2024)}]{Keithley428}%
  \BibitemOpen
  \bibfield  {author} {\bibinfo {author} {\bibnamefont {{Keithley Instruments}}},\ }\href {https://www.tek.com/en/manual/model-428-prog-programmable-current-amplifier-instruction-manual} {\emph {\bibinfo {title} {Model 428-PROG Programmable Current Amplifier Instruction Manual}}},\ \bibinfo {organization} {Tektronix} (\bibinfo {year} {2024}),\ \bibinfo {note} {accessed: 2026-02-21}\BibitemShut {NoStop}%
\bibitem [{\citenamefont {Yelin}\ \emph {et~al.}(2021)\citenamefont {Yelin}, \citenamefont {Chakrabarti}, \citenamefont {Vilan},\ and\ \citenamefont {Tal}}]{Otal21}%
  \BibitemOpen
  \bibfield  {author} {\bibinfo {author} {\bibfnamefont {T.}~\bibnamefont {Yelin}}, \bibinfo {author} {\bibfnamefont {S.}~\bibnamefont {Chakrabarti}}, \bibinfo {author} {\bibfnamefont {A.}~\bibnamefont {Vilan}},\ and\ \bibinfo {author} {\bibfnamefont {O.}~\bibnamefont {Tal}},\ }\href {https://doi.org/10.1039/D1NR05680H} {\bibfield  {journal} {\bibinfo  {journal} {Nanoscale}\ }\textbf {\bibinfo {volume} {13}},\ \bibinfo {pages} {18434} (\bibinfo {year} {2021})}\BibitemShut {NoStop}%
\bibitem [{\citenamefont {Kumar}, \citenamefont {Sethu},\ and\ \citenamefont {van Ruitenbeek}(2015)}]{Kumar15}%
  \BibitemOpen
  \bibfield  {author} {\bibinfo {author} {\bibfnamefont {M.}~\bibnamefont {Kumar}}, \bibinfo {author} {\bibfnamefont {K.~K.~V.}\ \bibnamefont {Sethu}},\ and\ \bibinfo {author} {\bibfnamefont {J.~M.}\ \bibnamefont {van Ruitenbeek}},\ }\href {https://doi.org/10.1103/PhysRevB.91.245404} {\bibfield  {journal} {\bibinfo  {journal} {Phys. Rev. B}\ }\textbf {\bibinfo {volume} {91}},\ \bibinfo {pages} {245404} (\bibinfo {year} {2015})}\BibitemShut {NoStop}%
\bibitem [{\citenamefont {Horowitz}\ and\ \citenamefont {Hill}(2015)}]{horowitz2015art}%
  \BibitemOpen
  \bibfield  {author} {\bibinfo {author} {\bibfnamefont {P.}~\bibnamefont {Horowitz}}\ and\ \bibinfo {author} {\bibfnamefont {W.}~\bibnamefont {Hill}},\ }\href@noop {} {\emph {\bibinfo {title} {The Art of Electronics}}},\ \bibinfo {edition} {3rd}\ ed.\ (\bibinfo  {publisher} {Cambridge University Press},\ \bibinfo {year} {2015})\BibitemShut {NoStop}%
\bibitem [{\citenamefont {Moore}\ \emph {et~al.}(2009)\citenamefont {Moore}, \citenamefont {Davis}, \citenamefont {Coplan},\ and\ \citenamefont {Greer}}]{moore2009building}%
  \BibitemOpen
  \bibfield  {author} {\bibinfo {author} {\bibfnamefont {J.~H.}\ \bibnamefont {Moore}}, \bibinfo {author} {\bibfnamefont {C.~C.}\ \bibnamefont {Davis}}, \bibinfo {author} {\bibfnamefont {M.~A.}\ \bibnamefont {Coplan}},\ and\ \bibinfo {author} {\bibfnamefont {S.~C.}\ \bibnamefont {Greer}},\ }\href@noop {} {\emph {\bibinfo {title} {Building Scientific Apparatus}}},\ \bibinfo {edition} {4th}\ ed.\ (\bibinfo  {publisher} {Cambridge University Press},\ \bibinfo {address} {Cambridge},\ \bibinfo {year} {2009})\BibitemShut {NoStop}%
\bibitem [{\citenamefont {Ornago}(2023)}]{Ornago2023}%
  \BibitemOpen
  \bibfield  {author} {\bibinfo {author} {\bibfnamefont {L.}~\bibnamefont {Ornago}},\ }\emph {\bibinfo {title} {Complexity of Electron Transport in Nanoscale Molecular Junctions}},\ \href {https://research.tudelft.nl/en/publications/complexity-of-electron-transport-in-nanoscale-molecular-junctions} {\bibinfo {type} {Dissertation (tu delft)}},\ \bibinfo  {school} {Delft University of Technology}, \bibinfo {address} {Delft, Netherlands} (\bibinfo {year} {2023}),\ \bibinfo {note} {supervisors: H.S.J. van der Zant, F.C. Grozema}\BibitemShut {NoStop}%
\bibitem [{\citenamefont {Pellicer}\ and\ \citenamefont {Sabater}(2025)}]{Pellicer}%
  \BibitemOpen
  \bibfield  {author} {\bibinfo {author} {\bibfnamefont {G.}~\bibnamefont {Pellicer}}\ and\ \bibinfo {author} {\bibfnamefont {C.}~\bibnamefont {Sabater}},\ }\href {https://www.mdpi.com/2227-9717/13/7/2061} {\bibfield  {journal} {\bibinfo  {journal} {Processes}\ }\textbf {\bibinfo {volume} {13}} (\bibinfo {year} {2025})}\BibitemShut {NoStop}%
\bibitem [{\citenamefont {{de Ara}}\ \emph {et~al.}(2022)\citenamefont {{de Ara}}, \citenamefont {Sabater}, \citenamefont {Borja-Espinosa}, \citenamefont {Ferrer-Alcaraz}, \citenamefont {Baciu}, \citenamefont {Guijarro},\ and\ \citenamefont {Untiedt}}]{DeAra22}%
  \BibitemOpen
  \bibfield  {author} {\bibinfo {author} {\bibfnamefont {T.}~\bibnamefont {{de Ara}}}, \bibinfo {author} {\bibfnamefont {C.}~\bibnamefont {Sabater}}, \bibinfo {author} {\bibfnamefont {C.}~\bibnamefont {Borja-Espinosa}}, \bibinfo {author} {\bibfnamefont {P.}~\bibnamefont {Ferrer-Alcaraz}}, \bibinfo {author} {\bibfnamefont {B.~C.}\ \bibnamefont {Baciu}}, \bibinfo {author} {\bibfnamefont {A.}~\bibnamefont {Guijarro}},\ and\ \bibinfo {author} {\bibfnamefont {C.}~\bibnamefont {Untiedt}},\ }\href {https://doi.org/https://doi.org/10.1016/j.matchemphys.2022.126645} {\bibfield  {journal} {\bibinfo  {journal} {Materials Chemistry and Physics}\ }\textbf {\bibinfo {volume} {291}},\ \bibinfo {pages} {126645} (\bibinfo {year} {2022})}\BibitemShut {NoStop}%
\bibitem [{\citenamefont {Sabater}\ \emph {et~al.}(2012)\citenamefont {Sabater}, \citenamefont {Untiedt}, \citenamefont {Palacios},\ and\ \citenamefont {Caturla}}]{Sabater12}%
  \BibitemOpen
  \bibfield  {author} {\bibinfo {author} {\bibfnamefont {C.}~\bibnamefont {Sabater}}, \bibinfo {author} {\bibfnamefont {C.}~\bibnamefont {Untiedt}}, \bibinfo {author} {\bibfnamefont {J.~J.}\ \bibnamefont {Palacios}},\ and\ \bibinfo {author} {\bibfnamefont {M.~J.}\ \bibnamefont {Caturla}},\ }\href {https://doi.org/10.1103/PhysRevLett.108.205502} {\bibfield  {journal} {\bibinfo  {journal} {Phys. Rev. Lett.}\ }\textbf {\bibinfo {volume} {108}},\ \bibinfo {pages} {205502} (\bibinfo {year} {2012})}\BibitemShut {NoStop}%
\bibitem [{\citenamefont {Pan}\ \emph {et~al.}(2023)\citenamefont {Pan}, \citenamefont {Matthews}, \citenamefont {Lawson},\ and\ \citenamefont {Kamenetska}}]{Kamenetska23}%
  \BibitemOpen
  \bibfield  {author} {\bibinfo {author} {\bibfnamefont {X.}~\bibnamefont {Pan}}, \bibinfo {author} {\bibfnamefont {K.}~\bibnamefont {Matthews}}, \bibinfo {author} {\bibfnamefont {B.}~\bibnamefont {Lawson}},\ and\ \bibinfo {author} {\bibfnamefont {M.}~\bibnamefont {Kamenetska}},\ }\href {https://doi.org/10.1021/acs.jpclett.3c02172} {\bibfield  {journal} {\bibinfo  {journal} {The Journal of Physical Chemistry Letters}\ }\textbf {\bibinfo {volume} {14}},\ \bibinfo {pages} {8327} (\bibinfo {year} {2023})},\ \bibinfo {note} {pMID: 37695735},\ \Eprint {https://arxiv.org/abs/https://doi.org/10.1021/acs.jpclett.3c02172} {https://doi.org/10.1021/acs.jpclett.3c02172} \BibitemShut {NoStop}%
\bibitem [{\citenamefont {Pascual}\ \emph {et~al.}(1993)\citenamefont {Pascual}, \citenamefont {M\'endez}, \citenamefont {G\'omez-Herrero}, \citenamefont {Bar\'o}, \citenamefont {Garc\'{\i}a},\ and\ \citenamefont {Binh}}]{Pascual1993}%
  \BibitemOpen
  \bibfield  {author} {\bibinfo {author} {\bibfnamefont {J.~I.}\ \bibnamefont {Pascual}}, \bibinfo {author} {\bibfnamefont {J.}~\bibnamefont {M\'endez}}, \bibinfo {author} {\bibfnamefont {J.}~\bibnamefont {G\'omez-Herrero}}, \bibinfo {author} {\bibfnamefont {A.~M.}\ \bibnamefont {Bar\'o}}, \bibinfo {author} {\bibfnamefont {N.}~\bibnamefont {Garc\'{\i}a}},\ and\ \bibinfo {author} {\bibfnamefont {V.~T.}\ \bibnamefont {Binh}},\ }\href {https://doi.org/10.1103/PhysRevLett.71.1852} {\bibfield  {journal} {\bibinfo  {journal} {Phys. Rev. Lett.}\ }\textbf {\bibinfo {volume} {71}},\ \bibinfo {pages} {1852} (\bibinfo {year} {1993})}\BibitemShut {NoStop}%
\bibitem [{\citenamefont {Krans}\ \emph {et~al.}(1993)\citenamefont {Krans}, \citenamefont {Muller}, \citenamefont {Yanson}, \citenamefont {Govaert}, \citenamefont {Hesper},\ and\ \citenamefont {van Ruitenbeek}}]{Krans93}%
  \BibitemOpen
  \bibfield  {author} {\bibinfo {author} {\bibfnamefont {J.~M.}\ \bibnamefont {Krans}}, \bibinfo {author} {\bibfnamefont {C.~J.}\ \bibnamefont {Muller}}, \bibinfo {author} {\bibfnamefont {I.~K.}\ \bibnamefont {Yanson}}, \bibinfo {author} {\bibfnamefont {T.~C.~M.}\ \bibnamefont {Govaert}}, \bibinfo {author} {\bibfnamefont {R.}~\bibnamefont {Hesper}},\ and\ \bibinfo {author} {\bibfnamefont {J.~M.}\ \bibnamefont {van Ruitenbeek}},\ }\href {https://doi.org/10.1103/PhysRevB.48.14721} {\bibfield  {journal} {\bibinfo  {journal} {Phys. Rev. B}\ }\textbf {\bibinfo {volume} {48}},\ \bibinfo {pages} {14721} (\bibinfo {year} {1993})}\BibitemShut {NoStop}%
\bibitem [{\citenamefont {Krans}, \citenamefont {{van Ruitenbeek}},\ and\ \citenamefont {{de Jongh}}(1996)}]{Krans96}%
  \BibitemOpen
  \bibfield  {author} {\bibinfo {author} {\bibfnamefont {J.}~\bibnamefont {Krans}}, \bibinfo {author} {\bibfnamefont {J.}~\bibnamefont {{van Ruitenbeek}}},\ and\ \bibinfo {author} {\bibfnamefont {L.}~\bibnamefont {{de Jongh}}},\ }\href {https://doi.org/https://doi.org/10.1016/0921-4526(95)00601-X} {\bibfield  {journal} {\bibinfo  {journal} {Physica B: Condensed Matter}\ }\textbf {\bibinfo {volume} {218}},\ \bibinfo {pages} {228} (\bibinfo {year} {1996})}\BibitemShut {NoStop}%
\bibitem [{\citenamefont {Instruments}(2024)}]{DatasheetLog104}%
  \BibitemOpen
  \bibfield  {author} {\bibinfo {author} {\bibfnamefont {T.}~\bibnamefont {Instruments}},\ }\href {https://www.ti.com/product/LOG104} {\enquote {\bibinfo {title} {Precision logarithmic and log ratio amplifier with scale factor of 0.5$-$v/decade},}\ }\bibinfo {howpublished} {Data Sheet Log 104 Texas Instruments} (\bibinfo {year} {2024}),\ \bibinfo {note} {consulted 6 october 2025}\BibitemShut {NoStop}%
\bibitem [{\citenamefont {{National Instruments}}(2022)}]{NIPCI6363Specs}%
  \BibitemOpen
  \bibfield  {author} {\bibinfo {author} {\bibnamefont {{National Instruments}}},\ }\href {https://www.ni.com/pdf/manuals/375561j.pdf} {\emph {\bibinfo {title} {NI PCIe-6363/PXIe-6363 Device Specifications}}},\ \bibinfo {organization} {National Instruments},\ \bibinfo {address} {Austin, Texas, USA} (\bibinfo {year} {2022}),\ \bibinfo {note} {16-Bit, 2 MS/s, 32 AI Channels, 4 AO Channels Multifunction IO}\BibitemShut {NoStop}%
\bibitem [{\citenamefont {{National Instruments}}(2016{\natexlab{a}})}]{ni6363}%
  \BibitemOpen
  \bibfield  {author} {\bibinfo {author} {\bibnamefont {{National Instruments}}},\ }\href {https://www.ni.com/pdf/manuals/370784g.pdf} {\emph {\bibinfo {title} {NI X Series Data Acquisition User Manual}}},\ \bibinfo {organization} {National Instruments},\ \bibinfo {address} {Austin, Texas, USA} (\bibinfo {year} {2016}{\natexlab{a}}),\ \bibinfo {note} {accessed: 2026-04-17}\BibitemShut {NoStop}%
\bibitem [{\citenamefont {{National Instruments}}(2007)}]{NIPCI6289Specs}%
  \BibitemOpen
  \bibfield  {author} {\bibinfo {author} {\bibnamefont {{National Instruments}}},\ }\href {https://www.ni.com/pdf/manuals/371290h.pdf} {\emph {\bibinfo {title} {PCI-6289 Specifications}}},\ \bibinfo {organization} {National Instruments},\ \bibinfo {address} {Austin, Texas, USA} (\bibinfo {year} {2007}),\ \bibinfo {note} {18-Bit, 625 kS/s, 32-Channel M Series Multifunction DAQ}\BibitemShut {NoStop}%
\bibitem [{\citenamefont {{National Instruments}}(2016{\natexlab{b}})}]{NIPCI6289Manual}%
  \BibitemOpen
  \bibfield  {author} {\bibinfo {author} {\bibnamefont {{National Instruments}}},\ }\href {https://www.ni.com/pdf/manuals/371022l.pdf} {\emph {\bibinfo {title} {M Series User Manual}}},\ \bibinfo {organization} {National Instruments},\ \bibinfo {address} {Austin, Texas, USA} (\bibinfo {year} {2016}{\natexlab{b}})\BibitemShut {NoStop}%
\bibitem [{\citenamefont {{FEMTO Messtechnik GmbH}}(2024)}]{FemtoDLPCA200}%
  \BibitemOpen
  \bibfield  {author} {\bibinfo {author} {\bibnamefont {{FEMTO Messtechnik GmbH}}},\ }\href {https://www.femto.de/en/current-transimpedance-amplifiers/dc-up-to-500-khz-low-noise-variable-amplification-dlpca/} {\emph {\bibinfo {title} {DLPCA-200 Variable Gain Low Noise Transimpedance Amplifier (Current Amplifier) User Manual}}},\ \bibinfo {organization} {FEMTO Messtechnik GmbH},\ \bibinfo {address} {Berlin, Germany} (\bibinfo {year} {2024}),\ \bibinfo {note} {accessed: February 22, 2026}\BibitemShut {NoStop}%
\bibitem [{\citenamefont {{Vishay Dale}}(2022)}]{Vishay}%
  \BibitemOpen
  \bibfield  {author} {\bibinfo {author} {\bibnamefont {{Vishay Dale}}},\ }\href {https://www.vishay.com/docs/31019/ptf.pdf} {\emph {\bibinfo {title} {PTF Metal Film Resistors, High Stability, High Reliability, Ultra-Precision}}},\ \bibinfo {organization} {Vishay Intertechnology, Inc.},\ \bibinfo {address} {Columbus, NE, USA} (\bibinfo {year} {2022}),\ \bibinfo {note} {document Number: 31019, Rev. 24-Aug-2022}\BibitemShut {NoStop}%
\bibitem [{\citenamefont {{Analog Devices}}(2008)}]{OP27Datasheet}%
  \BibitemOpen
  \bibfield  {author} {\bibinfo {author} {\bibnamefont {{Analog Devices}}},\ }\href {https://www.analog.com/en/products/op27.html} {\emph {\bibinfo {title} {{OP27}: Low Noise, Precision Operational Amplifier}}},\ \bibinfo {organization} {Analog Devices} (\bibinfo {year} {2008}),\ \bibinfo {note} {rev. H}\BibitemShut {NoStop}%
\bibitem [{\citenamefont {{Analog Devices}}(2018)}]{LT1028Datasheet}%
  \BibitemOpen
  \bibfield  {author} {\bibinfo {author} {\bibnamefont {{Analog Devices}}},\ }\href {https://www.analog.com/media/en/technical-documentation/data-sheets/1028fd.pdf} {\emph {\bibinfo {title} {{LT1028}: Ultra Low Noise Precision High Speed Op Amps}}},\ \bibinfo {organization} {Analog Devices} (\bibinfo {year} {2018}),\ \bibinfo {note} {rev. D}\BibitemShut {NoStop}%
\bibitem [{\citenamefont {{National Instruments}}(2023)}]{labview}%
  \BibitemOpen
  \bibfield  {author} {\bibinfo {author} {\bibnamefont {{National Instruments}}},\ }\href {https://www.ni.com/labview} {\emph {\bibinfo {title} {LabVIEW: System Design Software}}},\ \bibinfo {organization} {National Instruments},\ \bibinfo {address} {Austin, TX} (\bibinfo {year} {2023}),\ \bibinfo {note} {versión 2023 Q3}\BibitemShut {NoStop}%
\bibitem [{\citenamefont {Dion}\ \emph {et~al.}(2004)\citenamefont {Dion}, \citenamefont {Rydberg}, \citenamefont {Schr{\"o}der}, \citenamefont {Langreth},\ and\ \citenamefont {Lundqvist}}]{Dion2004}%
  \BibitemOpen
  \bibfield  {author} {\bibinfo {author} {\bibfnamefont {M.}~\bibnamefont {Dion}}, \bibinfo {author} {\bibfnamefont {H.}~\bibnamefont {Rydberg}}, \bibinfo {author} {\bibfnamefont {E.}~\bibnamefont {Schr{\"o}der}}, \bibinfo {author} {\bibfnamefont {D.~C.}\ \bibnamefont {Langreth}},\ and\ \bibinfo {author} {\bibfnamefont {B.~I.}\ \bibnamefont {Lundqvist}},\ }\href {https://doi.org/10.1103/PhysRevLett.92.246401} {\bibfield  {journal} {\bibinfo  {journal} {Physical Review Letters}\ }\textbf {\bibinfo {volume} {92}},\ \bibinfo {pages} {246401} (\bibinfo {year} {2004})}\BibitemShut {NoStop}%
\bibitem [{\citenamefont {Sierra}\ \emph {et~al.}(2021)\citenamefont {Sierra}, \citenamefont {Fabian}, \citenamefont {Kawakami}, \citenamefont {Roche},\ and\ \citenamefont {Valenzuela}}]{Sierra2021}%
  \BibitemOpen
  \bibfield  {author} {\bibinfo {author} {\bibfnamefont {J.~F.}\ \bibnamefont {Sierra}}, \bibinfo {author} {\bibfnamefont {J.}~\bibnamefont {Fabian}}, \bibinfo {author} {\bibfnamefont {R.~K.}\ \bibnamefont {Kawakami}}, \bibinfo {author} {\bibfnamefont {S.}~\bibnamefont {Roche}},\ and\ \bibinfo {author} {\bibfnamefont {S.~O.}\ \bibnamefont {Valenzuela}},\ }\href {https://doi.org/10.1038/s41565-021-00936-x} {\bibfield  {journal} {\bibinfo  {journal} {Nature Nanotechnology}\ }\textbf {\bibinfo {volume} {16}},\ \bibinfo {pages} {925} (\bibinfo {year} {2021})}\BibitemShut {NoStop}%
\bibitem [{\citenamefont {Ahn}, \citenamefont {Choi},\ and\ \citenamefont {Kim}(2020)}]{Ahn2020}%
  \BibitemOpen
  \bibfield  {author} {\bibinfo {author} {\bibfnamefont {S.~J.}\ \bibnamefont {Ahn}}, \bibinfo {author} {\bibfnamefont {W.~S.}\ \bibnamefont {Choi}},\ and\ \bibinfo {author} {\bibfnamefont {K.}~\bibnamefont {Kim}},\ }\href {https://doi.org/10.1038/s41699-020-0152-0} {\bibfield  {journal} {\bibinfo  {journal} {npj 2D Materials and Applications}\ }\textbf {\bibinfo {volume} {4}},\ \bibinfo {pages} {35} (\bibinfo {year} {2020})}\BibitemShut {NoStop}%
\end{thebibliography}%

\clearpage

\setcounter{section}{0}
\setcounter{figure}{0}
\setcounter{table}{0}
\renewcommand{\thesection}{S\arabic{section}}
\renewcommand{\thefigure}{S\arabic{figure}}
\renewcommand{\thetable}{S\Roman{table}}

\begin{center}
    \LARGE \textbf{Supplemental Material}
\end{center}

\subsection{Electronic Diagrams}
\begin{figure*}[htb]
    \centering
    \includegraphics[width=0.9\linewidth]{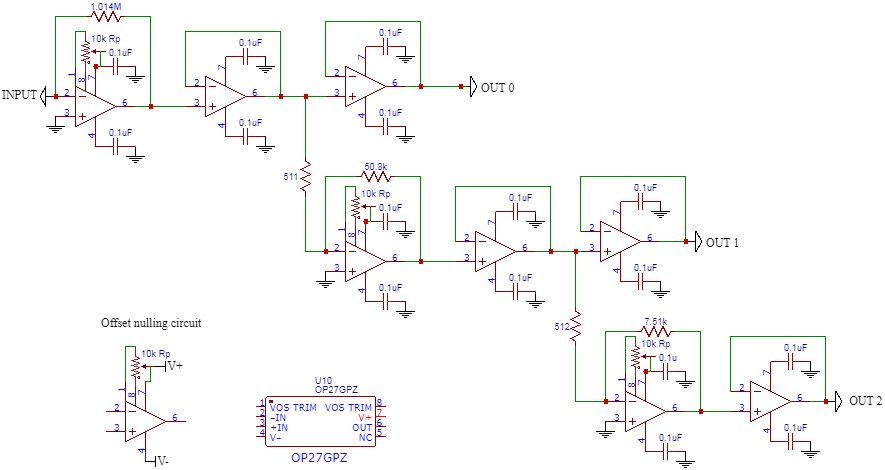}
    \caption{Electronic diagram of the MILAC.}
    \label{fig:3etapas}
\end{figure*}

A detailed electronic schematic of the MILAC architecture is shown in Fig. \ref{fig:3etapas}. The diagram illustrates the three-stage cascading sequence, including the primary transimpedance stage, subsequent voltage amplification blocks, and the signal conditioning components for offset nulling and power supply decoupling.

\subsection{Hardware of the \texorpdfstring{$I\text{--}V$}{I--V} amplifiers}
\begin{figure}[H]
    \centering
    \includegraphics[width=0.99\linewidth]{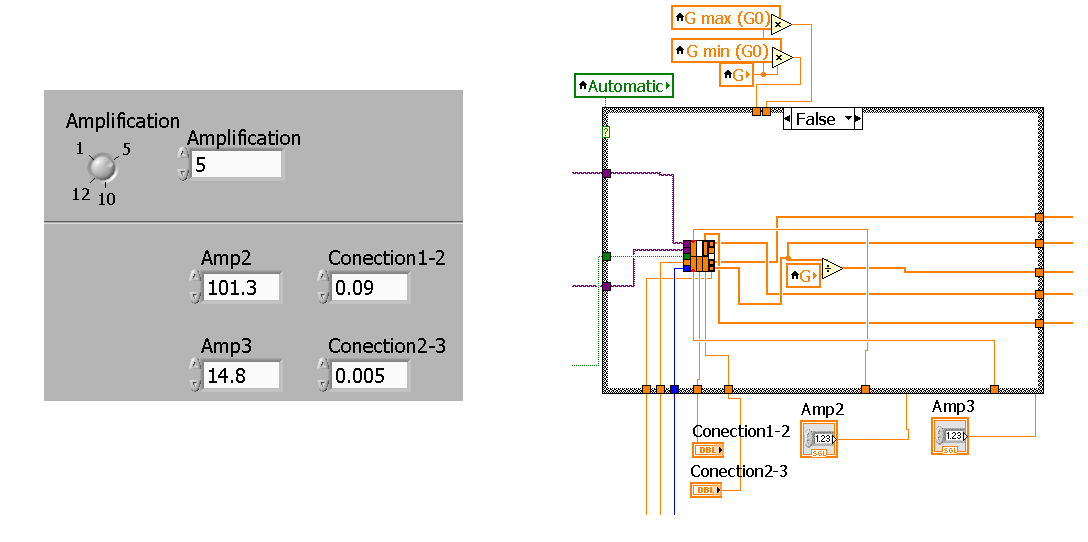}
    \caption{Front Panel and diagram \texttt{LabVIEW }'s sub-Iv that stitches the three segments of the signal.}
    \label{fig:MILCAStich}
\end{figure}

To stitch the three signal segments from the MILAC $V_{out}$ stages, the signals are captured via a DAQ system. The software, developed in LabVIEW, is illustrated in Fig. \ref{fig:MILCAStich}. The front panel allows for real-time gain adjustment based on the resistance ($R_\text{g}$) and provides the parameters necessary to seamlessly join the stages.

\subsection{Conversion \texorpdfstring{$V_\text{out}$}{V\_out} to G}
\begin{figure}[H]
    \centering
    \includegraphics[width=0.99\linewidth]{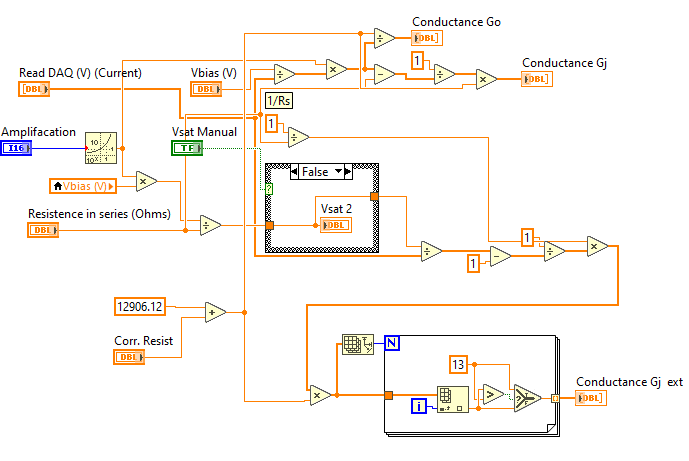}
    \caption{Snapshot of our \texttt{LabVIEW} script used to convert $V_{\text{out}}$ into conductance for the RILA setup.}
    \label{fig:labviewRILA}
\end{figure}

Figure \ref{fig:labviewRILA} shows the \texttt{LabVIEW} diagram used to convert $V_{\text{out}}$ into conductance for the RILA setup. The algorithm, implemented within an N-loop, calculates $G$ in units of $G_{0}$ directly from the voltage signals. This implementation also includes a fine-adjustment parameter for the series resistance ($R_{\text{correct}}$). This allows for in situ corrections of minor decimal variations in the resistor's value, significantly improving measurement resolution.

\subsection{Results}
\begin{figure}[H]
    \centering
    \includegraphics[width=0.99\linewidth]{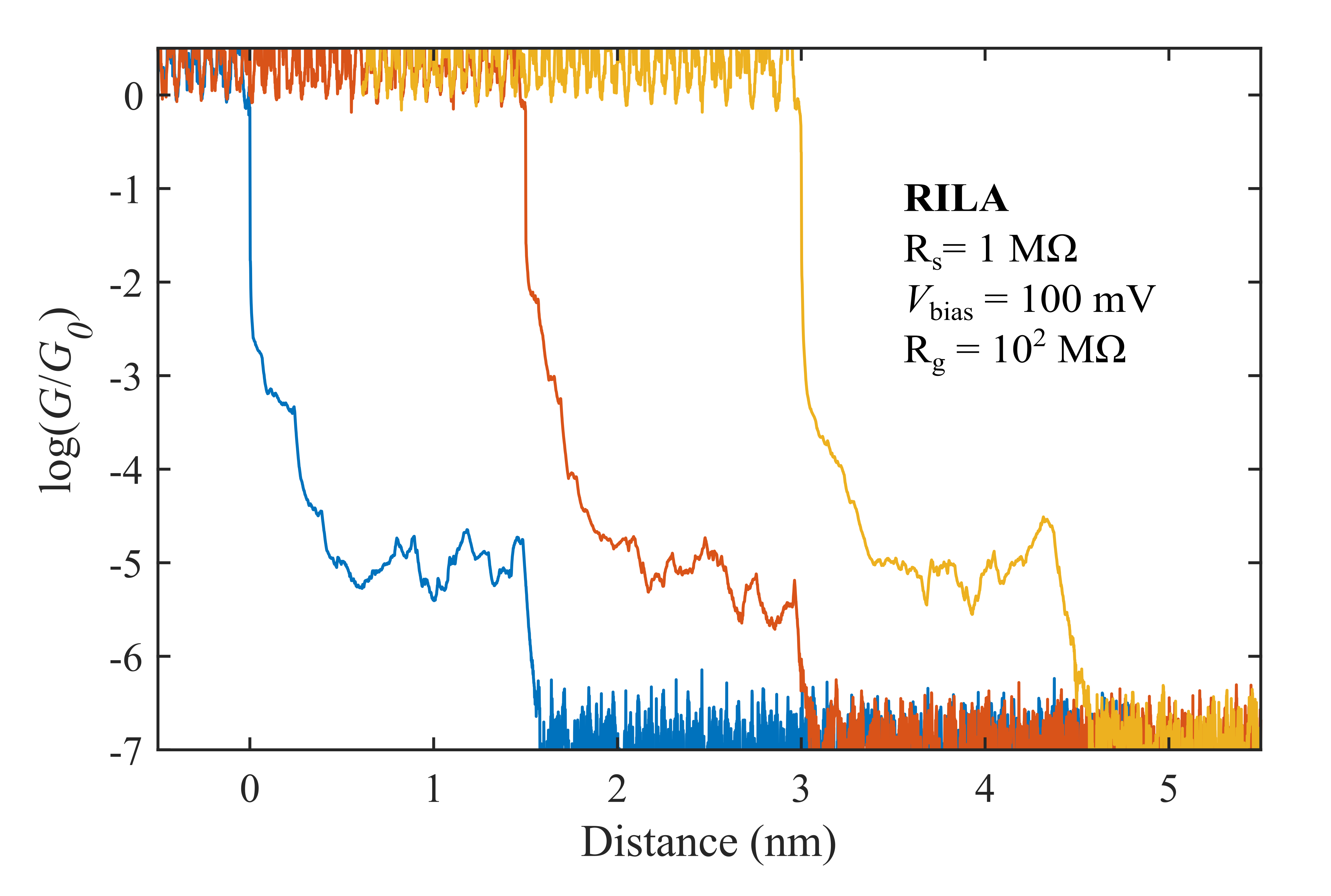}
    \caption{Representative traces measured using the RILA architecture optimized for measuring molecular junctions. Used parameters are shown in the top right part of the panel.}
    \label{fig:RILAUAM}
\end{figure}

As discussed in the general text, optimizing the operational parameters allows for a finer adjustment of the measurement range. For instance, by sacrificing the high-conductance regime, it is possible to improve the resolution in the tunneling region (vacuum conductance). However, this approach leads to a loss of detail in key areas of interest, such as the definition of conductance peaks at $1\,G_\text{0}$ and $2\,G_\text{0}$, which are essential for a broader discussion on molecular electronics. In Fig. \ref{fig:RILAUAM}, three representative traces are shown, demonstrating how the conductance range can be extended down to $10^{-6}\,G_\text{0}$ while maintaining the same $V_\text{bias}$ and $R_\text{s}$ values, provided that the $R_\text{g}$ gain is appropriately tuned.

\newpage
\subsection{Histograms MILAC}

\begin{figure*}[!h]
    \centering
    \includegraphics[width=0.99\linewidth]{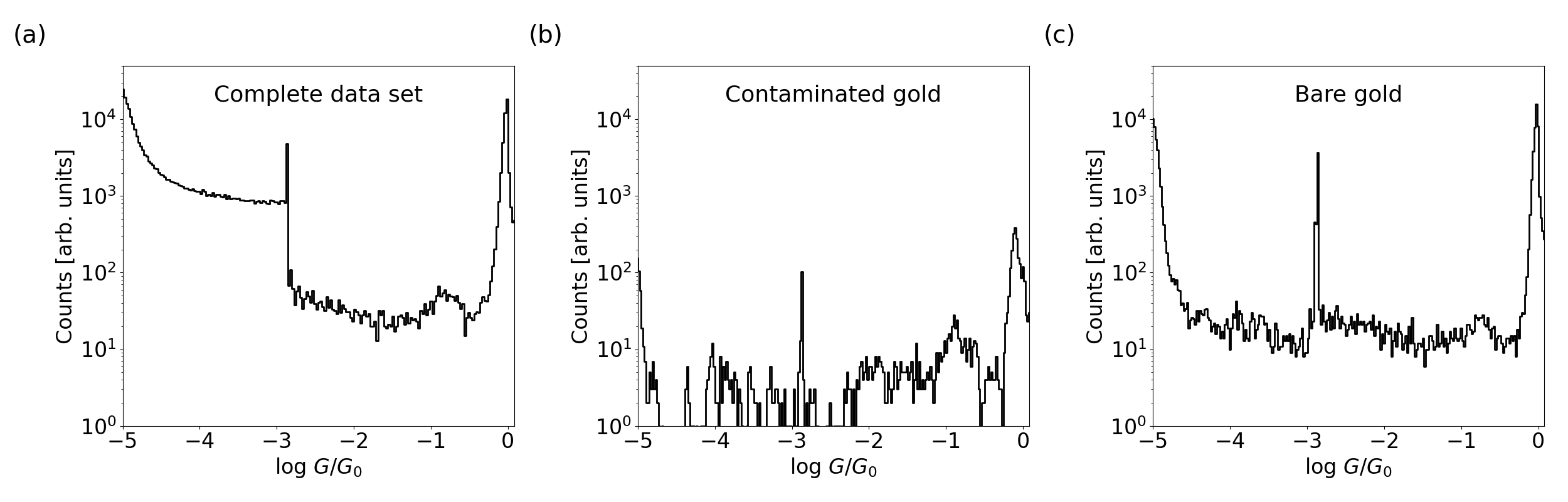}
    \caption{Conductance histograms in scale log-log. (a) Full raw dataset exhibiting capacitance and tunneling contributions that can screen potential contaminants. (b) Histogram of the traces classified as contamination. (c) Standard conductance histogram of clean gold after filtering. All panels exhibit a sharp-line near $10^{-3}\,G_0$ attributed to amplifier-induced overlap.}
    \label{fig:MILACraw}
\end{figure*}

As shown in the MILAC panel of Fig. \ref{fig:Histos}, the presented data correspond to traces that have already been processed using classification and filtering algorithms. The MILAC architecture, being a multi-stage design with high cumulative gain, exhibits an inherent sensitivity to RC time constants derived from parasitic capacitances in combination with the junction's tunneling resistance. Therefore, these effects must be corrected, as using the raw data would result in the histogram shown in Fig. \ref{fig:MILACraw}(a), represented on a log-log scale. In this histogram (Fig. \ref{fig:MILACraw}(a)), a sharp line is observed near $10^{-3}~G_0$, attributed to the merging process of the amplification stages, and another near $10^{-1}~G_0$, indicative of contamination. (See Fig. \ref{fig:MILCAStich} to check the connection overlap points). This region is particularly complex to analyze because the overlap zone of the multi-stage design coincides with typical levels of environmental contamination \cite{Pellicer}. To resolve this ambiguity, we have identified and subtracted the traces that exhibit a plateau around this range. These traces, classified as contaminated, represent approximately 5\% of the total dataset. The histogram in Fig. \ref{fig:MILACraw}(b) exclusively shows the traces identified as environmental contamination. Finally, the resulting histogram of the clean and filtered traces is presented in Fig. \ref{fig:MILACraw}(c), allowing for the isolation of the intrinsic junction behavior from electronic artifacts.

\newpage

\subsection{Tables}
\begin{table*}[!h]
\centering
\scriptsize
\caption{Measured Resistance (in \si{\kilo $\Omega$}) and quantified as conductance (in $G_\text{0}$) for different nominal resistors at varying $R_{\text{ref}}$.}
\label{Table:RRef}
\begin{tabular}{cc *{5}{c}}
\toprule
\multirow{2}{*}{\makecell{\textbf{Nominal} \\ \textbf{Resistance (\si{\kilo $\Omega$})}}} & 
\multirow{2}{*}{\makecell{\textbf{Nominal} \\ \textbf{Conductance (\(G_\text{0}\))}}} & 
\multicolumn{5}{c}{\textbf{Measured Conductance (\(G_\text{0}\)) at Reference Resistance}} \\
\cmidrule(lr){3-7}
& & \textbf{50 \si{\mega$\Omega$}} & \textbf{100 \si{\mega$\Omega$}} & \textbf{150 \si{\mega$\Omega$}} & \textbf{200 \si{\mega$\Omega$}} & \textbf{500 \si{\mega$\Omega$}} \\
\midrule
$(129 \pm 7.0) \times 10^{-3}$ & $100 \pm 6$ & $29.2 \pm 0.2$ & $16.99 \pm 0.03$ & $11.32 \pm 0.07$ & $(8.95 \pm 0.08) \times 10^{2}$ & $(3.63 \pm 0.03) \times 10^{2}$ \\
$1.29 \pm 0.07$ & $10.1 \pm 0.6$ & $10.1 \pm 0.2$ & $8.15 \pm 0.05$ & $6.10 \pm 0.05$ & $5.0 \pm 0.2$ & $2.77 \pm 0.02$ \\
$1.6 \pm 0.1$ & $8.0 \pm 0.5$ & $8.1 \pm 0.2$ & $7.72 \pm 0.06$ & $5.68 \pm 0.03$ & $4.6 \pm 0.3$ & $2.59 \pm 0.02$ \\
$3.3 \pm 0.2$ & $4.0 \pm 0.2$ & $4.04 \pm 0.07$ & $4.04 \pm 0.04$ & $4.05 \pm 0.05$ & $3.85 \pm 0.04$ & $2.01 \pm 0.02$ \\
$6.3 \pm 0.4$ & $2.1 \pm 0.1$ & $2.01 \pm 0.03$ & $2.0 \pm 0.02$ & $2.03 \pm 0.03$ & $2.0 \pm 0.1$ & $1.61 \pm 0.01$ \\
$12.9 \pm 0.8$ & $1.01 \pm 0.06$ & $10.1 \pm 0.02$ & $10.1 \pm 0.01$ & $1.004 \pm 0.009$ & $1.01 \pm 0.05$ & $1.01 \pm 0.04$ \\
$26.0 \pm 1.0$ & $0.50 \pm 0.03$ & $0.503 \pm 0.008$ & $0.503 \pm 0.008$ & $0.501 \pm 0.006$ & $0.51 \pm 0.03$ & $0.5 \pm 0.02$ \\
$39.0 \pm 2.0$ & $0.33 \pm 0.02$ & $0.337 \pm 0.006$ & $0.336 \pm 0.004$ & $0.335 \pm 0.002$ & $0.34 \pm 0.02$ & $0.34 \pm 0.01$ \\
$52.0 \pm 3.0$ & $0.25 \pm 0.01$ & $0.253 \pm 0.004$ & $0.252 \pm 0.02$ & $0.252 \pm 0.04$ & $0.25 \pm 0.02$ & $0.25 \pm 0.01$ \\
$(1.00 \pm 0.03) \times 10^{3}$ & $(1.30 \pm 0.04) \times 10^{-2}$ & $(1.32 \pm 0.02) \times 10^{-2}$ & $(1.31 \pm 0.01) \times 10^{-2}$ & $(1.32 \pm 0.02) \times 10^{-2}$ & $(1.3 \pm 0.1) \times 10^{-2}$ & $(1.29 \pm 0.05) \times 10^{-2}$ \\
$(1.00 \pm 0.03) \times 10^{4}$ & $(1.3 \pm 0.04) \times 10^{-3}$ & -- & $(1.31 \pm 0.02) \times 10^{-3}$ & $(1.30 \pm 0.03) \times 10^{-3}$ & $(1.31 \pm 0.04) \times 10^{-3}$ & $(1.30 \pm 0.03) \times 10^{-3}$ \\
$(1.00 \pm 0.03) \times 10^{5}$ & $(1.3 \pm 0.04) \times 10^{-4}$ & -- & -- & $(1.24 \pm 0.1) \times 10^{-4}$ & $(1.15 \pm 0.4) \times 10^{-4}$ & $(1.2 \pm 0.3) \times 10^{-4}$ \\
\bottomrule
\end{tabular}
\end{table*}

\begin{table*}[h]
\centering
\scriptsize
\caption{Measured Resistance (in \si{\kilo $\Omega$}) and quantified as conductance (in $G_\text{0}$) for different nominal resistors at varying $V_{\text{ref}}.$}
\label{tab:conductance_tableVref}
\begin{tabular}{cc *{6}{c}}
\toprule
\multirow{2}{*}{\makecell{\textbf{Nominal} \\ \textbf{Resistance (\si{\kilo $\Omega$}) }}} & 
\multirow{2}{*}{\makecell{\textbf{Nominal} \\ \textbf{Conductance (\(G_\text{0}\))}}} & 
\multicolumn{6}{c}{\textbf{Measured Conductance (\(G_\text{0}\)) at Reference Voltage}} \\
\cmidrule(lr){3-8}
& & \textbf{0.0001 V} & \textbf{0.001 V} & \textbf{0.01 V} & \textbf{0.1 V} & \textbf{1 V} & \textbf{10 V} \\
\midrule
$(129 \pm 7) \times 10^{-3}$ & $100 \pm 6$ & $(1.76 \pm 0.01) \times 10^{-2}$ & $(1.76 \pm 0.02) \times 10^{-1}$ & $2 \pm 2$ & $16.99 \pm 0.02$ & $95.4 \pm 0.7$ & $103.5 \pm 0.2$ \\
$1.29 \pm 0.07$ & $10.1 \pm 0.6$ & $(1.0 \pm 0.007) \times 10^{-2}$ & $(9.93 \pm 0.08) \times 10^{-2}$ & $9.0 \pm 0.4$ & $8.15 \pm 0.05$ & $10.12 \pm 0.08$ & $10.1 \pm 0.02$ \\
$1.6 \pm 0.1$ & $8.0 \pm 0.5$ & $(9.50 \pm 0.06) \times 10^{-3}$ & $(9.45 \pm 0.08) \times 10^{-2}$ & $9.0 \pm 0.3$ & $7.72 \pm 0.02$ & $8.08 \pm 0.08$ & $8.2 \pm 0.2$ \\
$3.2 \pm 0.2$ & $4.0 \pm 0.2$ & $(8.50 \pm 0.03) \times 10^{-3}$ & $(8.51 \pm 0.06) \times 10^{-2}$ & $8.0 \pm 0.2$ & $4.04 \pm 0.02$ & $4.0 \pm 0.5$ & $4.09 \pm 0.08$ \\
$6.2 \pm 0.4$ & $2.1 \pm 0.1$ & $(8.00 \pm 0.08) \times 10^{-3}$ & $(7.9 \pm 0.2) \times 10^{-2}$ & $(7.8 \pm 0.6) \times 10^{-1}$ & $2.003 \pm 0.024$ & $2.01 \pm 0.01$ & $2.01 \pm 0.08$ \\
$12.9 \pm 0.8$ & $1.01 \pm 0.06$ & $(7.70 \pm 0.07) \times 10^{-3}$ & $(7.7 \pm 0.2) \times 10^{-2}$ & $(7.5 \pm 0.5) \times 10^{-1}$ & $1.01 \pm 0.01$ & $1.02 \pm 0.01$ & $1.03 \pm 0.04$ \\
$26 \pm 1$ & $0.50 \pm 0.03$ & $(7.50 \pm 0.01) \times 10^{-3}$ & $(7.6 \pm 0.6) \times 10^{-2}$ & $0.45 \pm 0.06$ & $0.503 \pm 0.008$ & $0.505 \pm 0.005$ & -- \\
$39 \pm 2$ & $0.33 \pm 0.01$ & $(7.50 \pm 0.02) \times 10^{-3}$ & $(7.6 \pm 0.1) \times 10^{-2}$ & $0.31 \pm 0.04$ & $0.360 \pm 0.004$ & $0.340 \pm 0.003$ & -- \\
$52 \pm 3$ & $0.25 \pm 0.01$ & $(7.50 \pm 0.06) \times 10^{-3}$ & $(7.52 \pm 0.06) \times 10^{-2}$ & $0.23 \pm 0.03$ & $0.252 \pm 0.03$ & $0.253 \pm 0.004$ & -- \\
$(1.00 \pm 0.03) \times 10^{3}$ & $(1.30 \pm 0.04) \times 10^{-2}$ & $(2 \pm 1) \times 10^{-3}$ & $(2 \pm 1) \times 10^{-3}$ & $(1.23 \pm 0.1) \times 10^{-2}$ & $(1.32 \pm 0.01) \times 10^{-2}$ & $(1.32 \pm 0.02) \times 10^{-2}$ & -- \\
$(1.00 \pm 0.03) \times 10^{4}$ & $(1.30 \pm 0.04) \times 10^{-3}$ & $(2 \pm 1) \times 10^{-4}$ & $(8.5 \pm 4) \times 10^{-4}$ & $(1.23 \pm 0.1) \times 10^{-3}$ & $(1.30 \pm 0.02) \times 10^{-3}$ & $(1.31 \pm 0.05) \times 10^{-3}$ & -- \\
$(1.00 \pm 0.03) \times 10^{5}$ & $(1.30 \pm 0.04) \times 10^{-4}$ & -- & -- & $(8 \pm 5) \times 10^{-5}$ & $(1.2 \pm 0.1) \times 10^{-4}$ & -- & -- \\
\bottomrule
\end{tabular}
\end{table*}

\begin{table*}[h]
\centering
\caption{Measured conductance (in units of \(G_\text{0}\)) for each amplifier. Nominal values of resistance and conductance include a \(\pm 5\%\) error margin.}
\label{tab:conductance}
\scriptsize
\begin{tabular}{ccccc}
\toprule
\textbf{Nominal} & \textbf{Nominal} & \multicolumn{3}{c}{\textbf{Measured Conductance (\(G_\text{0}\))}} \\
\cmidrule(lr){3-5}
\textbf{Resistance (k\(\Omega\))} & \textbf{Conductance (\(G_\text{0}\))} & \textbf{ILA} & \textbf{ILOGA} & \textbf{RILA} \\
\midrule
$(129 \pm 7.0) \times 10^{-3}$ & $101 \pm 6$ & $13.5300 \pm 0.0007$ & $13.55 \pm 0.04$ & $10 \pm 80$ \\
$1.29 \pm 0.08$ & $10.1 \pm 0.6$ & $9.54 \pm 0.03$ & $10.17 \pm 0.05$ & $8 \pm 21$ \\
$1.3 \pm 0.1$ & $8.0 \pm 0.5$ & $7.69 \pm 0.02$ & $8.11 \pm 0.04$ & $12 \pm 9$ \\
$3.3 \pm 0.2$ & $4.0 \pm 0.2$ & $3.99 \pm 0.01$ & $4.13 \pm 0.04$ & $4.4 \pm 0.2$ \\
$6.3 \pm 0.4$ & $2.1 \pm 0.1$ & $1.960 \pm 0.006$ & $2.05 \pm 0.01$ & $2.04 \pm 0.05$ \\
$12.9 \pm 0.8$ & $1.01 \pm 0.06$ & $0.990 \pm 0.003$ & $1.01 \pm 0.013$ & $0.97 \pm 0.02$ \\
$26 \pm 2$ & $0.50 \pm 0.03$ & $0.4950 \pm 0.0001$ & $0.514 \pm 0.003$ & $0.490 \pm 0.003$ \\
$39 \pm 2$ & $0.33 \pm 0.01$ & $0.330 \pm 0.001$ & $0.34 \pm 0.01$ & $0.330 \pm 0.001$ \\
$52 \pm 3$ & $0.25 \pm 0.01$ & $0.25 \pm 0.01$ & $0.205 \pm 0.001$ & $0.240 \pm 0.001$ \\
$100 \pm 30$ & $0.0 \pm 4.0$ & $0.124 \pm 0.001$ & $1.310 \pm 0.008$ & $0.1280 \pm 0.0004$ \\
$(1.0 \pm 0.3) \times 10^{3}$ & $(1.30 \pm 0.04) \times 10^{-2}$ & $(1.80 \pm 0.06) \times 10^{-2}$ & $(1.321 \pm 0.008) \times 10^{-2}$ & $(1.300 \pm 0.004) \times 10^{-2}$ \\
$(1.0 \pm 0.3) \times 10^{4}$ & $(1.30 \pm 0.04) \times 10^{-3}$ & $(1.7 \pm 0.7) \times 10^{-3}$ & $(2 \pm 13) \times 10^{-3}$ & $(1.29 \pm 0.03) \times 10^{-3}$ \\
$(1.0 \pm 0.3) \times 10^{5}$ & $(1.30 \pm 0.04) \times 10^{-4}$ & $(2.3 \pm 0.9) \times 10^{-4}$ & $(1.55 \pm 0.01) \times 10^{-4}$ & $(1.3 \pm 0.3) \times 10^{-4}$ \\
$(1.0 \pm 0.3) \times 10^{6}$ & $(1.30 \pm 0.03) \times 10^{-5}$ & -- & $(1.39 \pm 0.03) \times 10^{-5}$ & $(1.39 \pm 2.60) \times 10^{-5}$ \\
$(5.0 \pm 0.1) \times 10^{6}$ & $(2.60 \pm 0.08) \times 10^{-6}$ & -- & $(3 \pm 3) \times 10^{-6}$ & $(3 \pm 30) \times 10^{-6}$ \\
\bottomrule
\end{tabular}
\end{table*}

\end{document}